\documentclass[a4paper,11pt]{article}
\pdfoutput=1 

\usepackage{jheppub} 
\usepackage{bm,amssymb,slashed,graphicx,multirow,soul,mathtools,xspace,array,tikz}
\usepackage[compat=1.1.0]{tikz-feynman}   
\usepackage{float}                   
\allowdisplaybreaks
\usepackage{bbold}
\usepackage{subfigure}
\usepackage{braket}
\usepackage{pdflscape} 
\usepackage{rotating} 

\definecolor{nicered}{rgb}{0.7,0.1,0.1}
\definecolor{nicegreen}{rgb}{0.1,0.5,0.1}
\definecolor{violet}{rgb}{0.7,0.3,0.3}
\hypersetup{colorlinks,citecolor= nicegreen,linkcolor= nicered}

\newcommand{\C}{{\cal C}}
\newcommand{\Q}{{\cal Q}}
\newcommand{\todo}[1]{{\color{red} \ifmmode\else[todo]\fi #1}}

\newcommand{\lp}{\left(}
\newcommand{\rp}{\right)}

\newcommand{\g}{\gamma}

\newcommand{\nn}{\nonumber}

\newcommand {\E}[1]{\times 10^{#1}}	
\newcommand {\e}[1]{\mathrm{~#1}}       
\newcommand{\re}[0]{\mathrm{Re}\,}
\newcommand{\im}[0]{\mathrm{Im}\,}
\newcommand{\mc}[1]{\mathcal{#1}}

\newcommand{\beq}{\begin{equation} }
\newcommand{\eeq}{\end{equation}} 
\newcommand{\bi}{\begin{itemize} }
\newcommand{\ei}{\end{itemize} }

\definecolor{Red}{rgb}{1.,0.,0.}
\definecolor{Grn}{rgb}{0.,0.75,0.}
\definecolor{Blu}{rgb}{0.,0.,1.}
\definecolor{Pink}{rgb}{1,0.08,0.58}
\definecolor{Orange}{rgb}{1., 0.5, 0.0}

\DeclareMathOperator{\diag}{diag}

\let\Im\relax
\DeclareMathOperator{\Im}{Im}

\newcommand{\lrPartial}{\negthickspace\stackrel{\leftrightarrow}{D}\negthickspace{}}
\newcommand{\lPartial}{\negthickspace\stackrel{\leftarrow}{D}\negthickspace{}}

\usepackage[T1]{fontenc} 

\usepackage{amsmath,amssymb,epsfig,ulem,color,slashed}
\allowdisplaybreaks  

\title{\boldmath  New Physics in CP Violating and Flavour Changing Quark Dipole Transitions}

\author[1,2]{Svjetlana Fajfer,}
\author[1,2]{Jernej F. Kamenik,}
\author[1,2]{Nejc Ko\v{s}nik,}
\author[3]{Aleks Smolkovi\v{c},}
\author[1]{Michele Tammaro,}

\affiliation[1]{Jo\v{z}ef Stefan Institute, Jamova 39, 1000 Ljubljana, Slovenia}
\affiliation[2]{Faculty of Mathematics and Physics, University of Ljubljana, Jadranska 19, 1000 Ljubljana, Slovenia}
\affiliation[3]{Albert Einstein Center for Fundamental Physics, Institute for Theoretical Physics, University of Bern, CH-3012 Bern, Switzerland}

\emailAdd{svjetlana.fajfer@ijs.si}
\emailAdd{jernej.kamenik@cern.ch}
\emailAdd{nejc.kosnik@ijs.si}
\emailAdd{smolkovic@itp.unibe.ch}
\emailAdd{michele.tammaro@ijs.si}

\abstract{We explore CP-violating (CPV) effects of heavy New Physics in flavour-changing quark dipole transitions, within the framework of Standard Model Effective Field Theory (SMEFT). First, we establish the relevant dimension six operators and consider the Renormalisation Group (RG) evolution of the appropriate Wilson coefficients. We investigate RG-induced correlations between different flavour-violating processes and electric dipole moments (EDMs) within the Minimal Flavour Violating and $U(2)^3$ quark flavour models. At low energies, we set bounds on the Wilson coefficients of the dipole operators using CPV induced contributions to observables in non-leptonic and radiative $B$, $D$ and $K$ decays as well as the neutron and electron EDMs. This enables us to connect observable CPV effects at low energies and general NP appearing at high scales. We present bounds on the Wilson coefficients of the relevant SMEFT operators at the high scale $\Lambda = 5~{\rm TeV}$, and discuss most sensitive CPV observables for future experimental searches.}

\begin{document} 

\maketitle


\newpage

%
\section{Introduction}
\label{sec:intro}
%

Violation of charge conjugation and parity (CP) is one of the most intriguing phenomena in particle physics. Experimentally to date it has only been observed in transitions among different quark flavours. In the past 60 years since its discovery~\cite{Christenson:1964fg}, many venues to observe CP violation (CPV) have been established in $b$-, $c$-flavoured hadron and kaon  decays as well as in neutral meson oscillations. Theoretically, within the Standard Model (SM) such CPV  originates from the CKM matrix, or more generally, from the quark Yukawa sector. Currently, all most precise experimental measurements of CPV from flavour factories are consistent with SM predictions controlled by a unique CP-odd phase of the CKM, see e.g. Ref.~\cite{Charles:2015gya}.

A second possible source of CP violation in the SM arises from QCD dynamics. This so-called strong CP violation is flavour universal and conventionally parameterized via the QCD $\bar \theta$ term. Current experimental limits on the neutron electric dipole moment (EDM) constrain $|\bar \theta|  \lesssim 10^{-10}$~\cite{Abel:2020pzs}, the smallness of which is difficult to understand and leads to the to the so-called strong CP problem of the SM. 
In the limit of vanishing $\bar \theta$, EDMs still arise in the SM due to the CP-odd phase of the CKM, but are severely suppressed: namely, quark EDMs  receive contributions only at the three-loop level, while in the case of leptons, they only arise at four loops (see e.g. Ref.~\cite{Kley:2021yhn}).
Numerically, the resulting neutron dipole moment was found to be $d_n \sim 10^{-32} \, e \cdot$cm~\cite{Khriplovich:1981ca}, six orders of magnitude below the current experimental limit. 

Beyond the SM, EDM measurements in several distinct systems allow to disentangle nonzero $\bar \theta$ contributions from other possible new physics (NP) sources of CP violation. Similarly, measurements of CP-odd observables in several distinct quark flavour transitions allow to search for new sources of CP violation in flavour beyond the CKM phase~\cite{Bonnefoy:2021tbt, Bakshi:2021ofj}.
However, while contributions to EDMs from flavour non-universal CP-odd phases, and conversely $\bar \theta$ contributions to CP-odd flavour observables, are practically negligible in the SM, this is not necessarily the case beyond SM. Consequently, the interplay between flavour-conserving and flavour-violating CP-odd observables in presence of NP requires specifying the flavour structure of the model (see e.g. Refs.~\cite{Mercolli:2009ns, Keren-Zur:2012buf}).

The SM gauge-kinetic sector is invariant under a global flavour symmetry
\beq\label{eq:flavourGroup}
G_F \equiv U(3)^5 = U(3)_q \times U(3)_u \times U(3)_d \times U(3)_l \times U(3)_e \,,
\eeq 
where  $q, \, u, \, d,\, l, \,e$ denote the left-handed quarks, right-handed up and down quarks and left- and right-handed leptons, respectively. The fermion Yukawa couplings to the Higgs ($Y_{u,d,e}$) act as the only (spurionic) sources of $G_F \to U(1)^4 = U(1)_B \times U(1)_e \times U(1)_\mu \times U(1)_\tau$ breaking in the SM\footnote{Here we are neglecting the breaking of lepton flavour numbers due to small neutrino masses, since they are irrelevant for the observables studied in this work.}. While in general, any flavour structure even beyond SM can be parameterized in terms of a finite power sum of SM Yukawas~\cite{Colangelo:2008qp}, the hierarchical structure of quark masses and nearly diagonal form of the CKM allow for an efficient flavour breaking expansion and power counting. Models beyond SM respecting this power counting, where formally the leading flavour breaking sources are proportional to the lowest powers of the SM Yukawas, are called Minimally Flavour Violating (MFV)~\cite{DAmbrosio:2002vsn}. While such a prescription completely determines the flavour breaking patterns even beyond SM, it does not preclude the appearance of new (flavour universal) sources of CP violation. Consequently, experimental searches for EDMs put strong constraints on such contributions~\cite{DAmbrosio:2002vsn, Kley:2021yhn}. To mitigate this, conventional definitions of MFV include the requirement/assumption that the SM Yukawas are also the only (spurionic) sources of CP violation even beyond SM, resulting in suppressed contributions to EDMs~\cite{Smith:2017dtz}. 

The idea of controlled spurionic breaking of $G_F$ can be developed beyond MFV and leads to so-called $U(2)^3$ flavour models~\cite{Barbieri:2011ci, Barbieri:2012uh}. These build upon the observation, that the top Yukawa coupling is the only $\mc{O}(1)$ flavour breaking in the SM, leading to a pattern $U(3)_q \times U(3)_u \to U(1)_t \times U(2)_q \times U(2)_u$, which is only further broken by small terms of the order of $|V_{ts}|$, $m_c/m_t$, etc. The hierarchical structure of the CKM furthermore dictates the simplest spurionic structure of such $U(1)_t \times U(2)_q \times U(2)_u$ breaking terms. The main difference of $U(2)^3$ flavour models with respect to MFV (related to both a smaller symmetry and an increased number of spurions) is the breaking of correlations between processes involving the third fermion generation versus those involving only the lighter two, including the appearance of additional non-universal CP-odd phases unrelated to the phase of the CKM. This in turn generally leads to significantly different predictions and correlations between EDMs and flavour violating CP-odd observables. 

Recently the authors in Ref.~\cite{Kley:2021yhn} analyzed EDMs of charged leptons and neutrons in presence of heavy NP within the framework Standard Model Effective Theory (SMEFT) at the one-loop level, including  both renormalization group (RG) running contributions and finite corrections. They considered both flavour universal ($U(3)^3$ symmetric) scenarios as well as MFV and $U(2)^3$ flavour models and derived bounds on the various spurionic contributions.  

In the present work we investigate complementary flavour- and CP-violating effects of heavy NP, and study correlations between different flavour-violating processes and EDMs both within MFV and $U(2)^3$ flavour models. To this end we consider dimension-6 effective operators within SMEFT that can contribute to CPV dipole transition observables at low energies, both via direct operator matching (and matrix element computation) as well as through RG mixing effects. We consider CP-violating observables in non-leptonic and radiative $B$, $D$ and kaon decays as well as the EDM of the electron and of the neutron. 
A particular attention is paid to flavour violating contributions to the latter due to long distance hadronic effects. These were first studied in Ref.~\cite{Mannel:2012qk}, where it was shown that the $c\to u$ chromo-electric dipole operator can induce the neutron EDM at the level of $10^{-31} \, e \cdot$cm via tree-level second-order weak interaction effects. We generalize these results to other flavour- and CP-violating dipole contributions and compute the relevant (heavy) baryon exchange contributions for $c \to u$, $s\to d$ and $b \to d$ transitions.

The purpose of our study is twofold. First, by studying the RG evolution of flavour- and CP-violating operators, we can establish a connection between observable CPV effects at low energies and general NP present at some high scale.
Secondly, our low-energy phenomenological analysis summarizes existing constraints on CPV dipole transition operators. It can serve as guidance for the experimental and theoretical improvements in the future. The paper is structured as follows: in Sec.~\ref{sec:SMEFT} we introduce the SMEFT operator basis relevant for low energy CPV dipole transitions and present the relevant RG evolution equations (RGEs). Sec.~\ref{sec:flavourExpansion} reviews the flavour expansion of SMEFT within MFV and $U(2)^3$ models. In Sec.~\ref{sec:FVandCPVconstraints} we analyse the relevant low energy phenomenology from flavour- and CP-violating dipole transitions.
The resulting constraints on particular SMEFT flavour scenarios are shown in Sec.~\ref{sec:results}. Finally in Sec.~\ref{sec:conclusions} we present our conclusions and prospects for the future.

%
\section{SMEFT Framework}
\label{sec:SMEFT}
%

In this Section we define the general framework for our analysis. First, we write out the relevant operators in the SMEFT basis, then we describe their interplay through RG running and one-loop matching to Low Energy Effective Field Theory (LEFT)~\cite{Jenkins:2017jig} at the weak scale.

\subsection{Operator Basis}
\label{subsec:OperatorBasis}
We work within the context of SMEFT valid at scales below any NP mass thresholds but much above the electroweak symmetry breaking (EWSB) Higgs condensate $v = 246$ GeV. The effective Lagrangian is\footnote{For brevity we omit the single dimension five lepton number violating Weinberg operator, which does not affect our analysis as well as possible hermitian operators ($\Q_i = \Q_i^\dagger$) since they do not contribute to CPV.}
\beq
{\cal L} = {\cal L}_{SM} + 
\frac{1}{\Lambda^2} \sum_{\Q_i \neq \Q_i^\dagger} (\C_i \Q_i  + {\rm h.c.})\,,
\eeq
where ${\cal L}_{SM}$ is the SM Lagrangian, $\Lambda$ is the relevant operator suppression scale, the sums run over the (non-hermitian) local dimension-six operators $\Q_i$, and $\C_i$ are the respective Wilson coefficients. We work with the s.c. ``Warsaw'' basis~\cite{Grzadkowski:2010es}, where leading heavy NP contributions to the quark dipole moments are parameterized by the operators
\begin{align}
	\Q_{\underset{pr}{uG}} &= \lp \bar q_p \sigma_{\mu\nu} T^A u_r \rp \tilde H G^{A\mu\nu}\,, &\quad \Q_{\underset{pr}{dG}} &= \lp \bar q_p \sigma_{\mu\nu} T^A d_r \rp H G^{A\mu\nu}\,, \nonumber\\
	\Q_{\underset{pr}{uB}} &= \lp \bar q_p \sigma_{\mu\nu} u_r \rp \tilde H B^{\mu\nu}\,, &\quad \Q_{\underset{pr}{dB}} &= \lp \bar q_p \sigma_{\mu\nu} d_r \rp H B^{\mu\nu}\,, \nonumber\\
	\Q_{\underset{pr}{uW}} &= \lp \bar q_p \sigma_{\mu\nu} u_r \rp \tau^a \tilde H W^{a\mu\nu}\,, &\quad \Q_{\underset{pr}{dW}} &= \lp \bar q_p \sigma_{\mu\nu} d_r \rp \tau^a H W^{a\mu\nu}\,. \label{eq:Basis:Gdipole}
\end{align} 
Here $H$ is the Higgs doublet, with $\tilde H = i \sigma_2 H^*$. The gauge bosons in the unbroken phase are $G,~W$ and $B$, with $\tau$ and $T$ the generators of $SU(2)_L$ and $SU(3)$, respectively. The left-handed quark  doublets are represented by $q$, while the right-handed singlets by $u$ and $d$; the quark generation indices are $p$ and $r$.
These operators generate dipole moments below the EW scale at tree-level, by integrating out the $H$ fields and rotating the $SU(2)_L$ bosons by the weak mixing angle. 

As we discuss in the next section, under one-loop RG evolution the above dipole operators mix into Yukawa-like operators
\beq\label{eq:operators:Yukawa}
\Q_{\underset{pr}{uH}} = \lp H^\dagger H \rp \lp \bar q_p u_r \tilde H \rp \,, \quad \Q_{\underset{pr}{dH}} = \lp H^\dagger H \rp \lp \bar q_p d_r H \rp \,.
\eeq
In addition, both the dipole and Yukawa operators receive RG running and EW matching contributions from additional sets of operators. First are the Higgs derivative operators 
\begin{align}
\Q_{\underset{rs}{Hq}}^{(1)} &= \lp H^\dagger i \lrPartial_\mu H \rp \lp \bar q_r \gamma^\mu q_s \rp\,, &\qquad \Q_{\underset{rs}{Hq}}^{(3)} &= \lp H^\dagger i \lrPartial_\mu^a H \rp \lp \bar q_r \tau^a\gamma^\mu q_s \rp\,, \nonumber \\
\Q_{\underset{rs}{Hu}} &= \lp H^\dagger i \lrPartial_\mu H \rp \lp \bar u_r \gamma^\mu u_s \rp\,, &\qquad 
\Q_{\underset{rs}{Hd}} &= \lp H^\dagger i \lrPartial_\mu H \rp \lp \bar d_r \gamma^\mu d_s \rp\,, \nonumber \\
\Q_{\underset{rs}{Hud}} &= i\lp \tilde H^\dagger \lrPartial_\mu H \rp \lp \bar u_r \gamma^\mu d_s \rp \,,\label{eq:operators:HD}
\end{align}
where $D^\mu = \partial^\mu  - i g_1 Y B^\mu - i g_2 \tau_a W^{a\mu} - i g_3 T_A G^{A\mu}$ is the SM covariant derivative, with $Y$ the $U(1)$ hypercharge and $g_1,~g_2,~g_3$ the gauge couplings of $U(1)$, $SU(2)_L$, $SU(3)$ respectively. The left-right derivative is defined as $H^\dagger i \lrPartial_\mu H =H^\dagger i D_\mu H - H^\dagger i \lPartial_\mu H $.
Second group consists of four-fermion operators
\begin{align}
\Q_{\underset{rspt}{quqd}}^{(1)} &= \lp \bar q_r^j u_s \rp \epsilon_{jk} \lp \bar q_p^k d_t \rp\,, &\qquad \Q_{\underset{rspt}{quqd}}^{(8)} &= \lp \bar q_r^j T^A u_s \rp \epsilon_{jk} \lp \bar q_p^k T^A d_t \rp\,,\nonumber \\
\Q_{\underset{rspt}{qu}}^{(1)} &= \lp \bar q_r \gamma_\mu q_s \rp \lp \bar u_p \gamma^\mu u_t \rp\,, &\qquad \Q_{\underset{rspt}{qu}}^{(8)} &= \lp \bar q_r T^A\gamma_\mu q_s \rp \lp \bar u_p T^A \gamma^\mu u_t \rp\,, \label{eq:operators:4fermions:LLRR}
\end{align}
where $j,k$ are $SU(2)_L$ indices and $\epsilon^{jk}$ is the Levi-Civita tensor.

On the other hand, here and in the following we neglect additional purely bosonic operators, e.g. $\Q_{H\tilde G} = H^\dagger H G_{\mu\nu}^A \tilde G^{A\mu\nu}$, as they do not depend on any flavour index. Namely, both their contributions to operator mixing via RG and to matching conditions at the weak scale vanish for flavour-changing transitions.

\subsection{Renormalization Group Evolution and Operator Mixing}
The basic assumption in this work is that any NP mass thresholds are much above the EW scale. The RG evolution of Wilson coefficients from the UV (NP) scale down to the IR (EWSB) scale is the first mechanism responsible for the interplay of different operators in low energy observables, due to operator mixing.

The complete set of one-loop RGEs for dimension-six operators in SMEFT was derived in Refs.~\cite{Jenkins_2013, Jenkins_2014, Alonso_2014}; furthermore, packages like \texttt{wilson}~\cite{Aebischer:2018bkb} and \texttt{DsixTools}~\cite{Celis:2017hod,Fuentes-Martin:2020zaz} have been developed to solve the system of RGEs numerically and compute the Wilson coefficients at any scale. We detail our approach for the full numerical analysis in Section~\ref{subsec:Numerics}. Here we outline the relevant mixing for the subset of operators listed in the previous Section, from Eqs.~\eqref{eq:Basis:Gdipole} to \eqref{eq:operators:4fermions:LLRR}. This is intended as a guideline to understand the behaviour of various operators with respect to quark dipoles, as these are the main focus of this manuscript.

We start with the dipole operators themselves. At one-loop order their RGE running and mixing consists of two types of terms, one from purely gauge interactions, the other from Yukawa vertices of the Higgs field. For the Wilson coefficients of the up-quark dipoles we can write
\beq\label{eq:mixing:dipoles:BWG}
\begin{split}
\dot \C_{\underset{rs}{u}G} &\propto g_{1(2)}g_3 \, \C_{\underset{rs}{u}B(W)} + \tilde g^2 \C_{\underset{rs}{u}G} + \left[ \lp \tilde Y_u^2 + \tilde  Y_d^2 \rp \C_{uG} \right]_{rs} - \left[ \C_{dG} \lp Y_u Y_d \rp \right]_{rs} \,, \\
\dot \C_{\underset{rs}{u}B(W)} &\propto g_{1(2)}g_3 \, \C_{\underset{rs}{u}G} + \tilde g^2 \C_{\underset{rs}{u}B(W)} + \left[ \lp \tilde Y_u^2 + \tilde  Y_d^2 \rp \C_{uB(W)}  \right]_{rs} - \left[ \C_{dB(W)} \lp Y_u Y_d \rp  \right]_{rs}\,,
\end{split}
\eeq
while the expressions for $\dot \C_{\underset{rs}{d}G}$ and $\dot \C_{\underset{rs}{d}B(W)}$ can be obtained from these by replacing $u \leftrightarrow d$.
The scale derivative is defined as $\dot\C\equiv (4\pi)^2 {\rm d}\C/{\rm d}\ln\mu$, whereas $r,s$ are flavour indices, $g_{1,2,3}$ are the gauge couplings, and $Y_u,Y_d$ are the quark Yukawa matrices. The factors $\tilde g^2$, $\tilde Y_u^2$, $\tilde Y_d^2$ indicate particular combinations of the gauge couplings and Yukawa matrices appearing in the full expressions, where we have omitted different contractions of flavour indices for the sake of brevity. At this point, the exact form of these factors is not relevant, as we are only interested in their parametric contribution to the mixing.\footnote{For details on the exact flavour structure of the above RGEs see e.g. ref.~\cite{Jenkins_2014}.}  

From Eq.~\eqref{eq:mixing:dipoles:BWG} it is clear that gauge interaction contributions dominate the dipole RGEs, as they preserve the exact flavour structure (see Sec.~\ref{sec:flavourExpansion} for the detailed expressions): any suppression from small flavour braking parameters cancels exactly between the left- and right-hand side of Eq.~\eqref{eq:mixing:dipoles:BWG}. 
Conversely, the relevance of both Higgs-exchange terms of the order $Y_{u,d}^2$ will depend on the flavour-scheme assumed. In particular, the mixing between up-quark and down-quark dipoles induced at this order and proportional to $Y_u Y_d$ might be phenomenologically relevant. Lastly, notice that the operators $\Q_{qH}$ do not enter the dipole RGEs at this order; that is, non-standard Yukawa-like interactions in the UV do not induce contributions to dipole transitions through their RG flow at one-loop (see also ref.~\cite{Fajfer:2021cxa}).

There is an additional contribution to the dipole RGEs, coming from chirality-flipping four-fermion operators $\Q_{quqd}^{(1)}$ and $\Q_{quqd}^{(8)}$, see Eq.~\eqref{eq:operators:4fermions:LLRR}. We have
\beq\label{eq:mixing:dipoles:BWG:QUQD}
\dot \C_{\underset{rs}{u}X} \propto  g_X \C_{\underset{psrt}{quqd}} \lp Y_d \rp_{pt}\,, \qquad \dot \C_{\underset{rs}{d}X} \propto  g_X \C_{\underset{rtps}{quqd}} \lp Y_u \rp_{pt}\,,
\eeq
where $X = B,~W,~G$ and $g_X$ is the respective gauge coupling. These equations show that the $\Q_{quqd}$ operators contribute to both up- and down-quark dipoles, proportionally to a single power of the down- and up-Yukawa, respectively. The effects on the latter ones in particular could thus be relatively enhanced by the top quark Yukawa contributions.

Next we consider all SMEFT operators whose RGEs are coupled to those of dipoles. This is the case for the Yukawa-like operators, where  $\dot \C_{qH}$ receive the dominant contributions of the form
\beq\label{eq:mixing:H}
\dot \C_{\underset{rs}{u}H} \propto \left[ \lp \tilde Y_u^2 + \tilde  Y_d^2 \rp \C_{uH} \right]_{rs} + \C_{\underset{rt}{d}H} \lp Y_u Y_d \rp_{ts} + \left[ \lp \tilde g^2 + \tilde Y_u^2 \rp \C_{uB(W)} \right]_{rs} + \left[\tilde Y_u^2 \C_{uG} \right]_{rs}\,,
\eeq
and similarly for $\dot \C_{\underset{rs}{d}H}$ by use of the replacement $u \leftrightarrow d$.
Note that the electroweak dipoles $\C_{uB(W)}$ contribute also through a gauge interaction term, not suppressed by small quark masses or CKM factors.

Two additional contributions to the right-hand side of \eqref{eq:mixing:H} come from the four-fermion operators, $\Q_{qu}$ and $\Q_{quqd}$. We have 
\beq\label{eq:mixing:H:QUQD}
\dot \C_{\underset{rs}{u}H} \propto \left[ \C_{quqd} \tilde Y_d^3 \right]_{rs} + \left[ \C_{qu} \tilde Y_u^3 \right]_{rs}\,, \qquad \dot \C_{dH} \propto \left[ \C_{quqd} \tilde Y_u^3 \right]_{rs}\,.
\eeq
Although the Yukawa suppression here is of the third power, it shows that these four-fermion operators can connect the Yukawa and dipole sectors: $\Q_{quqd}$ mixes to both via RGE, while $\Q_{qu}$ participates to the Yukawa RGE and then matches to dipoles at the electroweak scale (see next Section)\footnote{The operator $\Q_{qd}$ mixes as $\dot \C_{dH} \propto \C_{qd} \tilde Y_d^3$, which we find to be numerically negligible.}. This is important for the phenomenological analysis: on one hand, NP that generates only $\C_{qH}$ at the high scale cannot induce dipoles at lower scales at one-loop order. 
It is then unlikely to probe the effect of $\C_{qH}$ via flavour-violating dipole transitions at low energies. On the other hand, non-zero dipole operators $\C_{qB(W)[G]}$ at the high scale do RG mix into the Yukawa-like operators, and can in principle be probed through Higgs induced flavour- and CP-violating processes at low energies, such as neutral meson oscillations (see Sec.~\ref{subsec:MesonMixing}). Finally, high scale NP contributions to $\C_{quqd}$ and $\C_{qu}$ will generate low scale effects in both Yukawa-like and dipole operators. 

The two four-fermion operators of interest have a rich RG mixing structure; however any other operator that mixes with them but not with the Yukawa or dipole operators generates only a two-loop suppressed effect. Thus the relevant part of the RGE is
\beq\label{eq:mixing:FF}
\begin{split}
\dot \C_{\underset{prst}{qu}} &\propto \Big[ \C_{qu} \tilde Y_u^2 \Big]_{prst} + \Big[ \C_{quqd} Y_u Y_d \Big]_{prst}\,, \\
\dot \C_{\underset{prst}{quqd}} &\propto \Big[ \C_{qu} Y_u Y_d \Big]_{prst} +  \Big[ \C_{quqd} \lp \tilde Y_u^2 +  \tilde Y_d^2 \rp \Big]_{prst} \,.
\end{split}
\eeq

Finally, there is an additional set of operators, $\{ \Q_{Hq}^{(1,3)},~ \Q_{Hud},~ \Q_{Hu},~ \Q_{Hd} \}$, that mixes into $\Q_{qH}$. However, their contributions are always suppressed by small quark masses, thus we neglect them in the rest of this work.

\subsection{Matching at the electroweak scale}
\label{subsec:MatchingEWSB}
Around the EW scale the weak gauge bosons, the Higgs and the top quark become heavy degrees of freedom. At lower energies they are thus integrated out, and we consider a new low energy EFT -- LEFT, with particles lighter than $m_{t,h,W,Z}$ as the only dynamical degrees of freedom. We impose the appropriate matching conditions between the two theories above and below the matching scale $\mu_W$. 

The relevant LEFT Lagrangian describing flavour changing dipole transitions among quarks reads
\beq
\label{eq:FCdipoles}
 {\cal L}^\mathrm{eff}_{Fqq'} = c^{q q'}_{8G \pm} \mathcal Q^{qq'}_{8G \pm} + c^{q q'}_{7\gamma \pm} \mathcal Q^{qq'}_{7\gamma \pm} + \mathrm{h.c.} \,,
\eeq
where 
\begin{align}
\Q^{qq'}_{8G \pm} &= \frac{g_s}{32\pi^2} m_q \lp \bar q' \sigma_{\mu\nu}  T^A (1 \pm \gamma_5) q \rp G_A^{\mu\nu}\,, \nonumber \\
\Q^{qq'}_{7\gamma \pm} &= \frac{e}{32\pi^2} m_q \lp \bar q' \sigma_{\mu\nu} (1 \pm \gamma_5) q\rp F^{\mu\nu}\,,\label{eq:LEFT:c7g}
\end{align}
are mass dimension 6 operators, and the respective Wilson coefficients are dimensionful with mass dimension $-2$. Here $g_s=g_3$ and $ e = g_1 \cos \theta_W = g_2 \sin \theta_W$ are the QCD and EM gauge couplings, respectively. The sum over quark flavours is generation (and mass) ordered, so that $m_q > m_{q'}$, where $q$ denotes a quark field in its respective mass basis. 

The flavour violating Yukawa interaction can be described by the renormalizable Lagrangian
\beq\label{eq:YukawaOperator:LEFT}
\mathcal L^{\rm eff}_{Hqq'} = - c^{q'q}_{qY} \lp \bar q'_L q_R \rp h + \rm h.c.\,,  
\eeq
where the coefficients $c^{q' q}_{qY}$ are dimensionless.

The full one-loop matching conditions of SMEFT onto LEFT have been calculated in ref.~\cite{Dekens_2019}. Before proceeding, notice that our notation in Eq.~\eqref{eq:FCdipoles} differs from the one in the latter reference, where the dipoles are defined as dimension 5 operators through
\beq
\mathcal L^{\rm eff}_{qX} = c_{qX}^{rs}~ \bar q_{L,r} \lp \sigma\cdot X \rp q_{R,s} + \rm h.c. \,,
\eeq
and $c_{qX}^{rs}$, where $rs$ are flavour indices, has inverse mass dimension; here $X = F,G$ is a photon or gluon field strength. For simplicity we have suppressed gauge indices. The translation to the coefficients in  \eqref{eq:LEFT:c7g} (through identification $\mathcal L^{\rm eff}_{qX} ={\cal L}^\mathrm{eff}_{Fqq'}$) then reads
\begin{align}
c^{qq'}_{qG} &= \frac{g_s}{32\pi^2} 2 m_q \lp c^{qq'}_{8G -} \rp^*\,, & c^{q'q}_{qG} &= \frac{g_s}{32\pi^2} 2 m_q c^{qq'}_{8G +}\,, \nonumber \\
c^{qq'}_{qF} &= \frac{e}{32\pi^2} 2 m_q \lp c^{qq'}_{7\gamma -} \rp^*\,, & c^{q'q}_{qF} &= \frac{e}{32\pi^2} 2 m_q c^{qq'}_{7\gamma +}\,.
\end{align}

Our main focus are flavour-changing transitions, thus we first consider the case where $q\neq q'$. 
At this order, the SMEFT dipole operators can be matched onto LEFT by simply integrating out the heavy degrees of freedom in the broken EW phase, that is the Higgs $H$ and the weak gauge bosons $W$ and $Z$. We obtain
\begin{align}
c_{qF}^{rs} &= \frac{v}{\sqrt{2} \Lambda^2} \lp c_w \C_{\underset{rs}{q}B} - s_w \C_{\underset{rs}{q}W} \rp \,, \nonumber \\
c_{qG}^{rs} &= \frac{v}{\sqrt{2} \Lambda^2} ~ \C_{\underset{rs}{q}G}\label{eq:matchingDipole:Tree} \,,
\end{align}
where $s_w$ and $c_w$ are the sin and cosine of the weak mixing angle respectively. 
Similarly, for the Yukawa operator we have in a particular quark mass basis 
\beq\label{eq:matchingYukawa:Tree}
c^{rs}_{qY} = \frac{v^2}{\sqrt{2}\Lambda^2} ~ \C_{\underset{rs}{q}H}\,.
\eeq
Note that the above matching is independent of the matching scale $\mu_W$ and does not include any extra terms from other operators. 

At higher loop orders however, additional terms appear in the matching conditions in Eqs.~\eqref{eq:matchingDipole:Tree} - \eqref{eq:matchingYukawa:Tree}. Following the same approach as in the previous section, below we outline the main contributions, which are taken into account in the numerical analysis, see Sec.~\ref{subsec:Numerics}. 
The four-fermion operators $\Q^{(1,8)}_{qu}$ generate up-quark chirality-flipping operators via closed loops of massive top quarks\footnote{Similarly, the operator $\Q_{qd}$ can contribute to the down dipole coefficients, although with a factor $m_b$ from the bottom-quark loop.}; the resulting matching contributions are of the form
\begin{align}
c_{uF}^{rs} &\propto \frac{g_1 m_t}{\Lambda^2} \C_{\underset{r33s}{qu}} \,, \nonumber \\
c_{uG}^{rs} &\propto \frac{g_3 m_t}{\Lambda^2} \C_{\underset{r33s}{qu}} \,, \nonumber \\
c_{uY}^{rs} &\propto \frac{m_t^3}{v \Lambda^2} ~ \C_{\underset{3rs3}{qu}} \lp 1 + \ln\left[ \frac{m_t}{\mu_W} \right] \rp \,. \label{eq:matchingDipole:QU}
\end{align}
In the case of dipole coefficients this is a purely finite term, while the Yukawa receives both a finite and a renormalized contribution~\cite{Dekens_2019}. 

Similarly, the chirality-flipping four-fermion operators $\Q_{quqd}$ match onto dipoles and Yukawas at one-loop. Given their flavour structure, they generate down-quark operators by loop contracting up-quarks, and vice versa. Contrary to the previous case, the top-quark enhanced contributions are then
\begin{align}
c_{dF}^{rs} &\propto \frac{g_1 m_t}{\Lambda^2} \C_{\underset{r33s}{quqd}} \ln\left[ \frac{m_t}{\mu_W} \right] \,, \nonumber \\
c_{dG}^{rs} &\propto \frac{g_3m_t}{\Lambda^2} \C_{\underset{r33s}{quqd}} \ln\left[ \frac{m_t}{\mu_W} \right] \,, \nonumber \\
c_{dY}^{rs} &\propto \frac{m_t^3}{v\Lambda^2} ~ \C_{\underset{r33s}{quqd}} \lp 1 + \ln\left[ \frac{m_t}{\mu_W} \right] \rp \,. \label{eq:matchingYukawa:QUQD}
\end{align}
The corresponding contributions to the up-quark operators are instead suppressed by at least $m_b/v$. Note that here the matching contributions to dipoles are purely logarithmic with the coefficients matching those of the corresponding SMEFT RGEs, see Eq.~\eqref{eq:mixing:dipoles:BWG:QUQD}.

Finally, at the two-loop order, operators $\mathcal Q_{uH,dH}$ generate additional phenomenologically important contributions to Eq.~\eqref{eq:FCdipoles} as well as to leptonic dipole operators defined via 
\beq
\label{eq:FCdipolesl}
 {\cal L}^\mathrm{eff}_{\ell F}  = c^{rr}_{\ell F} \bar \ell_{L,r} \lp \sigma\cdot F \rp \ell_{R,r}  +\mathrm{h.c.} \,.
\eeq
The contributions proceed through Barr-Zee type diagrams~\cite{Brod:2013cka, Brod:2022bww}, where the SMEFT operator is inserted in the fermion loop. As before, here we give the parametric dependence of the leading matching terms:
\begin{align}
c_{dF}^{rr} &\propto {g_1 g_3^2} \frac{v}{\Lambda^2} \lp \frac{m_t}{m_h} \rp^2 \C_{\underset{rr}{dH}}\,,\label{eq:TWOLOOP:cdF}\\
c_{dG}^{rr} &\propto {g_3} \frac{v}{\Lambda^2} \left[ g_3^2 \lp \frac{m_t}{m_h} \rp^2 \C_{\underset{pp}{dH}} \delta^{rp} + g_1^2 \frac{m_{d_r}}{m_{q_p}}  \lp 1 + \log \lp \frac{m_h^2}{m_W^2} \rp \rp \C_{\underset{pp}{qH}} \lp 1 - \delta^{rp} \rp\right]\,,\label{eq:TWOLOOP:cdG}\\
c_{uF}^{rr} &\propto {g_1 g_3^2}{} \frac{v}{\Lambda^2} \lp \frac{m_t}{m_h} \rp^2 \C_{\underset{rr}{uH}} \,, \\
c_{uG}^{rr} &\propto {g_3}{} \frac{v}{\Lambda^2} \left[ g_3^2 \lp \frac{m_t}{m_h} \rp^2 \C_{\underset{pp}{uH}} \delta^{rp} + g_1^2 \frac{m_{u_r}}{m_{q_p}}  \lp 1 + \log \lp \frac{m_h^2}{m_W^2} \rp \rp \C_{\underset{pp}{qH}} \lp 1 - \delta^{rp} \rp\right]\,,\label{eq:TWOLOOP:cuG}\\
c_{\ell F}^{rr} &\propto {g_1^3}{} \frac{v}{\Lambda^2}   \frac{m_{\ell_r} }{m_h^2} \left[ m_t~\C_{\underset{33}{uH}} + \sum_{q=c,b}  12 m_q \C_{qH} Q_q^2 \lp \log^2 \frac{m_q^2}{m_h^2} + \frac{\pi^2}{3}\rp \right] \,, \label{eq:TWOLOOP:ceF}
\end{align}
where in the last expression $Q_q = 2/3 (-1/3)$ is the up (down) quark electric charge.
Quark dipoles receive the dominant matching contribution from the CP violating Higgs coupling of the same flavour, that is, by an insertion of the SMEFT operator on the external quark legs. The chromo-dipole coefficients instead receive the most relevant contributions from CP violating Higgs couplings to light quarks (up, down and strange). The second term in Eqs.~\eqref{eq:TWOLOOP:cdG} and \eqref{eq:TWOLOOP:cuG} is indeed the dominant one for $r=2,3$. Lastly, the two-loop matching to electron EDM is dominated by the top-quark loop, first term in Eq.~\eqref{eq:TWOLOOP:ceF}; however, large quadratic logarithms lead to contributions from the bottom and charm quarks that are only an ${\cal O}(5 - 10)$ smaller, and we show them for completeness.

%
\section{Flavour expansion in MFV and $U(2)^3$}
\label{sec:flavourExpansion}
%

When simultaneously considering SMEFT contributions to various flavour changing and flavour conserving processes, in addition to a consistent expansion in terms of operator dimension, gauge, and Yukawa couplings, one also needs to specify the flavour structure of the operators. 
In this Section we explore two schemes for the flavour-breaking patterns in SMEFT: MFV and a $U(2)^3$ symmetry based flavour expansion. In both cases we define the spurions that parameterize the explicit flavour symmetry breaking and carry out the expansion of operator coefficients to the relevant leading order.

\subsection{MFV}
\label{subsec:detailedMFV:spurions}
MFV models respect the flavour-breaking pattern of the SM, that is, the only spurions that break the flavour group, Eq.~\eqref{eq:flavourGroup}, are the Yukawa matrices. The spurion transformation properties under such group can be derived from the SM Lagrangian. The up- and down-quark Yukawa couplings are defined in terms of dimension-4 Lagrangian terms as
\beq\label{eq:YukLagrangian}
{\cal L}_Y = -\bar{q} Y_u u\, \tilde H - \bar{q} Y_d d\, H + {\rm h.c.}\,,
\eeq
where $\tilde H = i \tau^2 H^*$. Under the quark flavour subgroup, $\mathcal{G}_q = U(3)_q \otimes U(3)_u \otimes U(3)_d$, the quark fields transform as 
\beq\label{eq:MFV:quarkRot}
q \to U_q q \,, \quad u \to U_u u \,, \quad d \to U_d d \,,
\eeq
where $U_{q,u,d}$ are the $U(3)$ transformations in the fundamental representations, i.e.
\beq \label{eq:MFV:quarkRep}
q = (\bm 3,\bm 1,\bm 1) \,, \quad u=(\bm 1,\bm 3,\bm 1) \,, \quad d=(\bm 1,\bm 1,\bm 3) \,. 
\eeq 
In order to mantain the Lagrangian in \eqref{eq:YukLagrangian} formally invariant under the flavour symmetry, one promotes the Yukawa matrices to spurioniuc fields, transforming as $Y_u = (\bm 3, \bar{\bm3}, \bm 1)$, and $Y_d=(\bm3,\bm1,\bar{\bm3})$ under $\mathcal{G}_q$. Explicitly, they transform as
\beq\label{eq:MFV:YukawaRot}
    Y_u \to U_q Y_u U_u^\dagger\,,\qquad   Y_d \to U_q Y_d U_d^\dagger\,.
\eeq
In the MFV setting, the transformation properties of Yukawa matrices and quarks, Eqs.~\eqref{eq:MFV:YukawaRot} and \eqref{eq:MFV:quarkRot} respectively, are all the ingredients needed to write down the spurion expansion of various fermion bilinears. We derive explicitly the expansion for chirality flipping bilinears as an example.

The chirality flipping bilinears, present in the dipole operators, transform under $U(3)^3$ as
\beq
\bar q_r \Gamma u_s = \lp  \bar{\bm3}\otimes\bm1, \bm1\otimes\bm3,\bm1 \rp 
\eeq
where $\Gamma$ is the scalar or tensor Dirac structure, which we will omit in the following. To obtain a singlet under ${\cal G}_q$ we can contract this bilinear with $Y_u^{rs}$: the latter transforms as a $\bm3$ under $U(3)_q$, thus from the tensor product $\bm3\otimes \bar{\bm3}=\bm1\oplus \bm8$ we can obtain a singlet structure, by appropriately contracting the flavour indices. 
Similarly, $Y_u^{rs}$ is a $ \bar{\bm3}$ under $U(3)_u$ and we can obtain a singlet by contracting it with the fundamental representation. The simplest flavour structure is
\beq\label{eq:FuXexpansion:FirstOrder}
\C_{\underset{rs}{uX}} \Q_{\underset{rs}{uX}} = F_{uX}^{(1,0)} \lp \bar q_r u_s \rp Y_u^{rs} + {\rm h.c.} \,.
\eeq
Here we have introduced our notation for the flavour-independent expansion coefficients $F_{uX}^{(i,j)}$: the subscript indicates the associated operator, in this case $\Q_{uX}$ with $X=B,W,G,H$, while the superscript is the order of the respective contribution in $Y_u$ and $Y_d$.

Higher order terms in the expansion can be obtained via insertions of arbitrary powers of bilinear contractions, $( Y_u Y_u^\dagger )^n$ and $( Y_d Y_d^\dagger )^n$, in such a way that the transformation properties are respected. That is, whatever the structure of $\C_{uX}$ is, it should transform in the same way as $Y_u$, i.e. $\C_{uX} \to U_q \C_{uX} U_u^\dagger$. We then get the following expansion
\begin{align}
\label{eq:CuXExp}
    \C_{\underset{rs}{uX}} &= F_{uX}^{(1,0)} Y_u^{rs} + F_{uX}^{(3,0)} ([Y_u Y_u^\dagger] Y_u)^{rs} +  F_{uX}^{(5,0)} ([Y_u Y_u^\dagger]^2 Y_u)^{rs} +\cdots \\
    &\phantom{=} + F_{uX}^{(1,2)} (Y_d Y_d^\dagger Y_u)^{rs} +  F_{uX}^{(3,2)} (Y_d Y_d^\dagger [Y_u Y_u^\dagger] Y_u)^{rs} +  \tilde F_{uX}^{(3,2)} ( [Y_u Y_u^\dagger] Y_d Y_d^\dagger Y_u)^{rs}+\cdots \,, \nn
\end{align}
where the dots denote terms with higher orders of $Y_u Y_u^\dagger$.
Due to small down-quark masses, the expansion in $( Y_d Y_d^\dagger )^n$ converges rapidly (assuming perturbative $F^{(n,m)}_{X}$). On the other hand, the expansion in $( Y_u Y_u^\dagger )^n$ in general needs to be resummed due to the $y_t\sim 1$ entry.\footnote{Formally the expansion is always a polynomial of finite degree~\cite{Colangelo:2008qp}, and in the case of linear MFV only two coefficients are expected to be relevant.} For example, in the up-quark mass basis
 $I + a_1 [Y_u Y_u^\dagger] + a_2 [Y_u Y_u^\dagger]^2 +\cdots$  amounts to a matrix $\mathrm{diag}(1+ \mc{O}(y_u^2), 1 + \mc{O}(y_c^2), \mc{O}(y_t))$. Thus, we can effectively absorb contributions from higher orders of $Y_u Y_u^\dagger$ in the first line of Eq.~\eqref{eq:CuXExp} into a single term (denoted $F_{uX}^{(\infty,0)}$) as
\begin{equation}
   F_{uX}^{(1,0)} \begin{pmatrix}
   y_u + \mc{O}(y_u^3) &  &\\
   & y_c + \mc{O}(y_c^3) & \\
   & & \mathcal{O}(y_t)
   \end{pmatrix}_{rs}
   \simeq 
   F_{uX}^{(1,0)} Y^{rs}_{u} +
   F_{uX}^{(\infty,0)} \delta_{r3} \delta_{s3}.
\end{equation} 
The resummation thus reduces the number of independent spurions to two, following the flavour breaking pattern induced by the large top Yukawa: $U(3) \to U(2) \times U(1)$~\cite{Feldmann:2008ja, Kagan:2009bn}.
However, as we see, $F_{uX}^{(n,0)}$ are always flavour conserving in the up-quark basis, and are well constrained by EDM measurements~\cite{Kley:2021yhn}. Only terms including powers of $Y_d$ can entail explicit up-quark flavour violation. 
Focusing on flavour-changing and CP-violating transitions, the first relevant coefficient in our phenomenological analysis is then $F_{uX}^{(1,2)}$. More generally, the second line of Eq.~\eqref{eq:CuXExp} represents the leading contribution in $Y_d Y_d^\dagger$ insertions (and all orders in $Y_u Y_u^\dagger$). Again inserting the $( Y_u Y_u^\dagger )^n$ resummation, it reads
\beq
\begin{split}
   &F_{uX}^{(1,2)}\,  \left[\begin{pmatrix}
   1  &  &\\
   & 1  & \\
   & & \mathcal{O}(y_t) 
   \end{pmatrix}\,
   [V (Y_{d}^\mathrm{diag})^2 V^\dagger] \, \begin{pmatrix}
   y_u &  &\\
   & y_c  & \\
   & & \mathcal{O}(y_t)
   \end{pmatrix} \right]_{rs} \\
   &\simeq 
  F_{uX}^{(1,2)} \, y_b^2\, \begin{pmatrix}
   y_u |V_{ub}|^2 & y_c V_{ub} V_{cb}^{*} & \mathcal O (y_t) V_{ub} V_{tb}^* \\
   y_u V_{cb} V_{ub}^{*} & y_c |V_{cb}|^2 & \mathcal O (y_t) V_{cb} V_{tb}^*  \\
   y_u \mathcal O (y_t) V_{tb} V_{ub}^*\, & \,y_c \mathcal O (y_t) V_{tb} V_{cb}^*\, & \,\mathcal O (y_t) |V_{tb}|^2
  \end{pmatrix}_{rs} \\
   &\simeq 
   F_{uX}^{(1,2)} \, y_b^2\, V_{rb}V^*_{pb} Y_u^{ps} + F_{uX}^{(\infty 1,2)} \, y_b^2\, \delta_{r3} V_{tb}V^*_{pb} Y_u^{ps} \\
   &\phantom{=}+ F_{uX}^{(\infty 2,2)} \, y_b^2\, V_{rb}V^*_{tb} \delta_{s3} + F_{uX}^{(\infty 3,2)} \, y_b^2\, \delta_{r3} |V_{tb}|^2 \delta_{s3}\,,
\end{split}
\eeq
where $V$ is the CKM matrix and $Y_{d}^\mathrm{diag} = \diag(y_d,y_s,y_b)$. 
We have checked that $y_d, y_s$ can be safely neglected  in $V (Y_{d}^\mathrm{diag})^2 V^\dagger$ and we do not consider them. The resummation of the large top Yukawa again reduces the number of relevant spurions following the $U(3)\to U(2)\times U(1)$ breaking pattern. In practice, since the more general $U(2)$ flavour scheme (see next section) predicts exactly the same pattern of flavour breaking, we restrict our MFV phenomenological analysis in sec.~\ref{subsec:Results:MFV} to only the first term ($F_{uX}^{(1,2)}$).

The hierarchical structure of the CKM allows to perform the above resummation explicitly in any basis and with consistent results. For the down-type dipole coefficients we have, in the down-quark mass basis
\begin{align}\label{eq:CdXExp}
    \C_{\underset{rs}{dX}} &= F_{dX}^{(0,1)} Y_d^{rs} + F_{dX}^{(2,1 )} (Y_u Y_u^\dagger Y_d)^{rs} + F_{dX}^{(4,1)} ([Y_u Y_u^\dagger]^2 Y_d)^{rs} + \cdots\\
    &\simeq F_{dX}^{(0,1)} (Y_d^\mathrm{diag})_{rs} + F_{dX}^{(2,1)} \, y_t^2 \,(Y_d^\mathrm{diag})_{rp}  \,  V_{tp} V^{*}_{ts} + \cdots
    \,. \nn
\end{align}
As expected, now the leading contribution to flavour changing dipole processes among down-quarks will be proportional to $F_{dX}^{(2,1)}$. Note that here the large $Y_u$ resummation reduces the number of relevant terms to just the two leading ones. The leading charm Yukawa contribution is at least an order of magnitude below $y_t^2$ term in $s \to d$ processes.

It is clear that, once the order at which we truncate the ($Y_d$) expansion is fixed, all terms and respective coefficients in \eqref{eq:CuXExp} and \eqref{eq:CdXExp} will enter into the RG mixing, the threshold matching, and will also contribute to observables. The same line of reasoning can be applied to four-fermion operators, Eq.~\eqref{eq:operators:4fermions:LLRR}, with  additional complication that we need to consider the expansion of two bilinears. We report decompositions under the MFV flavour assumption of the SMEFT operators considered in this work in the following compact form: 
\begin{align}
    \C_{\underset{rs}{uX}} &= F_{uX}^{(1,2)} (Y_d Y_d^\dagger Y_u)^{rs}  + \dots\,,\\ 
    \C_{\underset{rs}{dX}} &= F_{dX}^{(2,1 )} (Y_u Y_u^\dagger Y_d)^{rs}  + \dots\,,\\
    \C_{\underset{rspt}{quqd}} &= F_{quqd}^{(1,1)} Y_u^{rs} Y_d^{pt} + \tilde F_{quqd}^{(1,1)} Y_u^{ps} Y_d^{rt}  + \dots\,.
\end{align}
Note, however, that all the terms in the expansion of $Q_{qu}$ under the MFV assumption are hermitian:
\begin{equation}\label{eq:MFVexpansion:QU}
     \C^\mathrm{hermitian}_{\underset{rspt}{qu}}  = \left[ F_{qu}^{(2,0)} (Y_u Y_u^\dagger)^{rs} + F_{qu}^{(0,2)} (Y_d Y_d^\dagger)^{rs} \right] \delta^{pt} + \hat F_{qu}^{(2,0)} (Y_u^\dagger)^{ps} Y_u^{rt}  + \dots\,,
\end{equation}
hence we do not consider them further.

\subsection{$U(2)^3$ symmetry}
\label{subsec:U2scheme}

The $U(2)^3$ flavour scheme is based on the idea to disentangle the third generation Yukawas from the first two. While in principle the top-quark Yukawa is the only ${\cal O}(1)$ flavour breaking parameter in the SM, the hierarchical structure of both up- and down-quark masses suggests the following breaking pattern of the quark flavour group
\beq 
U(3)^3 \to U(2)_q \times U(2)_u \times U(2)_d \times U(1)_t \times U(1)_b\,.
\eeq
That is, the flavour symmetry group distinguishes light (first two generations) and heavy (third generation) quarks. As a consequence, {in the limit of exact symmetry} flavour transitions among the light quarks are completely decorrelated from the heavy sector. Importantly however, the last two factors, $U(1)_t \times U(1)_b$, must also be explicitly broken by allowing for small mixing of the third generation with light quarks, which is necessary to reproduce the hierarchical non-diagonal structure of the CKM. This in turn leads to correlated effects in heavy-to-light quark transitions.

The spurions responsible for the leading breaking of $U(2)^3$ groups, denoted with $\Delta_u$ and $\Delta_d$, transform as
\beq 
\Delta_u = \lp \bm2, \bar{\bm2}, \bm1  \rp \,, \qquad \Delta_d = \lp \bm2, \bm1, \bar{\bm2}  \rp\,.
\eeq
The mixing of the third and light generations can be induced by adding another spurion charged under $U(1)_t$. It needs to be a doublet under $U(2)_q$, 
\beq 
V_q = \lp \bm2, \bm1, \bm1 \rp\,,
\eeq
and is responsible for breaking of $U(1)_t \times U(1)_b$. Following Refs.~\cite{Barbieri:2011ci} and \cite{Greljo:2022cah}, we start with the quark Yukawa matrices in the basis
\beq\label{eq:U2scheme:Yukawa:spurions}
Y_u' = \begin{pmatrix} \Delta_u & V_q \\ 0 & y_t \end{pmatrix}\,, \quad Y_d' = \begin{pmatrix} \Delta_d &  0 \\ 0 & y_b \end{pmatrix}\,,
\eeq
where 
\beq
V_q = \begin{pmatrix} 0 \\ \epsilon_q \end{pmatrix}\,, \quad \Delta_u = U_u^\dagger \begin{pmatrix} \delta'_u & 0 \\ 0 & \delta_u \end{pmatrix}\,, \quad \Delta_d = U_d^\dagger \begin{pmatrix} \delta'_d & 0 \\ 0 & \delta_d \end{pmatrix}\,,
\eeq
and
\beq
U_u = \begin{pmatrix} c_u & s_u \\ -s_u & c_u \end{pmatrix}\,, \quad U_d = \begin{pmatrix} c_d & e^{i\alpha_d}s_d \\ -e^{-i\alpha_d}s_d & c_d \end{pmatrix}\,,
\eeq
are unitary matrices respectively. Here and in the following we use abbreviated notation $c_x\equiv\cos x$ and $s_x\equiv \sin x$. 

All the parameters in the decomposition above are physical~\cite{Greljo:2022cah}, thus it is possible to directly connect them to CKM and mass factors. By requiring that it reproduces the structure of SM masses and mixings, we can estimate the size of the spurion parameters~\cite{faroughy2020flavour}. 
The parameters $\delta_{d}'$ and $\delta_d$ correspond to Yukawa couplings $y_d$ and $y_s$, respectively,
whereas the small parameter $\epsilon_q$ is responsible for mixing the third and light generations. Parameters $\delta_u'$ and $\delta_u$ correspond respectively to up-quark Yukawas $y_u$ and $y_c$ up to linear order in $\epsilon_q$.

In order to fix the values of the parameters in the spurion parameterization, and for easier matching onto the up/down-diagonal Warsaw basis, we diagonalise the Yukawa matrices in Eq.~\eqref{eq:U2scheme:Yukawa:spurions} via bi-unitary transformations 
\begin{equation}
\label{eq:U2:diag}
\begin{split}
Y_u^\prime &= W_u \hat{Y}_u V_u^\dagger\,, \\
Y_d^\prime &= W_d \hat{Y}_d V_d^\dagger\,,
\end{split}
\end{equation}
where $\hat{Y}_{u,d}$ denote the diagonal Yukawa matrices. The CKM matrix is then given as $V_{\rm CKM} = W_u^\dagger W_d$. From Eq.~\eqref{eq:U2scheme:Yukawa:spurions} it is straightforward to see that 
\beq\label{eq:rotationWd}
W_d = \begin{pmatrix} U_d^\dagger & 0 \\ 0 & 1 \end{pmatrix}\,, \qquad V_d = 1\,.
\eeq
The up-quark rotation matrices read
\begin{equation}
    W_u = \begin{pmatrix}
        c_u & -s_u & 0\\
        s_u & c_u & \tfrac{\epsilon_q}{y_t}\\
        -\tfrac{\epsilon_q s_u }{y_t} & -\tfrac{\epsilon_q c_u }{y_t} & 1
    \end{pmatrix}\,,\quad
    V_u = 1 + \tfrac{\epsilon_q}{y_t^2} \begin{pmatrix}
         0& 0 & \delta_u' s_u\\
         0&  0& \delta_u c_u \\
        -\delta_u' s_u  & -\delta_u c_u & 0
    \end{pmatrix}\,,
\end{equation}
where we have omitted terms of order $\mc{O}(\epsilon_q^2,\delta_u'^2,\delta_u^2)$. Note that the above matrices do not correspond to the CKM matrix in standard phase convention (as used in the PDG) and we extract mixing parameters $\theta_u$, $\theta_d$, $\alpha_d$ and $\epsilon_q$ by matching them to CKM rephasing invariants:
\beq\label{eq:CKMinvariants}
\begin{split}
 |V_{us}| &= \left(c_u^2 s_u^2 s_d^2 + (c_u s_d \cos\alpha_d - c_d s_u)^2\right)^{1/2}\,,\\
 |V_{cb}| &= \left|\epsilon_q c_u/y_t\right|\,,\quad
|V_{ub}| = \left|\epsilon_q s_u/y_t\right|\,,\\
 J &= -\frac{\epsilon_q^2\, s_d c_d\,s_u c_u\,\sin \alpha_d}{y_t^2}\,.
\end{split}
\eeq
The relations for CKM moduli are valid up to $\mc{O}(\epsilon_q,\delta_u^{(\prime)})$ while the expression for $J$ is valid up to $\mc{O}(\epsilon_q^2,\delta_u^{(\prime)})$. 
The values of the rephasing invariants are given in the appendix~\ref{app:NumericalInputs}.
In our calculations we use exact expressions for $W_{u,d}$, $V_{u,d}$, and for the rephasing invariants. One possible set of values of the spurion parameters that reproduce the physical information in CKM is $\theta_d = 2.9258$, $\theta_u = 0.085477$, $\epsilon_q = 0.032230$, $\alpha_d = 1.63147$, however this choice is only one of the many discrete possibilities which satisfy Eqs.~\eqref{eq:CKMinvariants}. Furthermore, additional quark field rephasings are required in order to obtain the CKM matrix, $V_\mathrm{CKM}$ in the standard PDG phase convention. For convenience, we explicitly give the rotation matrices $W_{u,d}$, $V_{u,d}$ that correspond to such a phase convention in appendix~\ref{app:NumericalInputs}.

Having the numerical sizes of these parameters, we can define a consistent power counting for the expansion. We truncate the series at ${\cal O}(10^{-4})$ level, that is, we expand up to factors of order ${\cal O}\lp V_q^2, \Delta_{u,d}\times V_q \rp$. Such expansions have been already performed in \cite{faroughy2020flavour,Greljo:2022cah}.

As a useful example to gain insights into this flavour scheme, here we report explicitly in matrix form the expansion of a LR bilinear. In the basis of $Y_{u,d}'$, we can conveniently write the flavour expansion in $U(2)^3$ for up-dipoles as~\cite{faroughy2020flavour,Greljo:2022cah}
\beq\label{eq:U2expansion:uX}
\C_{\underset{rs}{uX}} = \begin{pmatrix} F_{uX}^{(\Delta_u)} \delta_u' c_u &~~ - F_{uX}^{(\Delta_u)} \delta_u s_u &~~ 0  \\ F_{uX}^{(\Delta_u)} \delta_u' s_u &~~ F_{uX}^{(\Delta_u)} \delta_u c_u &~~ F_{uX}^{(V_q)} \epsilon_q \\ F_{uX}^{(V_q,\Delta_u)} \epsilon_q s_u \delta_u' &~~ F_{uX}^{(V_q,\Delta_u)} \epsilon_q c_u \delta_u &~~ F_{uX}^{(1)}  \end{pmatrix} + \dots\,,
\eeq
where, similarly as in the MFV case, we denote the expansion term corresponding to a spurion $A$ as $F_{uX}^{(A)}$. The term $A=1$ indicates no spurion insertions, $A=V_q$ one insertion of the $V_q$ spurion, and so on.
Independent combinations at the same order in the expansion will have different accents, e.g. $F_{uX}^{(A)}, \bar F_{uX}^{(A)}, \tilde F_{uX}^{(A)},\dots$. The pattern of flavour violation in Eq.~\eqref{eq:U2expansion:uX} can be easily understood by the structure of spurions entering the quark Yukawas in Eq.~\eqref{eq:U2scheme:Yukawa:spurions}. The zero-th order term, $F_{uX}^{(1)}$, appears in the 33 entry only. Transitions among the light quarks are instead dictated by a single insertion of $\Delta_u$, that is $F_{uX}^{(\Delta_u)}$. Transitions between the third and the light generations go through $V_q$, which at leading order mixes the third and the second generation, the $F_{uX}^{(V_q)}$ term. Finally, mixing of the third with the first generation appears only at order $\Delta_u\otimes V_q$ with $F_{uX}^{(V_q,\Delta_u)}$. 

We furthermore report the decompositions under the $U(2)^3$ flavour assumption of the SMEFT operators considered in this work in the following compact form:
\begin{align}
    \C_{\underset{rs}{uX}}&= F_{uX}^{(1)}\delta_{r3}\delta_{s3} + F_{uX}^{(V_q)}  \alpha_{r3} \delta_{s3} V_{q}^r +  F_{uX}^{(\Delta_u)} \alpha_{r3} \alpha_{s3} \Delta_u^{rs}  + F_{uX}^{(V_q,\Delta_u)} \delta_{r3}\alpha_{s3}(V_q^\dagger \Delta_u)^s\,, \\
    \C_{\underset{rs}{dX}}&= F_{dX}^{(1)}\delta_{r3}\delta_{s3} + F_{dX}^{(V_q)}  \alpha_{r3} \delta_{s3} V_{q}^r +  F_{dX}^{(\Delta_d)} \alpha_{r3} \alpha_{s3} \Delta_d^{rs}  + F_{dX}^{(V_q, \Delta_d)} \delta_{r3}\alpha_{s3}(V_q^\dagger \Delta_d)^s\,, \\
    \C_{\underset{rspt}{quqd}} &= 
    F_{quqd}^{(1)} \delta_{r3}\delta_{s3}\delta_{p3}\delta_{t3} +
    F_{quqd}^{(V_q)} \alpha_{r3} \delta_{s3} \delta_{p3} \delta_{t3} V_q^r + 
    \bar{F}_{quqd}^{(V_q)} \delta_{r3} \delta_{s3} \alpha_{p3} \delta_{t3} V_q^p \nonumber\\&+
    F_{quqd}^{(V_q^2)}\alpha_{r3} \delta_{s3} \alpha_{p3} \delta_{t3} V_q^r V_q^p +
    F_{quqd}^{(\Delta_u)} \alpha_{r3} \alpha_{s3} \delta_{p3} \delta_{t3} \Delta_u^{rs} +
    F_{quqd}^{(\Delta_d)} \delta_{r3} \delta_{s3} \alpha_{p3} \alpha_{t3} \Delta_d^{pt} \nonumber\\&+ 
    \bar{F}_{quqd}^{(\Delta_u)} \delta_{r3} \alpha_{s3} \alpha_{p3} \delta_{t3}\Delta_u^{ps}+
    \bar{F}_{quqd}^{(\Delta_d)}\alpha_{r3} \delta_{s3} \delta_{p3} \alpha_{t3} \Delta_d^{rt} \nonumber\\&+
    F_{quqd}^{(V_q \Delta_u)} \delta_{r3} \alpha_{s3} \delta_{p3} \delta_{t3}  (V_q^\dagger \Delta_u)^s +
    \bar{F}_{quqd}^{(V_q \Delta_u)} \alpha_{r3} \alpha_{s3} \alpha_{p3} \delta_{t3}  \Delta_u^{rs} V_q^{p} \nonumber\\&+ 
    F_{quqd}^{(V_q \Delta_d)} \delta_{r3} \delta_{s3} \delta_{p3} \alpha_{t3}  (V_q^\dagger \Delta_d)^t + 
    \bar{F}_{quqd}^{(V_q \Delta_d)} \alpha_{r3} \delta_{s3} \alpha_{p3} \alpha_{t3}  V_q^{r} \Delta_d^{pt} \nonumber\\&+
    \hat{F}_{quqd}^{(V_q \Delta_u)} \alpha_{r3} \alpha_{s3} \alpha_{p3} \delta_{t3}  V_q^{r} \Delta_u^{ps}  +
    \hat{F}_{quqd}^{(V_q \Delta_d)} \alpha_{r3} \delta_{s3} \alpha_{p3} \alpha_{t3}  \Delta_d^{rt} V_q^{p}\,, \label{eq:U2expansion:QUQD}\\ 
    \C_{\underset{rspt}{qu}} &= 
    F_{qu}^{(V_q)} \alpha_{r3} \delta_{s3} \alpha_{p3} \alpha_{t3} V_q^r \delta_{pt}+
    \bar{F}_{qu}^{(V_q)} \alpha_{r3} \delta_{s3} \delta_{p3} \delta_{t3} V_q^r +
    F_{qu}^{(\Delta_u)} \alpha_{r3} \delta_{s3} \delta_{p3} \alpha_{t3} \Delta_u^{rt} \nonumber\\& 
    + F_{qu}^{(V_q \Delta_u)} \alpha_{r3} \alpha_{s3} \delta_{p3} \alpha_{t3} (V_q^\dagger)^{s} \Delta_u^{rt}  +
    \bar{F}_{qu}^{(V_q \Delta_u)} \alpha_{r3} \alpha_{s3} \delta_{p3} \alpha_{t3} \delta_{rs}(V_q^\dagger\Delta_u)^{t} \nonumber\\& 
    +\hat{F}_{qu}^{(V_q \Delta_u)} \delta_{r3} \delta_{s3} \delta_{p3} \alpha_{t3} (V_q^\dagger\Delta_u)^{t}\,,\label{eq:U2expansion:QU}
\end{align}
where $\alpha_{rs} \equiv (1-\delta_{rs})$ is used to reflect the fact that $V_q, \Delta_u, \Delta_d$ only act on the first and second generation fermions, and there is no summation over repeated indices. Note that, contrary to the MFV case, some of the terms in the expansion of $Q_{qu}$ are non-hermitian. For completeness we also list the Hermitian part of $Q_{qu}$
\begin{equation}
\begin{split}
    \C^\mathrm{hermitian}_{\underset{rspt}{qu}} &= F_{qu}^{(1)} \alpha_{r3} \alpha_{s3} \alpha_{p3} \alpha_{t3} \delta_{rs} \delta_{pt}+ 
    \bar F_{qu}^{(1)} \alpha_{r3} \alpha_{s3} \delta_{p3} \delta_{t3} \delta_{rs}+ 
    \hat F_{qu}^{(1)} \delta_{r3} \delta_{s3} \alpha_{p3} \alpha_{t3} \delta_{pt} \\& \phantom{=}
    +\tilde F_{qu}^{(1)} \delta_{r3}\delta_{s3}\delta_{p3}\delta_{t3} 
    +F_{qu}^{(V_q^2)} \alpha_{r3} \alpha_{s3} \alpha_{p3} \alpha_{t3} V_q^r (V_q^\dagger)^s \delta_{pt}\\
    &\phantom{=}+
    \bar{F}_{qu}^{(V_q^2)} \alpha_{r3} \alpha_{s3} \delta_{p3} \delta_{t3} V_q^r (V_q^\dagger)^s \,,
\end{split}
\end{equation}
which is not considered further in this work. All of these expansions are done in the basis of $Y_u^\prime, Y_d^\prime$ in Eq.~\eqref{eq:U2scheme:Yukawa:spurions}. To match onto the up- or down-diagonal quark mass basis we perform the following rotations of the quark fields with the rotation matrices defined in Eq.~\eqref{eq:U2:diag}: $u \to V_u u$, $d \to V_d d$, and $q \to W_{u/d} q$ for the up/down-diagonal mass basis, respectively. Explicit forms of the rotation matrices that also reproduce the CKM matrix in standard phase convention are given in appendix~\ref{app:NumericalInputs}.

%
\section{Low energy constraints on flavour and CP violating operators}
\label{sec:FVandCPVconstraints}
%

In this Section we present the phenomenology of flavour-changing and CP-violating dipole operators in Eq.~\eqref{eq:LEFT:c7g}, in low energy observables. 
We first reiterate the contributions of flavour diagonal dipole operators and CP-violating Higgs Yukawas to electron (eEDM) and neutron (nEDM) electric dipole moments, Sec.~\ref{subsec:EDM}. 
We then estimate the additional effects on the nEDM from flavour changing vertices, Sec.~\ref{sec:FVnEDM}. Moving to flavour changing observables, we evaluate the most relevant meson decays in the $D$, $K$, and $B$ sector, in Sec.~\ref{sec:CPVpheno:Dmeson}, \ref{sec:CPVpheno:Kmeson} and \ref{sec:CPVpheno:Bmeson} respectively. Finally, in Sec.~\ref{subsec:MesonMixing}, we collect the bounds on CP-violating Yukawa coupling to quarks coming from neutral meson oscillation measurements. The complete set of bounds considered in our analysis is summarized in Table~\ref{table:bounds_summary}.

\begin{table}[!h]
\begingroup
\setlength{\tabcolsep}{8pt} 
\renewcommand{\arraystretch}{1.3} 
\centering
\begin{tabular}{c  c  c} 
 \hline\hline
 Process & Bound &  Sec.  \\ 
 \hline
   nEDM &  $\left| \tilde c_{7\g}^u - 8.2 ~ \tilde c_{7\g}^d + 0.56~ \tilde c_{7\g}^s - 11.~ \tilde c_{8G}^u - 48.~ \tilde c_{8G}^d + 0.16~ \tilde c_{8G}^s \right| < 0.69 ~{\rm TeV}^{-2}$ & \ref{subsec:EDM}
  \\
    eEDM & $\left| 0.16~ \Im\lp c_{dY}^{33} \rp + 0.27~ \Im\lp c_{uY}^{22} \rp + \Im\lp c_{uY}^{33} \rp \right| < 7.9\times10^{-4}$ & \ref{subsec:EDM}
  \\
  \hline
 nEDM ($c\to u\g$) & $\left|\im \left(c_{7\gamma+}^{cu} - c_{7\gamma-}^{cu} \right)\right| \lesssim 90\,\mathrm{TeV}^{-2}$ & \ref{sec:FVnEDM}  
 \\
 nEDM ($b\to d\g$) & $\left|\cos \gamma \,\im \left(c_{7\gamma+}^{bd*} - c_{7\gamma-}^{bd*} \right) - \sin \gamma \, \re \left(c_{7\gamma+}^{bd*} - c_{7\gamma-}^{bd*} \right) \right| \lesssim 6.8\times10^3\,\mathrm{TeV}^{-2}$ & \ref{sec:FVnEDM}  
 \\
  nEDM ($s\to d\g$) & $\left|\im \left(c_{7\gamma+}^{sd} - c_{7\gamma-}^{sd} \right)\right| \lesssim 2.3\times10^3\,\mathrm{TeV}^{-2}$ & \ref{sec:FVnEDM}  
  \\
  \hline
  $\Delta A_{{\rm CP}}(D^0)$ & $\left|\im \lp c_{8G\pm}^{cu} (\mu_W) \rp \right| \lesssim 0.05\,\mathrm{TeV}^{-2}$ & \ref{sec:CPVpheno:Dmeson}  
  \\
  $A_{{\rm CP}} (D^0 \to \rho \gamma)$ & $\left|\im \lp c_{7\g\pm}^{cu} (\mu_W) \rp \right| \lesssim 40\,\mathrm{TeV}^{-2}$ & \ref{sec:CPVpheno:Dmeson}  
  \\
    $\epsilon'/\epsilon$ & $ - 0.87 \,{\rm TeV}^{-2} < \,{\rm Im} (c^{sd}_{8G +} (\mu_{\rm EW}) - {\rm Im} (c^{sd}_{8G -} (\mu_{\rm EW}))  < 0.47 \,{\rm TeV}^{-2}$ & \ref{sec:CPVpheno:Kmeson}  
  \\
      ${\cal B}(K_L\to\pi^0 e^+e^-)$ & $ |{\rm Im} (c^{sd}_{7\gamma +}(\mu_{EW}) + c^{sd}_{7\gamma -}(\mu_{EW}))| < 4.86 \, {\rm TeV}^{-2} $ & \ref{sec:CPVpheno:Kmeson}  
  \\
 \hline
   \multirow{2}{7em}{$A_{{\rm CP}}(B\to X\g)$, ${\cal B}(B\to X\g)$} &  $|{\rm Im} (c^{bs}_{7\gamma +}(\mu_{EW}))| \lesssim 0.3 \, {\rm TeV}^{-2}$ & \multirow{2}{1.3em}{\ref{sec:CPVpheno:Bmeson} }
 \\ 
  & $|{\rm Im} (c^{bs}_{8G +}(\mu_{EW}))| \lesssim 0.5 \, {\rm TeV}^{-2}$ & 
 \\
   \multirow{2}{7em}{${\cal B}(B\to X\g)$} &  $|{\rm Im} (c^{bs}_{7\gamma -}(\mu_{EW}))| \lesssim 0.5 \, {\rm TeV}^{-2}$ & \multirow{2}{1.3em}{\ref{sec:CPVpheno:Bmeson} }
 \\ 
  & $|{\rm Im} (c^{bs}_{8G -}(\mu_{EW}))| \lesssim 2 \, {\rm TeV}^{-2}$ & 
 \\ \hline
   \multirow{2}{4em}{$K^0 - \bar K^0$} &  ${\rm Re}(({c_{dY}^{12}})^2),~{\rm Re}((c_{dY}^{21})^2) \in[-5.9,5.6] \times 10^{-10}$ & \multirow{2}{1.3em}{\ref{subsec:MesonMixing} }
 \\ 
  & ${\rm Im}((c_{dY}^{12})^2),~{\rm Im}((c_{dY}^{21})^2) \in [-2.9,1.6] \times 10^{-12}$ & 
 \\
    $D^0 - \bar D^0$ & $|c_{uY}^{12}|^2,~|c_{uY}^{21}|^2 < 5.0 \times 10^{-9}$ & \ref{subsec:MesonMixing}
  \\
    $B_d^0 - \bar B_d^0$ & $|c_{dY}^{13}|^2,~|c_{dY}^{31}|^2 < 2.3 \times 10^{-8}$ & \ref{subsec:MesonMixing} 
  \\  
    $B_s^0 - \bar B_s^0$ & $|c_{dY}^{23}|^2,~|c_{dY}^{32}|^2 < 1.8 \times 10^{-6}$ & \ref{subsec:MesonMixing} 
  \\  
 \hline\hline
\end{tabular}
\caption{Summary of low energy processes considered in this work (1st column), the respective bounds on the low energy coefficients (2nd column), and more details given in sections listed in the last column. The bounds from radiative $B$ meson decays are given separately for $c^{bs}_{7\gamma \pm}$ and $c^{bs}_{8G \pm}$ as an approximate indication of their range, while the proper bound is obtained by their combined fit to observables, see Sec.~\ref{sec:CPVpheno:Bmeson} for details. 
}\label{table:bounds_summary}
\endgroup
\end{table}

\subsection{Electron and Neutron EDMs}
\label{subsec:EDM}

The electric and chromoelectric dipole moments of a fermion $f$ are defined via the effective Lagrangian of $f$ interacting with appropriate field strengths as (see e.g.~\cite{Pospelov:2005pr})
\beq
\mc{L}_\mathrm{dipole} = -\frac{i d_f}{2} \lp \bar f \sigma^{\mu\nu} \gamma_5 f \rp  F_{\mu\nu}  - \frac{i \tilde d_f g_s}{2} \lp \bar f \sigma^{\mu\nu} G_{\mu\nu} \gamma_5 f \rp  \,,
\eeq
where $F_{\mu\nu}$ and $G_{\mu\nu}$ are the electromagnetic and chromomagnetic field strengths, respectively. In particular, the above definition of an electric dipole moment $d_f$ implies a nonrelativistic Hamiltonian $H_\mathrm{nr} = - (2 d_f \bm{S})\cdot \bm{E}$ for a fermion $f$ in a static electric field, in agreement with~\cite{nEDM:2020crw}. Conversely, $\tilde d_f$ is the chromoelectric dipole moment and can be present only for colored fermions. The correspondence between flavour diagonal EDM operators and dipole moments at low energy scale is 
\beq
d_f = -2 \Im c^{ff}_{fF} = -\frac{e m_f}{8\pi^2} \Im  \tilde c^{f}_{7\g} \,,\qquad  \tilde d_f = -2 \Im c^{ff}_{fG} = -\frac{m_f}{8\pi^2} \Im \tilde c^{f}_{8G}\,,
\eeq
where $\tilde c^{f}_{7\g(8G)} \equiv c^{ff}_{7\g(8G)+} - c^{ff}_{7\g(8G)-}$.

Null results from electron and neutron EDM measurements put stringent limits on their size. Namely, we have~\cite{Pignol:2021uuy, ACME:2018yjb}
\begin{align}
    |d_n| &< 1.8\times10^{-26}~e~{\rm cm}\,,  \\
    |d_e| &< 1.1\times10^{-29}~e~{\rm cm}\,.
\end{align}
The low-energy dipole operators induce potentially large shifts in these quantities, resulting in strong bounds on CP-violating contributions from new interactions. 

A strong bound could in principle be also obtained by considering the EDM of mercury, see the discussion in Ref.~\cite{Brod:2022bww}. The total EDM can be expressed as the contribution of unpaired neutrons and protons in the nucleus, plus the additional pion-nucleon interactions; the latter could then lead to slightly stronger bounds on the chromo-dipole coefficients. However, the prediction of the total mercury EDM is not completely under control, as it strongly depends on hadronic parameters and pion-nucleon couplings, which are affected by large uncertainties. Thus we refrain from including bounds from nuclear EDMs in this work and only consider the more established neutron EDM (nEDM).

At the hadronic scale, $\mu_{\rm had}\sim2$ GeV, the nEDM is obtained as a combination of light quark dipoles and chromo-dipoles, where we are neglecting small contributions from the Weinberg operator.
We translate this expression, given in Ref.~\cite{Brod:2022bww} at the hadronic scale, to the weak scale $\mu_W$, by rescaling the dipole coefficients via their QCD running~\cite{Isidori:2012yx}. We can write in general
\beq\label{eq:Bounds:nEDM:rescaled}
\left| \sum_{q=u,d,s} \lp a_n^q ~ \Im\left[\tilde c_{7\g}^q(\mu_W) \right] + b_n^q ~ \Im\left[\tilde c_{8G}^q(\mu_W) \right] \rp\right| < |d_n|\,,
\eeq
where $a_n^q,b_n^q$ are numerical coefficients which, in addition to the QCD rescaling, depend on hadronic and electric dipole operators matrix elements. For simplicity, we normalize all coefficients to have $a_n^u = 1$ and express the limit in TeV$^{-2}$. 
At the weak scale we obtain the bounds reported in Table~\ref{table:bounds_summary},
where we have omitted the scale dependence of the coefficients.

Flavour conserving quark dipole operators in SMEFT match onto the $c_{7\g(8G)}$ coefficients at the tree-level, while four-fermion operators enter at the one-loop level, as discussed in Section~\ref{subsec:MatchingEWSB}. A complete analysis of such operators and the respective contributions has been performed in Ref.~\cite{Kley:2021yhn}. As a consistency check, using Eq.~\eqref{eq:Bounds:nEDM:rescaled} and the method detailed in Section~\ref{subsec:Numerics}, we reproduce their results for the leading terms in the MFV and $U(2)^3$ expansions\footnote{Note that in the convention of Ref.~\cite{Kley:2021yhn} the Yukawa matrices in the $U(2)^3$ scheme are defined with the heavy quark Yukawa couplings factored out and absorbed in the respective Wilson coefficient. In our definition, Eq.~\eqref{eq:U2scheme:Yukawa:spurions}, these couplings are instead kept explicitly. The bound on coefficients involving the spurions $\Delta_u$ and $\Delta_d$ thus differ by a factor of $y_t^{-1}$ and $y_b^{-1}$ respectively.}. At the two-loop order, CP-violating Higgs couplings also enter nEDM via Barr-Zee type diagrams~\cite{Brod:2013cka, Brod:2022bww}. The parametric expressions of these contributions are given in Eqs.~\eqref{eq:TWOLOOP:cdF} to \eqref{eq:TWOLOOP:cuG}.

Lastly, the leading contributions to the electron EDM (eEDM) come solely from the two-loop matching of CP-violating Higgs couplings. As shown in Eq.~\eqref{eq:TWOLOOP:ceF}, only heavy quarks (top, bottom and charm) have a sizeable effect. In the low energy basis, the running of the $c^{ee}_{\ell F}$ coefficient is dictated by QED and thus negligible. We can then perform the matching directly with SMEFT operators at the weak scale and obtain 
\beq
\left| 0.16~ \Im\lp c_{dY}^{33} \rp + 0.27~ \Im\lp c_{uY}^{22} \rp + \Im\lp c_{uY}^{33} \rp \right| < 7.9\times10^{-4}\,.
\eeq
Note that in the low-energy running to the hadronic scale, bottom and charm coefficients receive Next-to-Leading-Log (NLL) corrections~\cite{Brod:2023wsh}; these amount to $\sim10\%$ and $\sim30\%$ corrections with respect to our fixed order computation for bottom and charm quark coefficients evaluated at the weak scale, respectively. 

As shown in Ref.~\cite{Brod:2022bww}, the constraints on the modified Higgs coupling contributions to eEDM strongly depend on the assumption about the electron Yukawa coupling with the Higgs. Allowing for modified electron-Higgs couplings could lead to cancellations and consequently to more relaxed constraints. In this work however, we assume that the Higgs coupling to electrons is purely dictated by the SM Yukawa $y_e = 2.9\times10^{-6}$.

\subsection{Flavour changing dipole contributions to neutron EDM}
\label{sec:FVnEDM}
\begin{figure}
\centering
         \includegraphics[scale=0.45]{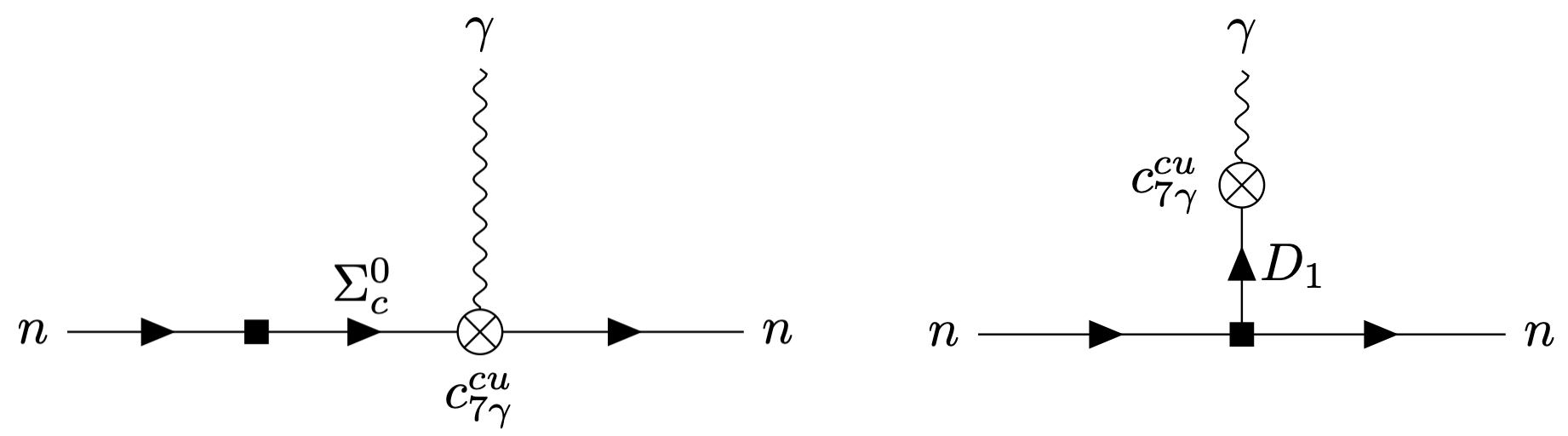} 
    \caption{Estimate of a long-distance contribution of CP-violating $c^{uc}_{7\gamma\pm}$ to the neutron EDM. The left diagram contributes to $d_n$ via a $\Sigma_c \to n \gamma$ insertion accompanied by the charged weak interaction. There is a second diagram that involves conjugated vertices: $n \to \Sigma_c \gamma$ vertex followed by $\Sigma_c^0 \to n$ mixing. The right diagram represents the $D_1$ axial-meson exchange that should be accompanied by the conjugated vertices with a $\bar D_1$ exchange.
    }
    \label{fig:nEDM-SigmaC}
\end{figure}

Here we focus on the flavour changing dipole operators that contribute to the electric dipole moment of the neutron. 
The flavour changing dipole operators $\mathcal{Q}^{qq'}_{7\gamma\pm}$ can contribute to the low energy CP-violating and flavour conserving amplitude $n \to n \gamma$, that results in neutron EDM. This can be achieved by mixing the neutron with the nearest baryonic state $B$ with matching quantum numbers, and a subsequent CPV dipole transition of the state $B$ into $n\gamma$. Another possibility is to insert the lightest axial-vector $1^+$ meson as an intermediate state that can mix with the external photon via the CPV dipole operator. In Fig.~\ref{fig:nEDM-SigmaC} we show the two diagrams contributing in the specific case of $c \to u \gamma$ electric dipole operator insertion. A similar approach was taken in Ref.~\cite{Mannel:2012qk} to estimate the contribution of CKM phases to neutron EDM. In the following we estimate the total contribution to the nEDM due to the $c\to u \g$ operator and then generalize it to $b\to d \g$ and $s\to d \g$ dipole transitions. Note that $b \to s \gamma$ does not concern any of the valence quarks of the neutron and thus cannot be constrained from nEDM.

\paragraph{$\bm{c \to u \gamma}$:}

First, we focus on the left-hand diagram in Fig.~\ref{fig:nEDM-SigmaC}. The box vertex corresponds to the SM weak mixing between a neutron and the intermediate $(1/2)^+$ baryon octet state, namely $\Sigma_c^0$. Such mixing is proportional to the CKM factor $V_{cd} {V_{ud}}^*$. For the cases of $s\to d\gamma $ and $b\to d\gamma$, the neutron mixes with the $\Lambda$ and $\Lambda_b$ baryon states, with CKM suppressions of $V_{ud} {V_{us}}^*$ and $V_{ud} {V_{ub}}^*$, respectively. The CP-violating operator then induces an effective vertex between the intermediate baryon state and the neutron, thus leading to dipole transitions $\Sigma_c^0 \to n \gamma$ ($\Lambda \to n \gamma$ and $\Lambda_b \to n \gamma$ for $s \to d \gamma$ and $b \to s \gamma$ dipoles, respectively). The latter dipole transitions are denoted by a crossed circle vertex. A recent study also pointed out a possibility of electric dipole operators leading to enhanced CP asymmetries of various radiative charmed-baryon decays~\cite{Adolph:2022ujd}.

Heavy to light quark transition $\mc{Q}_{7\pm}^{cu}$ induces the transition between the neutron and the lowest lying octet state $\Sigma_c^0$, comprised of valence quarks $cdd$. The form factors of $\Lambda_c \to p$ have been computed on the lattice~\cite{Meinel:2017ggx} and we use these results as an approximation to the $\Sigma_c \to n$ matrix element. The form factors of the $\sigma^{\mu\nu} q_\nu \gamma_5$ operator are defined as
\beq\label{eq:FF-LambdaN}
\begin{split}
    \Braket{n(p')|\bar u i\sigma^{\mu\nu} q_\nu \gamma_5 c| \Sigma_c^0 (p)} &= -\bar u(p') \gamma_5 \left[\tilde h_+(q^2) \frac{q^2}{s_-} \left(p+p' - (m_\Sigma^2-m_n^2) \frac{q}{q^2}\right)^\mu +\right.\\ 
    &\quad +  \left.\tilde h_\perp(q^2) (m_\Sigma-m_n)\left(\gamma + \frac{2 m_n}{s_-} p - \frac{2 m_\Sigma}{s_-} p'\right)^\mu \right] u(p)\,,
\end{split}
\eeq
where $q = p-p'$ and $s_- = (m_{\Sigma_c}-m_n)^2-q^2$. In the limit of a real photon with imposing $q^2 \to 0$ and $\epsilon \cdot q \to 0$, only the $\tilde h_\perp$ form factor contributes:
\beq\label{eq:Sigmac:realphotonlimit}
\begin{split}
     \Braket{n(p')|\bar u i\sigma^{\mu\nu} q_\nu \gamma_5 c| \Sigma_c^0 (p)}_{q^2 \to 0} &= -\bar u(p') \gamma_5 \tilde h_\perp (0) \left[(m_\Sigma-m_n) \gamma^\mu - (p+p')^\mu  \right] u(p) \\
     &=\tilde h_\perp (0) \,\bar u(p') i \sigma^{\mu\nu} q_\nu \gamma_5  u(p)\,.
\end{split}
\eeq
The lattice calculation~\cite{Meinel:2017ggx} found $\tilde h_\perp(0) \approx 0.5$ for $\Lambda \to p$, the value of which we adopt also for $\Sigma_c \to n$ transition.
 
The weak mixing vertex (see Fig.~\ref{fig:nEDM-SigmaC}) between $\Sigma_c$ and $n$ can be parameterized as
\begin{equation}\label{eq:Sigmac-n:weakmixing}
    \Braket{\Sigma_c(p) | (\bar c_L \gamma^\nu d_L)(\bar d_L \gamma_\nu u_L) | n (p) } = \bar u_\Sigma(p) \left[g_S + g_P \gamma_5\right] u_n(p)\,,
\end{equation}
where $g_S$ and $g_P$ are the scalar and pseudoscalar mixing parameters. Notice that only the parity-conserving term $g_S$ will contribute to the neutron EDM, since parity-violation is already contained in the matrix element of $\Sigma_c \to n \gamma$ transition, see Eq.~\eqref{eq:Sigmac:realphotonlimit}. Lacking a proper estimate of this matrix element, we resort to dimensional analysis; this suggests that $g_S \sim 4 \Lambda_\mathrm{h}^3$, where 4 is a combinatorial factor accounting for the two valence $d$-quark contained both in $n$ and $\Sigma_c$. We estimate that the hadronic scale $\Lambda_\mathrm{h}$ can take any value in the range between $\Lambda_\mathrm{QCD}\sim200$ MeV and $m_{\Sigma_c} = 2.45$ GeV. Finally, we can assemble the final result by combining together the two vertices, Eqs.~\eqref{eq:Sigmac:realphotonlimit} and \eqref{eq:Sigmac-n:weakmixing}, and including the $\Sigma_c$ propagator. The contribution of the left-hand side diagram in Fig.~\ref{fig:nEDM-SigmaC} to the neutron EDM then amounts to
\begin{equation}
\label{eq:nEDM:ctougamma}
    d_n^{\Sigma_c} = e\,\frac{G_F  g_S \tilde{h}_\perp(0) m_c}{8\sqrt{2}\pi^2  (m_n+m_{\Sigma_c})}\,\im \left[(c_{7\gamma+}^{cu} - c_{7\gamma-}^{cu} ) V_{cd} V_{ud}^*\right]\,,
\end{equation}
where $m_c = 1.27$ GeV is the charm quark mass.

Now let us focus on the right-hand side diagram in Fig.~\ref{fig:nEDM-SigmaC}. The box vertex again indicates the neutron coupling with axial-vector $1^+$ meson, $D_1$. The vertex $n \to n D_1$ can be written in the factorization approximation as the product
\beq\label{eq:fact}
\begin{split}
\Braket{n(p') D_1(q, \epsilon) | (\bar c_L \gamma^\nu d_L) (\bar d_L \gamma_\nu u_L) | n(p)} & \approx
\Braket{n | \bar d_L \gamma_\nu d_L | n}\,\Braket{D_1(q, \epsilon)| \bar c_L \gamma^\nu u_L |0} \\ 
& \propto   \epsilon^{*\nu}(q) \Braket{n | \bar d_L \gamma_\nu d_L | n}\,.
\end{split}
\eeq
where $\epsilon(q)$ is polarization vector of the intermediate $D_1$. 
Parity violating terms generated by the axial operator $\bar d \gamma_\nu \gamma_5 d$ result in axial and induced-pseudoscalar form factors~\cite{Green:2017keo}, none of which can contribute to the EDM of the neutron. However, working beyond the factorization approximation we can have an induced EDM transition:
\begin{equation}
\Braket{n(p') D_1(q, \epsilon) | (\bar c_L \gamma^\nu d_L) (\bar d_L \gamma_\nu u_L) | n(p)}_\mathrm{NF} = 
-i d_{nnD}(q^2) \epsilon^{*\mu} \bar u(p') \sigma_{\mu\nu} q^{\nu} \gamma_5 u(p)\,.
\end{equation}
By simple dimensional analysis and accounting for combinatorial factors due to $d$ field contractions with external states, we find that the form factor is at most $d_{nnD}(q^2 \to 0) \lesssim 4 m_{D_1}$. The axial $D_1$ meson then mixes into the photon via insertion of \eqref{eq:FCdipoles} and the transverse decay constant:
\begin{equation}
    \Braket{0 | \bar u \sigma^{\mu\nu} q_\mu \gamma^5 c | D_1 (\epsilon, q)} = i f_{D_1}^T m_{D_1}^2 \epsilon^\nu\,,
\end{equation}
where $\epsilon$ is the polarization vector of $D_1$. Results on the lattice report similar values for transverse and vector decay constants, within $10\%$ accuracy, in the case of $1^-$ mesons~\cite{Jansen:2009hr} (see also \cite{Pullin:2021ebn}), thus we use the results of the 
$f_{D^*}^V$~\cite{Becirevic:2012ti} and approximate $f_{D_1}^T \simeq 0.3 \e{GeV}$. In fact, the meson exchange diagram also gets contributions by $\bar D_1$ state with an opposite sign. The total meson exchange contribution then reads:
\begin{equation}
\label{eq:nEDM-D1}
      d_n^{D_1} = -e\,\frac{G_F  g_{nnD}  m_c f_{D_1}^T}{8\sqrt{2}\pi^2}\,\im \left[(c_{7\gamma+}^{cu} - c_{7\gamma-}^{cu} ) V_{cd} V_{ud}^*\right]\,.
\end{equation}
The upper bound on the neutron EDM~\cite{PhysRevD.98.030001}, $d_n/e < 9.1\times 10^{-13}~\mathrm{GeV}^{-1}$, then leads to
\begin{equation}\label{eq:nEDM:ctougamma:bound}
    \left|\im \left(c_{7\gamma+}^{cu} - c_{7\gamma-}^{cu} \right)\right| \lesssim 3.7\textrm{-}90\,\mathrm{TeV}^{-2}\,.
\end{equation}
The range for the upper bound reflects our ignorance of the hadronic parameters $d_{nnD}$ and $g_S$. Taking maximal (minimal) allowed values of those parameters leads to the strictest (most conservative) bound. Future progress on non-perturbative matrix elements entering this calculation will result in more precise estimates. In our further analysis we will use the most conservative upper bound, namely $90\,\mathrm{TeV^{-2}}$. Note that this bound is numerically dominated by the $d_n^{D_1}$ contribution with lowest possible $g_{nnD_1} = 4\Lambda_{QCD} \approx 1.2\e{GeV}$.

\paragraph{$\bm{b \to d \gamma}$:}
In this case the electric dipole operator $\mc{Q}_{7\pm}^{bd}$ is inserted between the neutron and $\Lambda_b$ ($udb$). Form factors of the heavy-to-light transition $H \to N$, where $H$ is the heavy baryon and $N$ the nucleon, can be parameterized using the velocity-dependent heavy baryon bispinor $u(v)$~\cite{Jenkins:1990jv}:
\begin{equation}
\Braket{N(p')|\bar u i\sigma^{\mu\nu} q_\nu \gamma_5 c| H (p)} =
\sqrt{m_H}\,\bar u(p') \left[F_1(v\cdot p') + F_2(v\cdot p') \slashed{v} \right] i\sigma^{\mu\nu} q_\nu \gamma_5 u(v)\,.
\end{equation}
Here $p'$ is the nucleon moment and $p = m_H v + \mathcal{O}(\Lambda_{QCD})$, where the heavy quark mass scale is factored out~\cite{Manohar:2000dt}. The form factors $F_1$ and $F_2$ are now functions of $v \cdot p'$. We can relate these form factors to the $\tilde h_\perp(0)$ factor, see Eq.~\eqref{eq:FF-LambdaN}, as
\begin{equation}
\tilde h^{H \to N}_\perp(0) = \sqrt{m_{H}} \left(-F_1+ 2 F_2  \right)|_{v \cdot p' = (m_H^2+m_N^2)/(2 m_H)},    
\end{equation}
where we have already neglected subleading $m_N/m_H$ terms in the heavy quark expansion. Using the above expansion, we can express form factors for $\Lambda_b \to n$ in terms of the respective $\Lambda_c \to p$ form factors as
\beq
\begin{split}
  \tilde h^{\Lambda_b \to n}_\perp(0) &= \sqrt{m_{\Lambda_b}}  \left(-F_1+ 2 F_2  \right)|_{v \cdot p' = (m_{\Lambda_b}^2+m_N^2)/(2 m_{\Lambda_b})}\\
  &=\sqrt{\frac{m_{\Lambda_b}}{m_{\Lambda_c}}} \tilde h^{\Lambda_c \to p}_\perp \left(m_{\Lambda_c}^2 (1-m_{\Lambda_b}/m_{\Lambda_c})+m_N^2 (1-m_{\Lambda_c}/m_{\Lambda_b})\right) \\
  &= 1.57\,\tilde h_\perp^{\Lambda_c \to p} (q^2=-7.1\,\mathrm{GeV}^2) \,.
\end{split}
\eeq
This point is deep in the space-like region for $\Lambda_c \to p$ transition, while lattice results only reach $q^2=-0.36\,\mathrm{GeV}^2$~\cite{Meinel:2017ggx}, thus rendering in practice impossible to extrapolate from them. We use $\tilde h^{\Lambda_b \to n}_\perp(0) \approx 0.3$ as a crude estimate. 
 
Similarly to the previous case, for the weak mixing $n \to \Lambda_b$ transition we have $g_S^{n \to \Lambda_b} = 2 \Lambda_h^3$, where the combinatorial factor of 2 comes from two $d$-quarks in neutron. The resulting contribution to $d_n$ is 
 \begin{equation}
    d_n^{\Lambda_b} = e\,\frac{G_F  g_S^{n \to \Lambda_b} \tilde{h}^{\Lambda_b \to n}_\perp(0) m_b}{8\sqrt{2}\pi^2  (m_n+m_{\Lambda_b})}\,\im \left[(c_{7\gamma+}^{bd} - c_{7\gamma-}^{bd} ) V_{ud} V_{ub}^*\right]\,,
 \end{equation}
where $m_b = 4.18$ GeV and $m_{\Lambda_b} = 5.62$ GeV are the bottom quark and $\Lambda_b$ baryon mass respectively.

The $B_1(5721)$ axial meson exchange contribution can be easily adapted from~\eqref{eq:nEDM-D1},
\begin{equation}
\label{eq:nEDM-B1}
      d_n^{B_1} = -e\,\frac{G_F  g_{nnB_1}  m_b f_{B_1}^T}{8\sqrt{2}\pi^2}\,\im \left[(c_{7\gamma+}^{bd} - c_{7\gamma-}^{bd} ) V_{ud} V_{ub}^*\right]\,,
\end{equation}
where now $g_{nn B_1} \in [\Lambda_{QCD}, m_{B_1}]$ without a combinatorial factor, since there is only one $u$-quark in the neutron. The tensor decay constant has been calculated using light-cone sum rules in~\cite{Pullin:2021ebn} resulting in $f_{B_1}^T = 0.28\pm 0.06$.

The resulting constraint depends on both the real and imaginary parts of the flavour-violating coefficient $c_{7\g\pm}^{bd}$ due to the interplay with the CKM phase in $V_{ub} = |V_{ub}| \exp(-i \gamma)$, namely
 \begin{equation}
        \left|\cos \gamma \,\im \left(c_{7\gamma+}^{bd} - c_{7\gamma-}^{bd} \right) + \sin \gamma \, \re \left(c_{7\gamma+}^{bd} - c_{7\gamma-}^{bd} \right) \right| \lesssim 35\textrm{-}6800\,\mathrm{TeV}^{-2}\,.
 \end{equation}
 Again, the most aggressive (lowest) bound is dominated by $d_n^{\Lambda_b}$ when we set the scale to $\Lambda_h = m_{\Lambda_b}$. On the other hand, we will be using the conservative $6800\e{TeV^{-2}}$ which is determined by $d_n^{B_1}$ with $g_{nnB_1} = \Lambda_{QCD} \approx 0.3\e{GeV}$.

\paragraph{$\bm{s \to d \gamma}$:}
To evaluate the $d_n^\Lambda$ contribution, driven by $Q^{sd}_{7\gamma\pm}$, we can adapt the expression for $\Sigma_c \to n \gamma$ contribution to $d_n$:
\begin{equation}
    d_n^\Lambda = e\,\frac{G_F  g_S^{\Lambda \to n} \tilde{h}^{\Lambda \to n}_\perp(0) m_s}{8\sqrt{2}\pi^2  (m_n+m_\Lambda)}\,\im \left[(c_{7\gamma+}^{sd} - c_{7\gamma-}^{sd} ) V_{ud} V_{us}^*\right]\,.
\end{equation}
Here we express the mixing parameter as $g_S = 2 \Lambda_h^3$, where the 2 stems from two possible contractions with the $d$-quark in the neutron. The electric dipole transition form factor $\tilde{h}^{\Lambda \to n}_\perp(0)$, to the best of our knowledge, has not been determined theoretically and cannot be simply extracted from the experimental value of $\mathrm{Br}(\Lambda \to n \gamma)$~\cite{Larson:1993ig}. Due to lack of better estimate we will vary $\tilde{h}^{\Lambda \to n}_\perp(0)$ in the range $0.2\textrm{-}0.4$, around the value for $\tilde h^{\Sigma_c \to n} = 0.3$.

The axial-meson $K_1(1270)$ exchange contribution
reads:
\begin{equation}
      d_n^{K_1} = -e\,\frac{G_F  g_{nnK_1}  m_s f_{K_1}^T}{8\sqrt{2}\pi^2}\,\im \left[(c_{7\gamma+}^{sd} - c_{7\gamma-}^{sd} ) V_{us}^* V_{ud}\right]\,,
\end{equation}
with $g_{nn K_1} \in [\Lambda_{QCD}, m_n]$. The tensor current decay constant has been estimated in the light-cone sum rule approach to be $f_{K_1}^T = (145 \pm 15)\e{MeV}$~\cite{Yang:2007zt}. The resulting constraint is 
 \begin{equation}
    \left |\im \left(c_{7\gamma+}^{sd} - c_{7\gamma-}^{sd} \right)\right| \lesssim 760\textrm{-}2300\,\mathrm{TeV}^{-2}.
 \end{equation}

\subsection{Direct CP violation in singly Cabibbo suppressed $D$ meson decays}
\label{sec:CPVpheno:Dmeson}

The LHCb collaboration has reported the first observation of CP violation in the charm sector. In particular, they have measured~\cite{LHCb:2019hro} 
\begin{equation}
    \Delta A_{CP} \equiv A_{CP}(K^+ K^-) - A_{CP} (\pi^+ \pi^-) = (-1.54 \pm 0.29) \times 10^{-3}\,,
    \label{eq:ACP}
\end{equation}
where 
\begin{equation}
    A_{CP}(f) \equiv \frac{\Gamma(D^0 \to f) - \Gamma(\bar D^0 \to f)}{\Gamma(D^0 \to f) + \Gamma(\bar D^0 \to f)}\,.
\end{equation}
While at present the SM predictions for this observable cannot be firmly established, they could possibly account for the observed value~\cite{Brod:2011re, Brod:2012ud, Grossman:2019xcj, Cheng:2019ggx}. In the following we will thus assume NP can at most saturate the measured value in Eq.~\eqref{eq:ACP}. One can parameterize such effects in terms of a NP effective low energy Lagrangian as~\cite{Isidori:2011qw}
\begin{equation}
    \Delta A_{CP}^{8G \pm} \simeq \frac{1}{\sqrt{2}\lambda_{cu}^s G_F} {\rm Im} (c^{c u}_{8G \pm}) {\rm Im} \left ( R^{8G \pm} \right)\,,
\end{equation}
where we have assumed approximate $U$-spin limit, $\lambda_{cu}^s = |V_{cs} V_{us}^*| \simeq 0.22$, and we have defined $ R^{8G \pm} =  R^{8G \pm}_{\pi\pi} + R^{8G \pm}_{K K}$, with
\beq
R^{8G \pm}_{P P} = \frac{ \langle P P |4 \mathcal Q^{uc}_{8G \pm}  | D  \rangle}{\langle P P | \bar u \gamma_\mu (1-\gamma_5) q \bar q \gamma^\mu (1-\gamma_5) c | D \rangle} \,,
\eeq
where $q=s,d$ for $P=K,\pi$, respectively. At the charm scale $\mu_{c} \simeq 2$\,GeV, $|{\rm Im}( R^{8G \pm}(\mu_{c}))| $ is expected to be $\mathcal O(1)$. Explicitly, at the leading perturbative order in the QCD factorization limit one obtains $|R^{8G \pm}(\mu_{c})| \simeq 0.23$~\cite{Grossman:2006jg}. Since large absorbtive (rescattering) effects are expected in these decays~\cite{Brod:2011re, Golden:1989qx}, in the following we use this value as a crude estimate of $|{\rm Im}( R^{8G \pm}(\mu_{c}))| $. The operator $\mathcal Q^{cu}_{8G \pm}$ however runs with QCD, so the corresponding Wilson coefficient needs to be renormalized to leading logarithmic order as~\cite{Isidori:2012yx} 
\beq
c^{uc}_{8G \pm} (\mu_{EW}) = \left[\frac{\alpha_s(\mu_b)}{\alpha_s (\mu_{\rm EW})} \right]^{14/23} \times \left[ \frac{\alpha_s(\mu_{c})}{\alpha_s (\mu_b)} \right]^{14/25} \times  c_{8G \pm} (\mu_{c})\,,
\eeq
or explicitely $c^{uc}_{8G \pm} (\mu_{EW}) \simeq 1.9 c_{8G \pm} (\mu_{c})$ for $\mu_{\rm EW} \simeq 160$ GeV. Comparing the above expression to the experimental value, we thus obtain the limit
\begin{equation}
    |{\rm Im} (c^{c u}_{8G \pm} (\mu_{\rm EW}))| \lesssim 0.05\, {\rm TeV}^{-2}\,.
\end{equation}
A more reliable theoretical estimate of ${\rm Im}( R^{8G \pm})$ would however be needed to establish this bound on a firm ground.

The EM dipole operators $\mathcal Q^{uc}_{7\gamma \pm}$ are probed by the direct CP asymmetry measured by Belle, $A_\mathrm{CP} (D^0 \to \rho \gamma) = 0.056\pm0.152\pm 0.006$~\cite{Belle:2016mtj}. Together with the measured branching fraction $\mc{B}(D^0 \to \rho \gamma) = (1.77\pm 0.30\pm 0.07)\times 10^{-5}$ the authors of Ref.~\cite{deBoer:2017que} derive that the NP modification of the amplitudes can be at most $|\delta A_7^{(\prime)}| \lesssim 0.5$, where $\delta A_7^{(\prime)} = \frac{v^2}{2}\delta c^{cu}_{7\pm} (m_c)$. Resulting constraint that follows is
\begin{equation}
    | c^{cu}_{7\g\pm}(\mu_{EW})| < 40\e{TeV}^{-2}.
\end{equation}
This limit will improve at Belle II through CP asymmetries in radiative $D$ meson decays~\cite{Isidori:2012yx}, allowing for improvement of the bound on $c_{7\pm}^{cu}$.

\subsection{Direct vs indirect CP violation in Kaon decays}
\label{sec:CPVpheno:Kmeson}

In kaon decays the observable $\epsilon'/\epsilon$ measures the size of direct CP violation in $K_L \to \pi\pi$ relative to the indirect CP violation described by $\epsilon_K$. The experimental world average for this quantity is~\cite{NA48:2002tmj, KTeV:2002qqy, KTeV:2010sng}
\begin{equation}
(\epsilon'/\epsilon)_{\rm exp} = (16.6 \pm 2.3) \times 10^{-4}\,.
\end{equation}
Theoretically, determination of the SM contributions to $\epsilon'/\epsilon$ has been a long lasting challenge, but recent Lattice QCD calculations find values consistent with the experimental result~\cite{RBC:2020kdj}
\begin{equation}
    (\epsilon'/\epsilon)_{\rm SM} = (21.7 \pm 8.4) \times 10^{-4} \,.
\end{equation}
The possible effects of heavy NP can be parameterized in terms of an effective low energy Lagrangian as
\begin{equation}
    \epsilon'/\epsilon = (\epsilon'/\epsilon)_{\rm SM} + \frac{P_{8g} g_s} {16\pi^2}  {\rm Im} (c^{sd}_{8G+} - c^{sd}_{8G-})\,,
\end{equation}
where $P_{8g}$ parameterizes the matrix element 
$\langle \pi \pi | 16 \pi^2 \mathcal Q^{sd}_{8G \pm} /g_s | K \rangle$, which has been computed in Refs~\cite{Constantinou:2017sgv, Buras:2018evv}. In particular, using the master formulae in Ref.~\cite{Aebischer:2018csl}, one obtains 
$P_{8g}(\mu_{\rm EW}) g_s(\mu_{\rm EW}) / 16\pi^2 \simeq 0.0026 \rm TeV^{2}$ for $\mu_{\rm EW} \simeq 160$GeV. Comparing the experimental world average with the SM expectation we obtain a bound of
\begin{align}
 - 0.87 \,{\rm TeV}^{-2} < & \,{\rm Im} (c^{sd}_{8G +} (\mu_{\rm EW}) - {\rm Im} (c^{sd}_{8G -} (\mu_{\rm EW}))  < 0.47 \,{\rm TeV}^{-2}\,.
\end{align}

The EM dipole operators $\mathcal Q^{sd}_{7\gamma \pm}$ contribute to rare semileptonic Kaon decays. In particular $\mathcal B (K_L \to \pi^0 e^+ e^-)$ is sensitive to the combination ${\rm Im}(c^{sd}_{7\gamma +} + c^{sd}_{7\gamma -})$~\cite{Mescia:2006jd}\footnote{We are neglecting the $c^{sd}_{8G \pm} $ contributions, since they are more tightly constrained by $\epsilon'/\epsilon$}
\beq
\mathcal B (K_L \to \pi^0 e^+ e^-) \simeq 4.6 \times 10^{-12}  \left[\frac{{\rm Im} (c^{sd\ast}_{7\gamma +} + c^{sd\ast}_{7\gamma -}) }{{\rm Im}(\lambda^t_{sd})} \frac{\sqrt 2}{8\pi G_F} \frac{m_s}{m_K} \frac{B_T(0)}{f_+(0)}\right]^2\,,
\eeq
where $\lambda^t_{sd} \equiv V_{ts} V_{td}^*$. Here we take $B_T(0)/f_+(0) \simeq 0.64$~\cite{Baum:2011rm} at the QCD renormalization scale of $\mu_c \simeq 2$\, GeV and we have neglected pure SM and interference contributions, which are negligible compared to the current experimental upper bound of $\mathcal B (K_L \to \pi^0 e^+ e^-) < 2.8 \times 10^{-10}$~\cite{KTeV:2003sls}. Including QCD RG evolution to the weak scale given by~\cite{Isidori:2012yx}
\beq
c^{sd}_{7\gamma \pm} (\mu_{EW}) = \left[\frac{\alpha_s(\mu_b)}{ \alpha_s (\mu_{\rm EW})} \right]^{16/23} \times \left[\frac{\alpha_s(\mu_{c})}{\alpha_s (\mu_b) }\right]^{16/25} \times c^{sd}_{7\gamma \pm} (\mu_{c})\,,
\eeq
or explicitly $c^{sd}_{7\gamma \pm} (\mu_{EW}) \simeq 2 c^{sd}_{7\gamma \pm} (\mu_{c}) $ for $\mu_{\rm EW} \simeq 160$\,GeV, we obtain a bound of
\beq
|{\rm Im} (c^{sd}_{7\gamma +}(\mu_{EW}) + c^{sd}_{7\gamma -}(\mu_{EW}))| < 4.86 \, {\rm TeV}^{-2}\,.
\eeq

\begin{figure}[!h]
    \centering
    \includegraphics[width=0.5\linewidth]{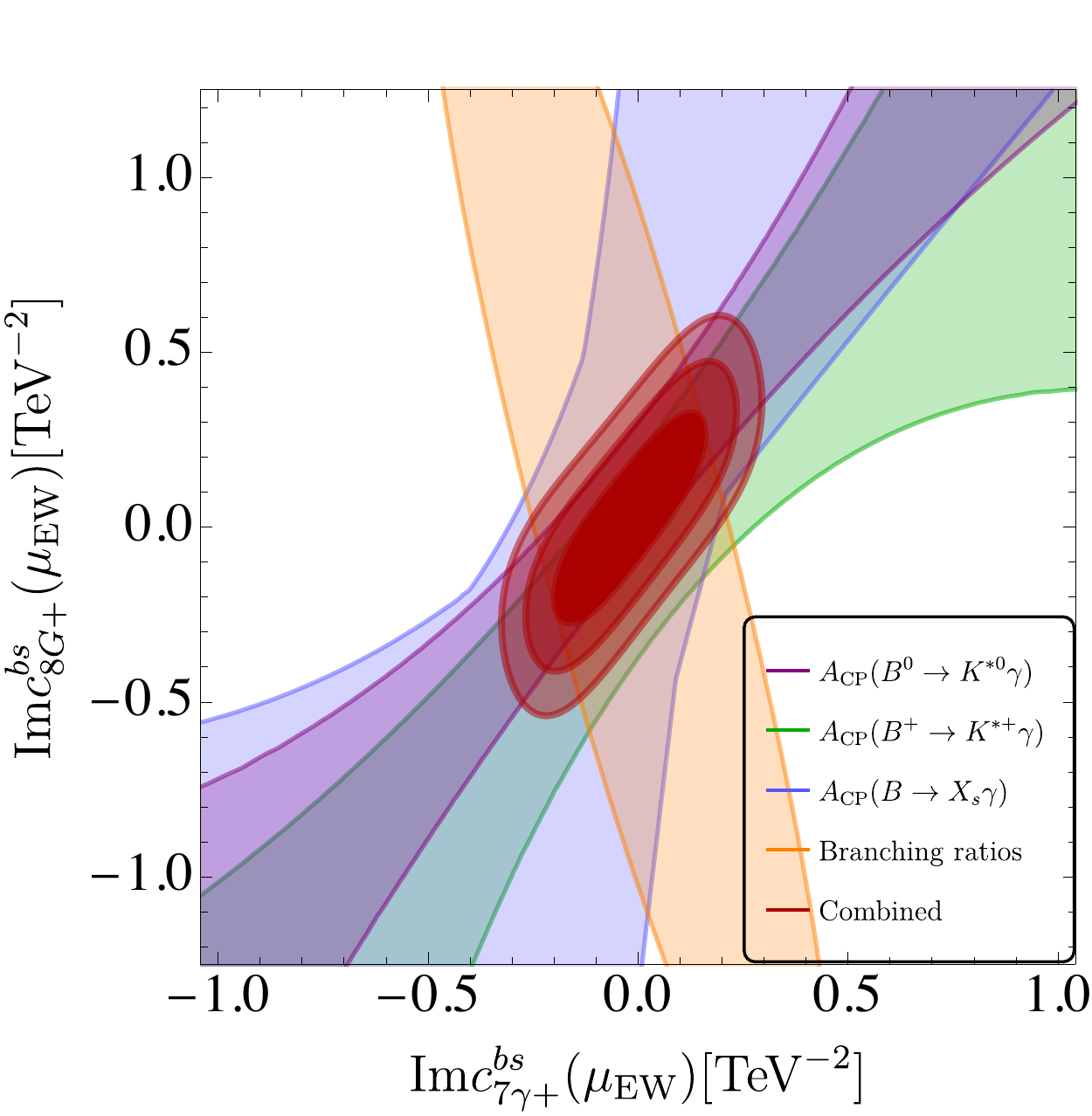}~
    \includegraphics[width=0.48\linewidth]{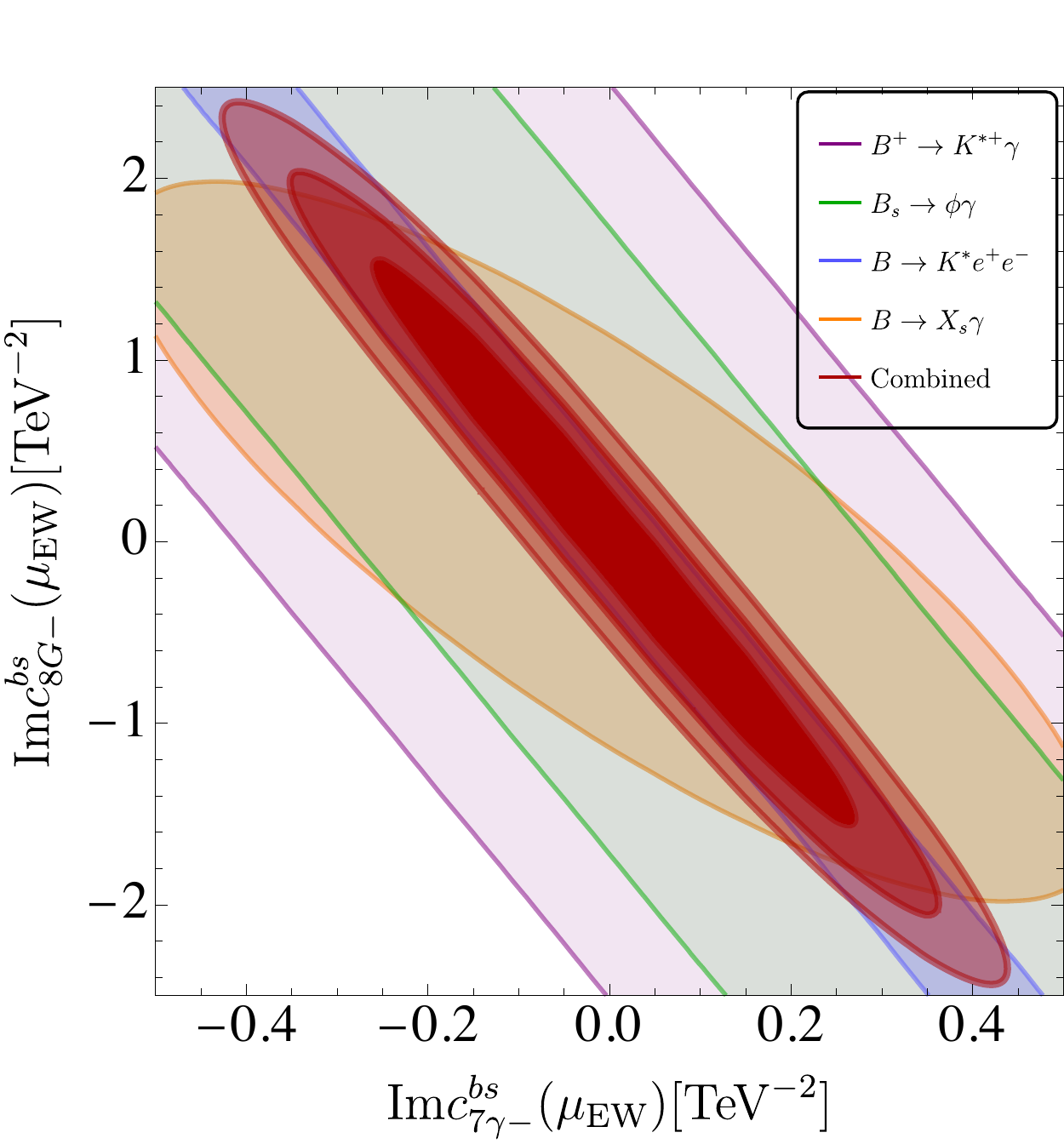}
    \caption{Bounds in the $({\rm Im} c^{bs}_{7\gamma +},{\rm Im} c^{bs}_{8G +})$ and $({\rm Im} c^{bs}_{7\gamma -},{\rm Im} c^{bs}_{8G -})$ planes from radiative $B$ meson decays. The allowed $1\sigma$ contours of single contributions are shown, whereas the allowed $1$, $2$ and $3\sigma$ regions are shown for the combination of all constraints. }
    \label{fig:bs-bounds}
\end{figure}

\subsection{Radiative $B$ meson decays}
\label{sec:CPVpheno:Bmeson}
The $\mathcal Q^{bs}_{7\gamma \pm}$ and $\mathcal Q^{bs}_{8G \pm}$ operators contribute to radiative and rare semileptonic $B$ meson decays. We use \texttt{flavio}~\cite{Straub:2018kue} to predict the CPV observables $A_{\mathrm{CP}}(B^{(0,+)} \to K^{*(0,+)} \gamma)$, as well as the branching ratios $\mathcal B(B^+\to K^* \gamma)$, $\mathcal B(B_s\to\phi\gamma)$ and $\mathcal B(B\to X_s \gamma)$, and the photon polarisation observables measured by LHCb in $B\to K^\ast e^+ e^-$ at very low $q^2$~\cite{LHCb:2020dof}. Furthermore, we consider the inclusive CP asymmetry $A_{\mathrm{CP}}(B \to X_s \gamma)$~\cite{Benzke:2010tq}. As for the experimental values, we use the latest averages by the HFLAV collaboration~\cite{HFLAV:2019otj}. 

The obtained bounds on the Wilson coefficients ${\rm Im} \lp c^{bs}_{7\gamma \pm}\rp$ and ${\rm Im} \lp c^{bs}_{8G \pm}\rp$ at the scale $\mu_{\rm EW} = 160$\,GeV are shown on Fig.~\ref{fig:bs-bounds}, where we assume NP modifies only the imaginary parts of the Wilson coefficients. The considered CP asymmetries are only sensitive to ${\rm Im}\lp c^{bs}_{7\gamma +}\rp$ and ${\rm Im}\lp c^{bs}_{8G +}\rp$~\cite{Paul:2016urs}, hence they only appear on the left plot of Fig.~\ref{fig:bs-bounds}. On the left plot we combine all of the considered branching ratios into a single ellipse (orange), whereas on the right plot we show their sensitivities one-by-one. On the right plot we also show the constraints from the photon polarisation observables in $B\to K^\ast e^+ e^-$, which offer excellent sensitivity, however information from $B\to X_s \gamma$ is crucial to achieve a closed combined fit in the $({\rm Im} c^{bs}_{7\gamma -},{\rm Im} c^{bs}_{8G -})$ plane.

\subsection{Neutral Meson Oscillations}
\label{subsec:MesonMixing}

Flavour (and CP) violating Higgs boson couplings induced by $\mathcal Q_{uH}$ and $\mathcal Q_{dH}$ operators lead to Higgs-mediated FCNCs in the quark sector and are currently best constrained by neutral meson oscillation measurements~\cite{Harnik:2012pb}. 
Writing the effective off-diagonal Higgs boson couplings to $q=u,d$ quarks as in Eq.~\eqref{eq:YukawaOperator:LEFT} and the tree level matching condition in Eq.~\eqref{eq:matchingYukawa:Tree}, we can identify the effective off-diagonal Yukawa couplings with the SMEFT Wilson coefficients in the corresponding quark mass basis at the Higgs boson mass scale $\mu_H \simeq 125$\,GeV via
$c_{qY}^{pr} = \mathcal C_{{}^{qH}_{pr}}(\mu_H) {v^2}/{\sqrt 2 \Lambda^2} $. The resulting bounds on $c_{qY}^{pr}$ from neutral meson oscillation measurements~\cite{Harnik:2012pb} are compiled in Table~\ref{table:bounds_summary}.

%
\section{Results}\label{sec:results}
\begin{figure}[!h]
    \centering
    \includegraphics[width=0.8\linewidth]{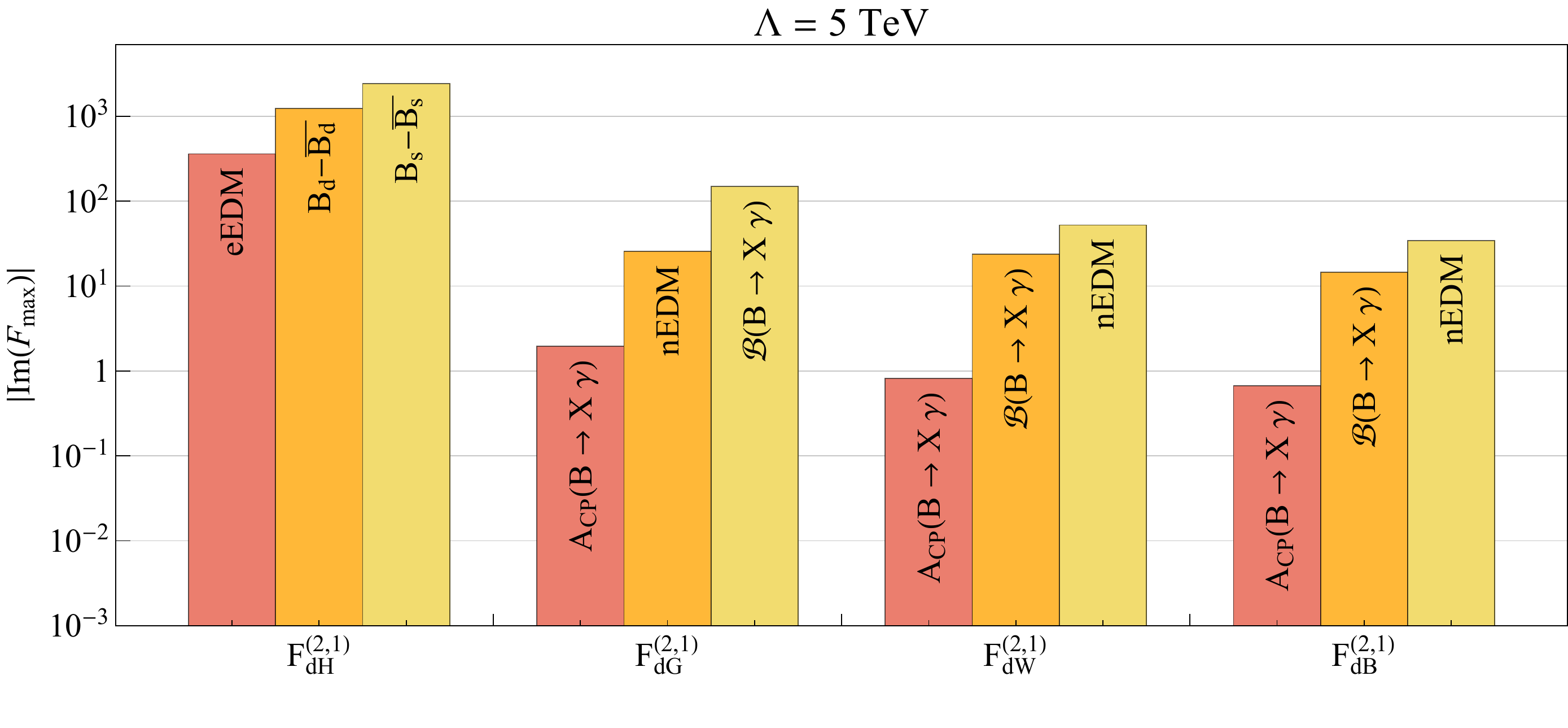}
    \includegraphics[width=0.8\linewidth]{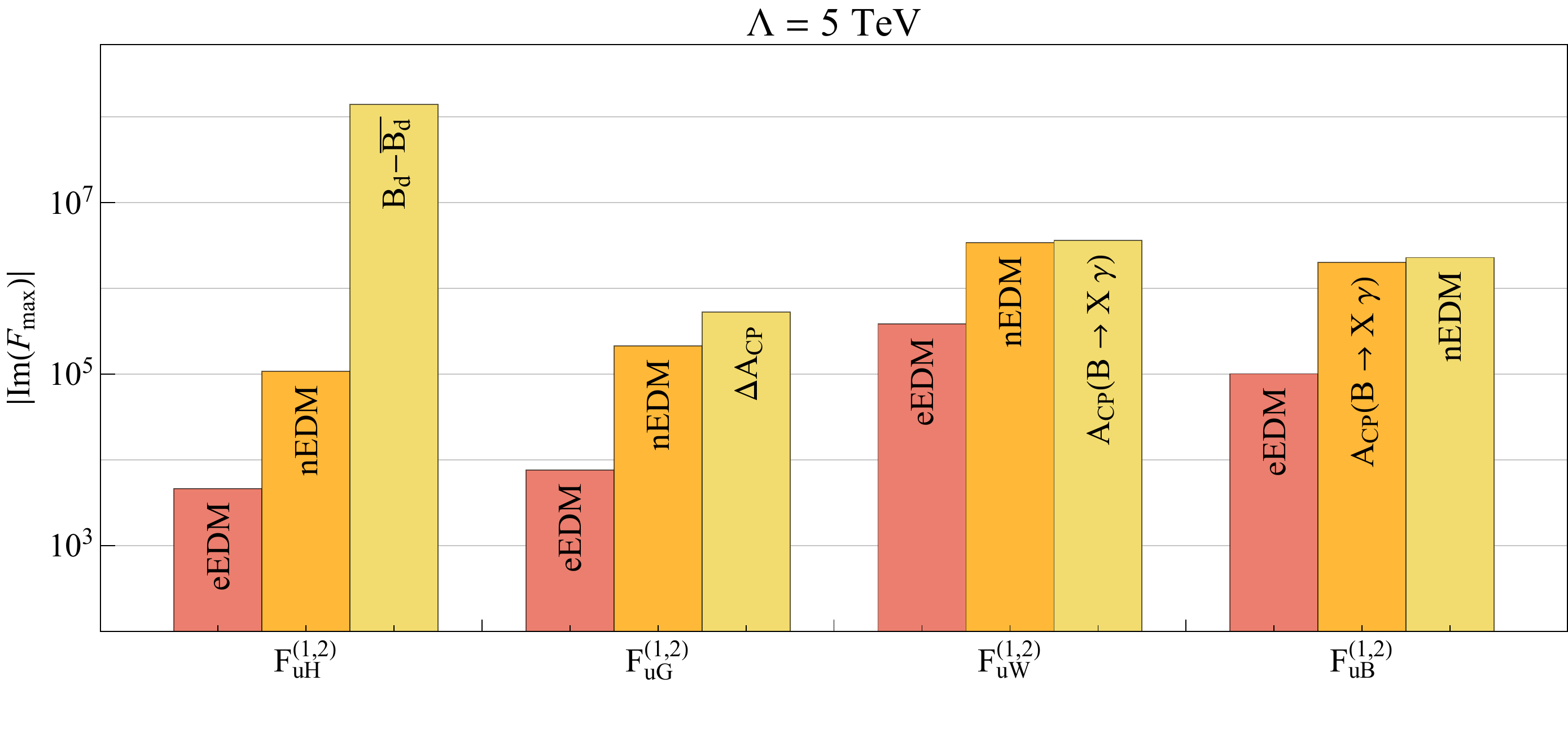}
    \caption{{\bf Top}: Upper bounds on the imaginary part of down-quark Yukawa-like (first set) and dipole (second to fourth set) coefficients in the MFV flavour scheme, fixing the New Physics scale to $\Lambda = 5$ TeV. We show the three strongest bounds for each coefficient as red, orange and dark yellow for first, second and third respectively. {\bf Bottom}: same for up-quark coefficients. The specific process that provides the bound is indicated by the label in the bar, as summarized in Table~\ref{table:bounds_summary}.}
    \label{fig:MFV_dipoles_down}
\end{figure}

%
In this Section we present the main results of this work, the bounds obtained on SMEFT coefficients at the $\Lambda = 5$ TeV scale in the two flavour schemes considered, namely MFV and $U(2)^3$. We first outline the numerical methodology to compute the effects in low energy observables in Section~\ref{subsec:Numerics}. In Section~\ref{subsec:Results:MFV} we discuss the results in the MFV scheme. The bounds on Yukawa and dipole operators are shown in Fig.~\ref{fig:MFV_dipoles_down}, while the full list of bounds is presented in Table~\ref{table:BestBounds:MFV}. In Section~\ref{subsec:Results:U2} we discuss the results in the $U(2)^3$ scheme. The bounds on Yukawa and dipole operators are shown in Figs.~\ref{fig:U2_dipoles_up} and \ref{fig:U2_dipoles_down}, while the full list of bounds is given in Tables~\ref{table:BestBounds:U2:UpDipoles}, \ref{table:BestBounds:U2:DownDipoles}, \ref{table:BestBounds:U2:QUQD1},
\ref{table:BestBounds:U2:QUQD8}, and
\ref{table:BestBounds:U2:QU}

\subsection{Numerical method}
\label{subsec:Numerics}
We fix the high energy scale to $\Lambda = 5$~TeV in our analysis, and present our results in terms of upper bounds on individual SMEFT Wilson coefficients, expanded in either the MFV or $U(2)^3$ scheme, separately. The schematic of our procedure is as follows:
\begin{enumerate}
    \item Fix flavour scheme (e.g. MFV);
    \item Fix one flavour invariant coefficient to be purely imaginary (e.g. ${\rm Im}[F_{dG}^{(2,1)}] = c$), and all the other coefficients to vanish at the scale $\Lambda$;
    \item Perform RG running from $\Lambda$ to the weak scale, which we fix to $\mu_W = 160$~GeV;
    \item Perform matching to LEFT;
    \item Compute effect on observables;
    \item Repeat the process by scanning over different values of $c$.
\end{enumerate}
The first two steps of our procedure are the only assumptions made on the imprint of heavy New Physics on SMEFT: we assume how the SM flavour symmetry is broken and we assume that a single type of operator is generated at the scale $\Lambda$ by integrating out the heavy degrees of freedom. Furthermore, we take the non-vanishing flavour invariant coefficients to be purely imaginary, as our work is focused on new CP-odd phases; in general the CP-even parts can be best bounded by a different set of observables. 

The RG running in step 3 is taken care of by the software {\tt wilson}~\cite{Aebischer:2018bkb}. Two inputs require special attention. Firstly, we need to set the operator basis as either `Warsaw up' or `Warsaw down', that is the Warsaw operator basis~\cite{Grzadkowski:2010es}, with either the up or down quarks rotated to their respective mass basis. 
We fix the basis to `Warsaw down' for all initial conditions.\footnote{We have checked, that our results, both in the MFV  and $U(2)^3$ schemes, are invariant under the change of the SMEFT basis to `Warsaw up', as expected. }
Individual observables are of course computed in the appropriate mass basis so in general a rotation to the chosen basis of SMEFT coefficients in the unbroken phase is required. 
Secondly, we need to input as the RGE initial condition the full spurionic term for the desired coefficient. Taking $F_{dG}^{(2,1)}$ as an example, the full initial condition at scale $\Lambda$ is $\C_{\underset{rs}{dG}}(\Lambda) = i \, c \times(Y_u Y_u^\dagger Y_d)^{rs}(\Lambda)$, where the flavour indices are $r,s = 1,2,3$. Assuming that the running of CKM matrix elements is negligible, all the scale dependence of $Y_{u,d}$ is then given by the quark masses, which thus need to be evaluated at the high scale. We list explicitly their numerical values in Appendix~\ref{app:NumericalInputs}.

The matching condition in step 4 also needs some additional care, as different conventions, normalizations and methods are used in the literature. We use the notation for low energy operators as in Eq.~\eqref{eq:LEFT:c7g}, and the matching conditions from \cite{Dekens_2019}; the latter includes the full one-loop matching condition in the Warsaw basis. We find that in most cases adding loop contributions to the matching does not give a sizable effect, thus we only keep the tree level part, except for two particular cases. Firstly, in the analysis for the four-fermion operator $\Q_{qu}$. This operator provides purely finite top-loop enhanced contribution to the matching to dipoles; that is, these do not mix with dipoles under RG running. The one-loop term is thus the dominating piece in the interplay with the low energy observables considered. Secondly, the CPV Higgs boson couplings can induce relatively large shifts in the matching contributions via two-loop Barr-Zee diagrams. After running our SMEFT coefficients to the electroweak scale, we use the results for the two-loop effects to the matching with the low energy basis presented in Refs.~\cite{Brod:2022bww,Brod:2018pli}, and schematically in Eqs.~\eqref{eq:TWOLOOP:cdF} to \eqref{eq:TWOLOOP:ceF}. We do not include other possible two-loop effects in our analysis; a full calculation of two-loop running and matching in SMEFT is not available in the literature yet and is beyond the scope of this work. Furthermore, we expect these additional terms to be suppressed.

Finally in step 6, we scan the flavour invariant coefficients in the range $c \in [10^{-3},10^3]$, taking log-equidistant points in the interval. Bounds that exceed these limits are obtained by extrapolating the  results of our scan outside this region.\footnote{Note that {\tt wilson} gives numerically unstable results when the initial condition is close to unity, $\tilde c/(\Lambda/{\rm GeV})^2 \sim 1$, where $\tilde c$ indicates the numerical value of some initial condition.}

\subsection{Results in MFV scheme}
\label{subsec:Results:MFV}
In the MFV scheme there are only flavour universal NP phases, meaning that all quark transitions are correlated, with the relative magnitude dictated by ratios of CKM matrix elements and quark masses. Nevertheless, we expect that processes with the smallest CKM and mass suppression, which are typically known with better experimental precision, will dominate in constraining the flavour invariant coefficients. 

We show in the top plot of Fig.~\ref{fig:MFV_dipoles_down} the three strongest bounds on the imaginary part of down-quark Yukawa-like (first set) and dipole (second to fourth sets) coefficients, $F_{dH}^{(2,1)}$ and $F_{dX}^{(2,1)}$ respectively. The same bounds are reported in Table~\ref{table:BestBounds:MFV}. The striking result shown is the dominant importance of flavour-changing bounds for down-quark dipoles. In particular, the constraints from $ A_{CP}(B\to X\g)$ are in general $F_{dX}^{(2,1)}  \lesssim 1$ and are at least one order of magnitude stronger than nEDM bounds. 
Conversely, the Yukawa-like coefficient $F_{dH}^{(2,1)}$ is mostly bounded by the electron EDM induced by CPV Higgs couplings, with $F_{dH}^{(2,1)} \lesssim 5\times10^2$, while $B$ meson mixings can only reach ${\cal O}(10^3)$ limits. The weakness of the latter bounds is to be expected, as the RG mixing with dipoles is small and further suppressed by the MFV expansion, thus significant quark dipole transitions are not generated by Yukawa-like operators.
Secondly, the dominant meson mixing contribution is induced by two insertions of the SMEFT operators, leading to the heavy scale suppression of $\Lambda^{-4}$. 

The same line of reasoning outlined above for $F_{dH}^{(2,1)}$ and $F_{dX}^{(2,1)}$ can be applied to $F_{uH}^{(1,2)}$ and $F_{uX}^{(1,2)}$, with bounds shown in the bottom plot of Fig.~\ref{fig:MFV_dipoles_down} and listed in Table~\ref{table:BestBounds:MFV}. In this case the MFV suppression for the $c-u$ sector is even stronger, $\sim y_c y_b^2 V_{ub}V_{cb}^*$, thus the bounds are far less stringent. It is noticeable however that the CP violating Higgs couplings, contributing via the two-loop Barr-Zee diagrams, give the strongest bound on all the coefficients shown. 

Finally, we comment on bounds on the four-fermion operators, $\Q_{quqd}^{(1)}$ and $\Q_{quqd}^{(8)}$, listed in Table~\ref{table:BestBounds:MFV}. These match to down dipoles and Yukawa operators via top-quark loops and are most constrained by two-loop EDMs and $b-s$ processes. In the MFV scheme however, these bounds only reach ${\cal O}(50)$ level in the best case due to the severe CKM suppression.

\begin{figure}[!h]
    \centering
    \includegraphics[width=.8\linewidth]{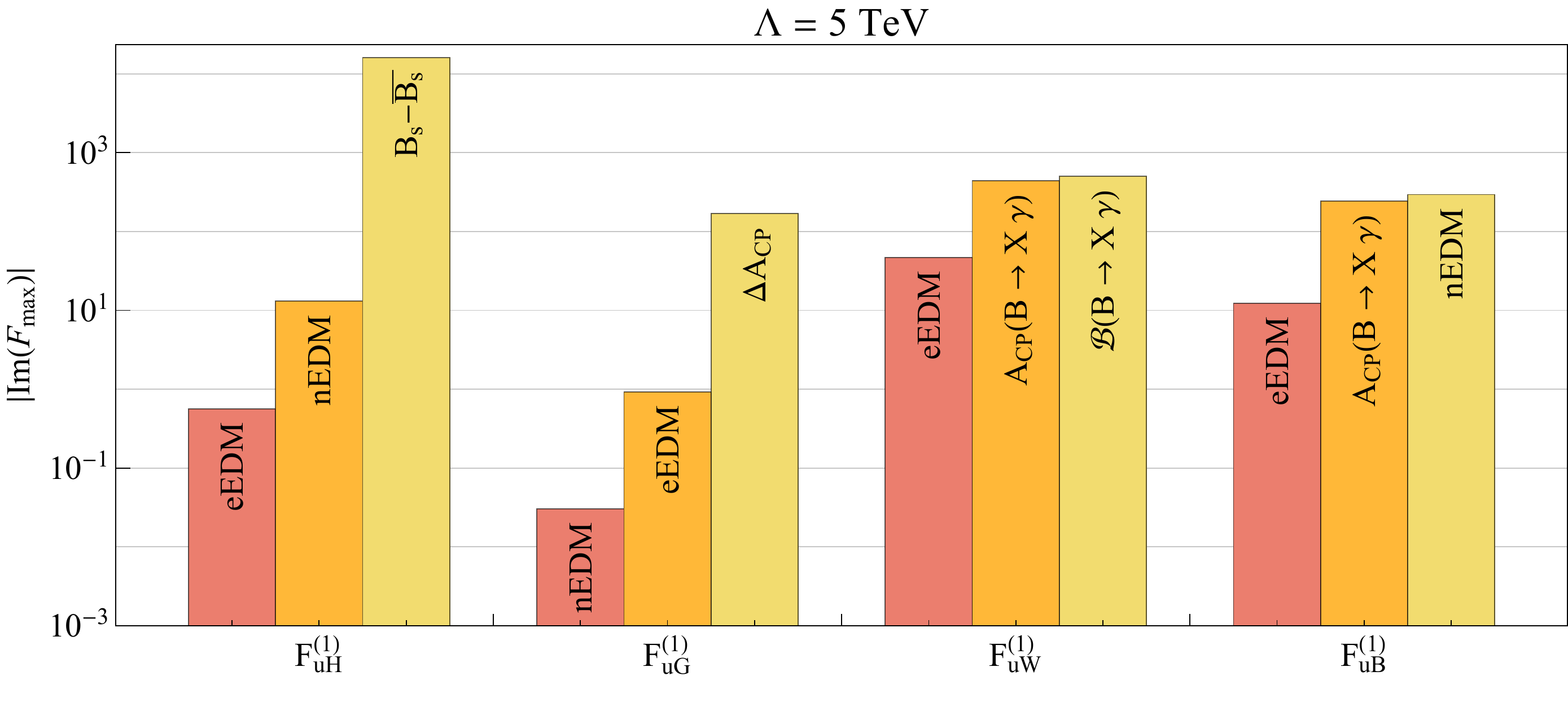}\\
    \includegraphics[width=.8\linewidth]{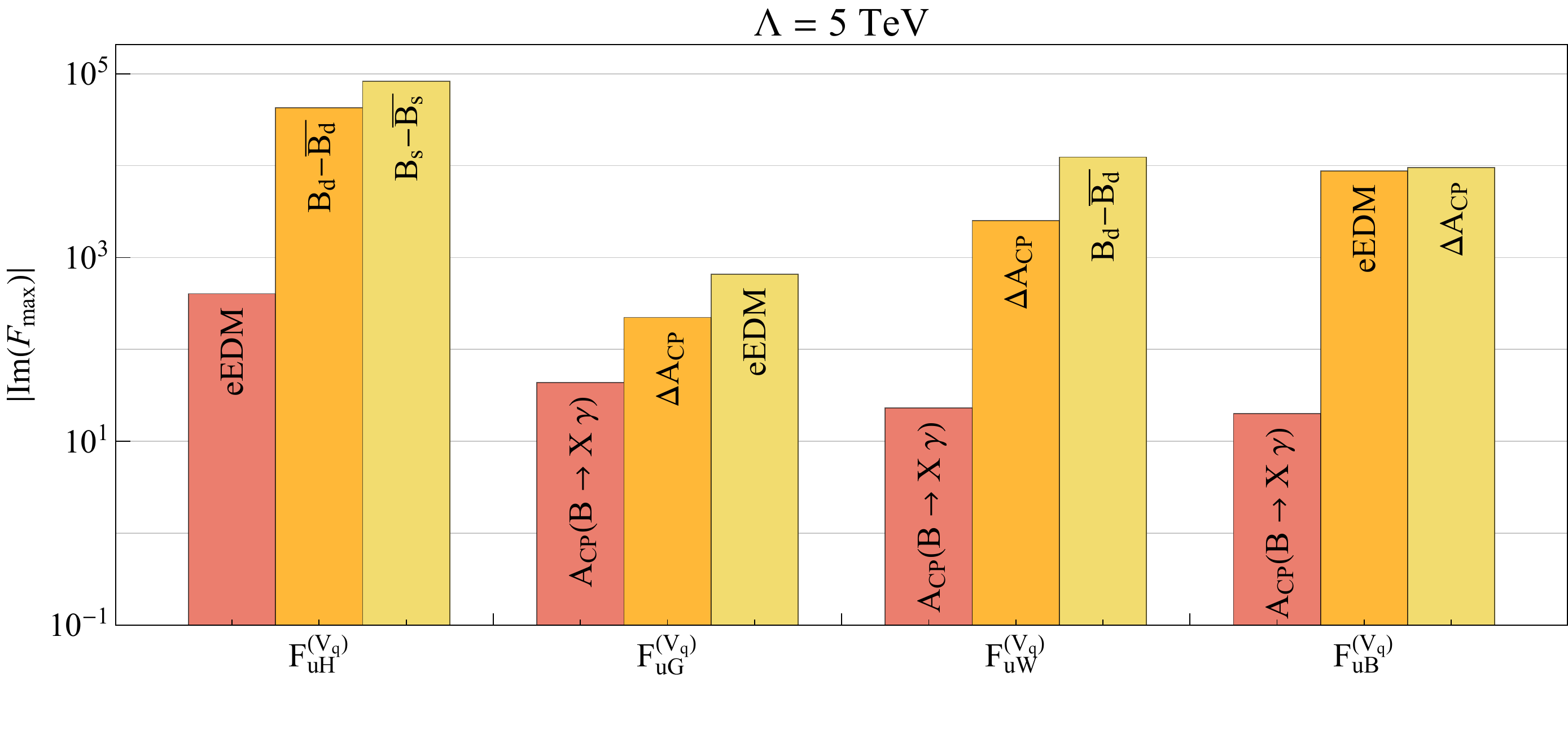}\\
    \includegraphics[width=.8\linewidth]{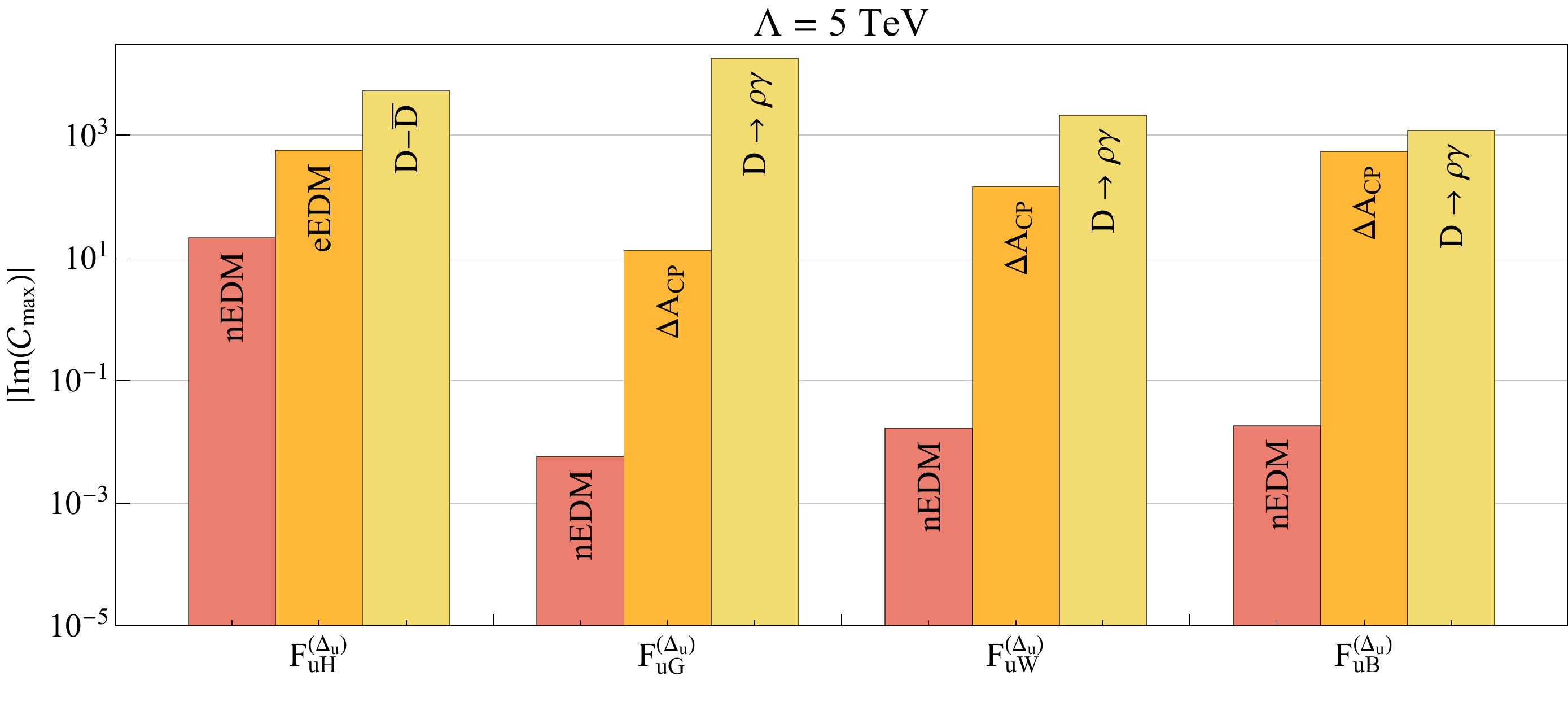}
    \caption{Upper bounds on the imaginary part of up-quark Yukawa (first set) and dipole (second to fourth set) coefficients in the $U(2)^3$ flavour scheme, fixing the New Physics scale to $\Lambda = 5$ TeV. See Eq.~\eqref{eq:U2expansion:uX} for the definition of the spurion expansion structures. The color code is the same as Fig.~\ref{fig:MFV_dipoles_down}. }
    \label{fig:U2_dipoles_up}
\end{figure}

\subsection{Results in $U(2)^3$ scheme}
\label{subsec:Results:U2}
In the $U(2)^3$ scheme we can disentangle transitions involving heavy quarks from the respective two light quark generations. The relevance of constraints for each coefficient will then in general reflect the spurion structure.

The dipole expansion down to ${\cal O}(10^{-4})$ contains four spurion structures, as shown in Eq.~\eqref{eq:U2expansion:uX}. We show in Fig.~\ref{fig:U2_dipoles_up} the results for up-quark dipoles, considering the $F_{uX}^{(1)}$ (top), $F_{uX}^{(V_q)}$ (center) and $F_{uX}^{(\Delta_u V_q)}$ (bottom) coefficients, respectively. The full list of bounds is reported in Table~\ref{table:BestBounds:U2:UpDipoles}, including bounds on the $F_{uX}^{(V_q,\Delta_u)}$ spurion. 

The first spurion, $F_{uX}^{(1)}$, induces NP contributions to operators involving exclusively third generation quarks. Consequently, any contributions to flavour-changing observables are purely due to RG running and LEFT matching. The chromo-electric dipole operator naturally generates a large shift in the nEDM, and is thus strongly bounded by it. Otherwise, the strongest bounds come from CPV Higgs couplings and their contributions to eEDM; these lead to $F_{uH(G)}^{(1)}\lesssim1$ and $F_{uW(B)}^{(1)}\lesssim10-10^2$. Additionally, the RG mixing in Eq.~\eqref{eq:mixing:dipoles:BWG} leads to large $b \to s$ effects at low energy, which in turn are probed by $B$-meson physics. 

The $V_q$ spurion is responsible for the heavy-light quark mixing. Furthermore, it does not match onto nEDM, thus CPV flavour-changing processes represent the only relevant probes. In a similar way as the $F_{uX}^{(1)}$ case, the RG evolution induces down-quark coefficients mixing heavy and light flavours, leading to $F_{uG,W,B}^{(V_q)}\lesssim20$, dominated by $A_\mathrm{CP}(B\to X \gamma)$. A competitive bound can only be achieved by $\Delta A_{CP}(D^0)$ in the case of the chromo-electric dipole, while other limits are two or more order of magnitude weaker. The $F_{uH}^{(V_q)}$ coefficient is again most stringently bounded by contributions to the eEDM, although only at ${\cal O}(10^2-10^3)$.

The $\Delta_u$ spurion dictates the light quark sector flavour expansion, and thus leads predominantly to contributions to flavour conserving terms; it is not surprising then that the EDMs dominate the budget of the constraints. We have $F_{uG,W,B}^{(\Delta_u)}\lesssim10^{-2}$ and $F_{uH}^{(\Delta_u)}\lesssim10$ from nEDM.
In this sector RG mixings with the respective down-quark counterparts are suppressed by mass terms, $\delta_d$ and $\delta_d'$, while the matching to low energy dipoles is purely tree level. This is reflected in the next-to-strongest bounds coming from the $D$ meson phenomenology, in particular from $A_{CP}(D^0)$; we have $F_{uG}^{(\Delta_u)}\lesssim10$ for chromo-electric dipole operator, while electromagnetic dipoles are only constrained to $F_{uW,B}^{(\Delta_u)}\lesssim10^2 - 10^3$.

The last spurion in the expansion shown in Eq.~\eqref{eq:U2expansion:uX}, $\Delta_u\otimes V_q$, generates higher order terms for the heavy - light generation mixing. These terms receive an additional $V_{ts}$ suppression with respect to $\Delta_u$, thus leading to similar but weaker limits, see Table~\ref{table:BestBounds:U2:UpDipoles}. Indeed, the strongest bounds come from $A_{CP}(D^0)$ but at the ${\cal O}(10^2-10^3)$ level.

\begin{figure}[t]
    \centering
    \includegraphics[width=.8\linewidth]{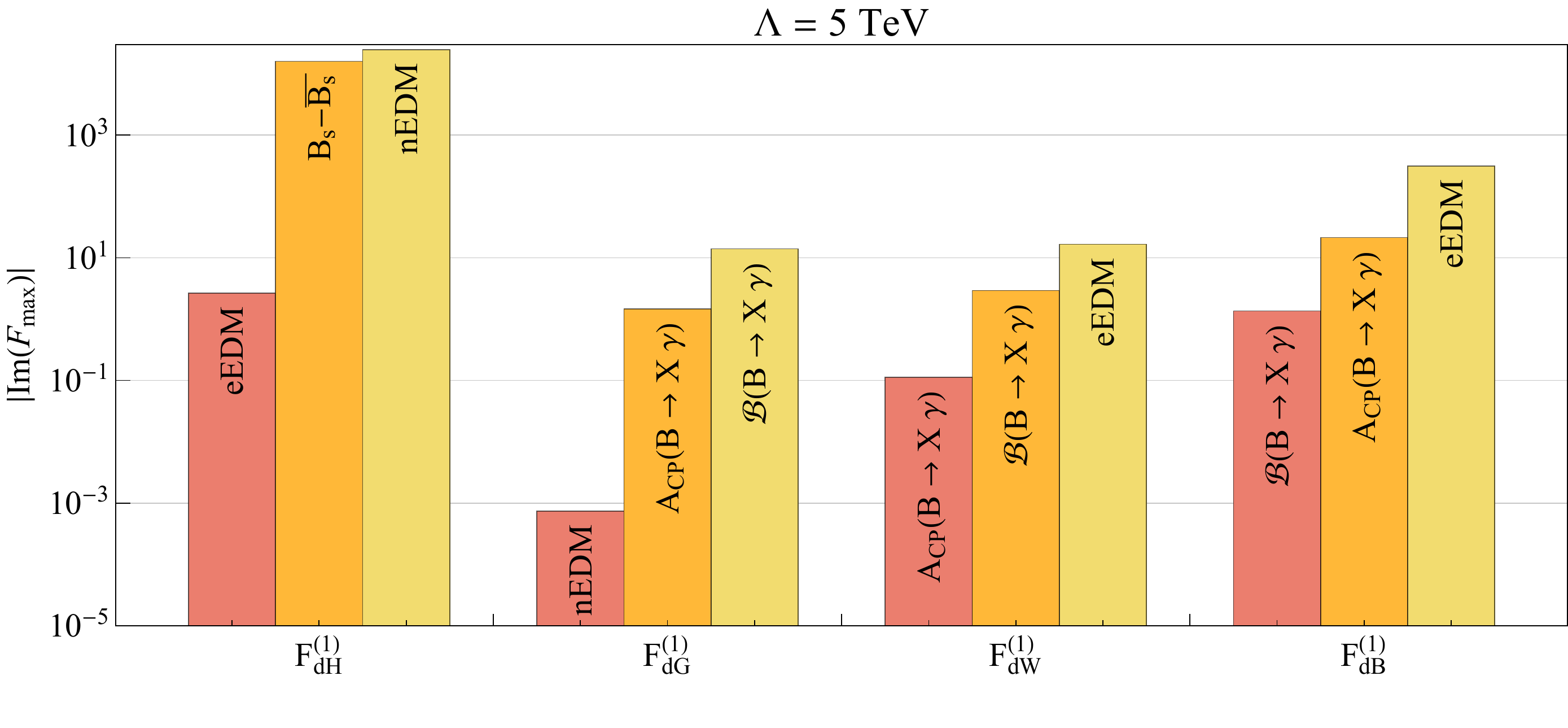}\\
    \includegraphics[width=.8\linewidth]{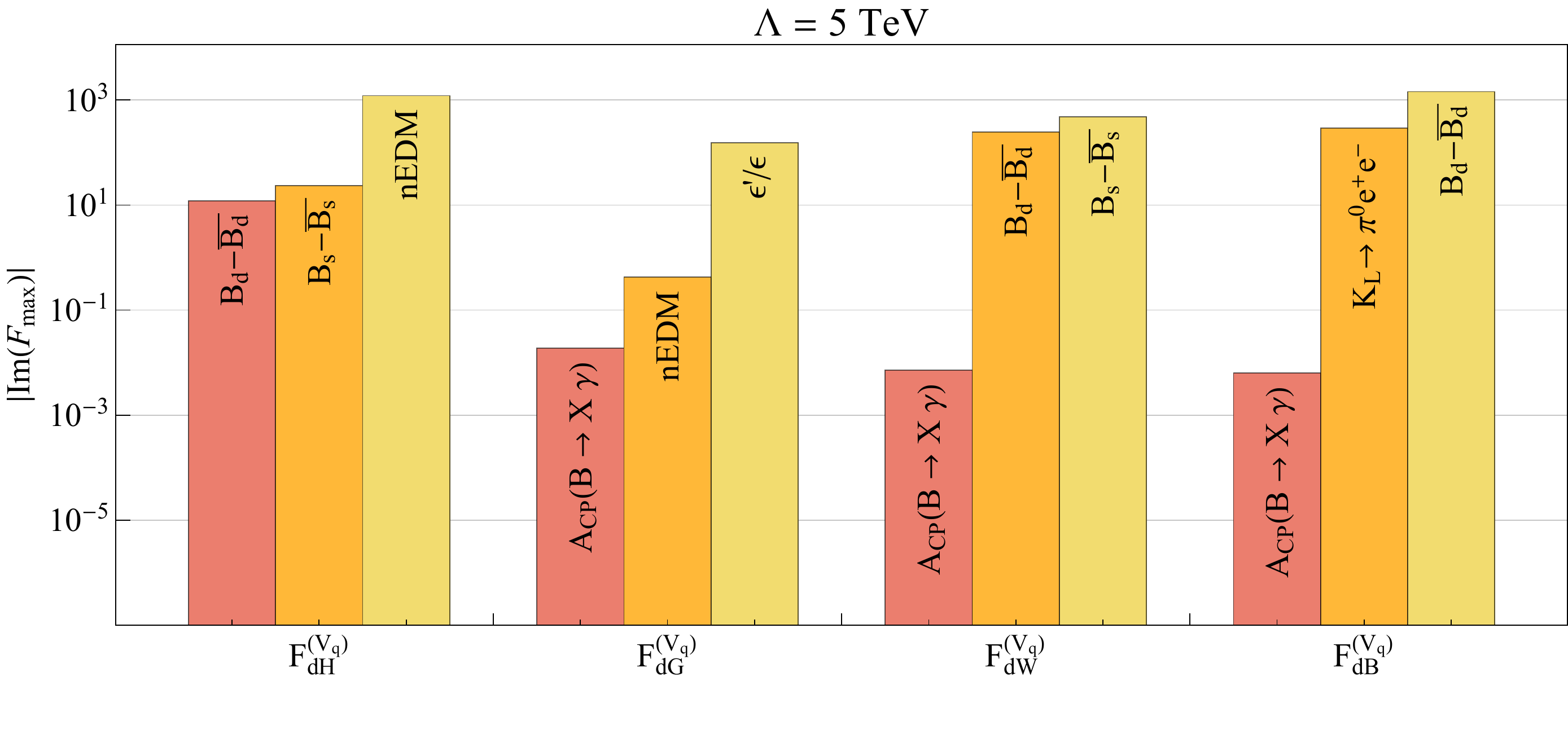}\\
    \includegraphics[width=.8\linewidth]{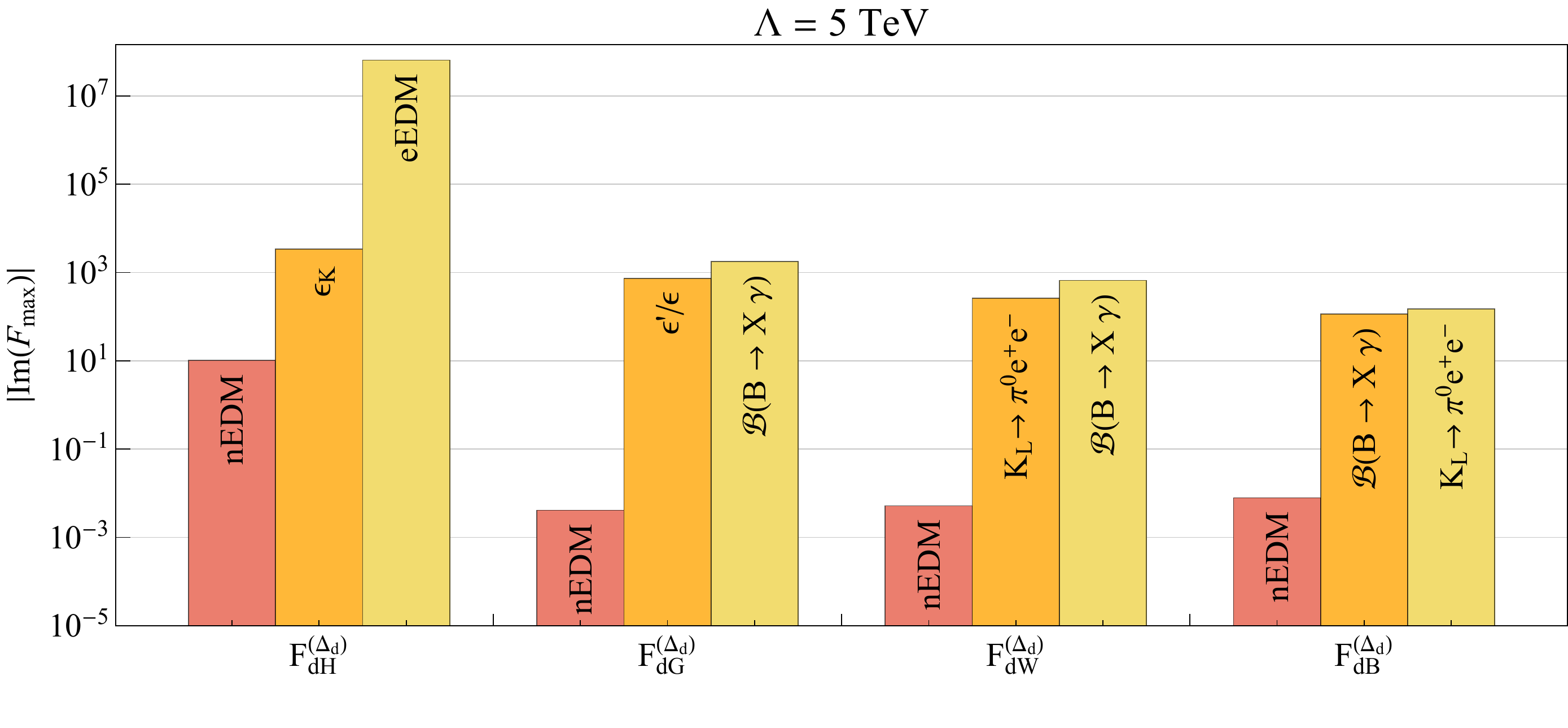}
    \caption{Same as Fig.~\ref{fig:U2_dipoles_up} for down-quark dipoles in the $U(2)^3$ scheme. }
    \label{fig:U2_dipoles_down}
\end{figure}

The analysis in the case of down-quark Wilson coefficients goes along the same lines of the up-quark case. We show the results in Fig.~\ref{fig:U2_dipoles_down} and Table~\ref{table:BestBounds:U2:DownDipoles}. In this sector $B$ and $K$ processes are generated directly, instead of via RG evolution. This is already clear in the results for $F_{dX}^{(1)}$, where $B\to X\g$ leads to ${\cal O}(0.1-1)$ bounds for all dipoles coefficients. These represent the tightest constraints for $W$ and $B$, while chromo-dipoles and Yukawa coefficients induce large contributions to EDMs, thus $F_{dG}^{(1)}\lesssim10^{-3}$ from nEDM and $F_{dH}^{(1)}\lesssim1$ from eEDM. 

Similarly, bounds on the $F_{dG}^{(V_q)}$ coefficients are dictated by $B$ meson physics, as $b-s$ mixings are generated at tree level. We have $F_{dG,W,B}^{(1)}\lesssim10^{-2}$ from $A_{{\rm CP}}(B\to X\g)$ and $F_{dH}^{(1)}\lesssim10$ from $B_d$ mixings. 

On the other hand, the $F_{dG}^{(\Delta_d)}$ spurions induce $d$ and $s$ processes directly and thus large shifts in EDMs, giving $F_{dG,W,B}^{(1)}\lesssim5\times10^{-3}$ and $F_{dH}^{(1)}\lesssim10$. The respective bounds from $s-d$ transitions, as $K$ meson mixing and $K_L$ decay, can only reach the ${\cal O}(10^3)$ level.

Finally, we comment on four-fermion operators. In the $U(2)^3$ flavour scheme the number of independent spurion structures appearing in such operators is quite large, see Eqs.~\eqref{eq:U2expansion:QUQD} and \eqref{eq:U2expansion:QU}. However, similarly to the dipole case, their effects and resulting bounds can be understood through the quark sectors upon which they act. 
Bounds for chirality flipping operators, $\Q_{quqd}^{(1)}$ and $\Q_{quqd}^{(8)}$, are listed in Tables~\ref{table:BestBounds:U2:QUQD1} and \ref{table:BestBounds:U2:QUQD8}, respectively. As in the MFV case, the mixing into the $b-s$ and $b-d$ sectors is the most relevant source of constraints. This is the case of the $(1)$ and $V_q$ spurion structures, which are limited to be $\lesssim{\cal O}(1)$ by $A_{\rm CP}\lp B\to X\g \rp$ measurements. The $V_q^2$ spurion follows this pattern, albeit with a weaker bound, ${\cal O}(10^3)$, due to the additional suppression. The $\Delta_u$ and $\Delta_d$ spurions act in the light quark sectors, thus the $F_{quqd}^{(\Delta_u)}$ and $F_{quqd}^{(\Delta_d)}$ are again strongly constrained by nEDM, with $F_{quqd}^{(\Delta_d)}\lesssim10^{-3}$ versus $F_{quqd}^{(\Delta_u)}\lesssim10$, due to the contraction of top quarks in the former case. Similarly, the different index contractions induce $\bar F_{quqd}^{(\Delta_d)}\lesssim10^2$ from nEDM, while $\bar F_{quqd}^{(\Delta_u)}\lesssim10^4$ from $B$ meson mixing. Lastly, the $V_q\otimes\Delta_{u,d}$ spurion can only be probed at the $10^3 - 10^5$ order due to the stronger suppression from light quark masses.
For the case of operators $\Q_{qu}^{(1)}$ and $\Q_{qu}^{(8)}$, we only consider the non-hermitian terms in Eq.~\eqref{eq:U2expansion:QU}, with bounds reported in Table~\ref{table:BestBounds:U2:QU}. The effect of these operators on observables is induced solely by the one-loop matching to low energy coefficients, Eqs.~\eqref{eq:matchingDipole:QU}, thus it is expected that CPV probes have a very weak constraining power for the $\Q_{qu}^{(1)}$ and $\Q_{qu}^{(8)}$ spurions. Indeed, the strongest bounds shown in Table~\ref{table:BestBounds:U2:QU} are typically of ${\cal O}(10^3 - 10^4)$. The single exception is the spurion $F_{qu}^{(\Delta_u)}$, which generates light quark dipoles and thus large shifts in the nEDM, with $F_{qu}^{(\Delta_u)}\lesssim4$.

%
\section{Conclusions}\label{sec:conclusions}
%
In the present paper we analysed CP violating effects of heavy NP inducing flavour-violating quark dipole transitions. We parameterized NP at high scales within the dimension six SMEFT framework and defined the relevant (dipole, four-quark and Higgs Yukawa-like) operators.
To properly calculate the effect of NP on low energy CPV observables, we included the one-loop RG running of SMEFT at dimension six, together with one-loop (and some known relevant two-loop) matching conditions at the weak scale. Both effects induce mixing of dipoles among themselves, as well as mixing with Yukawa-like operators and four-fermion operators, which thus needed to be included in our phenomenological analysis. This interplay leads in principle to possible probes of dipoles from the Yukawa and four-quark sector and vice versa. The most significant impact comes from one-loop contributions involving top quarks in the matching conditions of four fermion chirality flipping operators to flavour changing dipoles, as well as two-loop contributions of Yukawa-like operators to flavour conserving dipoles.

We explored two different symmetry based approaches to specifying the flavour structure of  SMEFT: MFV and $U(2)^{3}$. The MFV scheme leads to flavour universal NP phases, while all the flavour breaking is dictated by the SM Yukawa matrices. flavour violating dipoles in the up (down) sector are thus induced by insertions of down (up) Yukawas; this translates to suppressed NP contributions due to insertions of multiple CKM (and/or small quark mass) factors. 
The $U(2)^{3}$ symmetry scheme instead disentangles the third quark generation from the first two by assuming separate flavour groups: the light quarks transform under a $U(2)$ group, and the flavour symmetry is broken in a similar way as in the SM. Third generation quarks instead transform under a different group, namely $U(1)_{t}\times U(1)_{b}$, broken only by small mixings with the light generations. This leads to flavour specific NP phases and a richer flavour structure than in the MFV case.

We performed a comprehensive phenomenological analysis of existing constraints on CPV dipole transitions at low energy, as summarised  in Table~\ref{table:bounds_summary}. 
Firstly, we considered bounds on EDMs of the neutron and electron. These observables are mostly sensitive to flavour conserving and CP violating dipole operators, however important one- and two-loop matching contributions are also obtained for chirality flipping four-quark and Yukawa operators, respectively. We also considered the effects of flavour violating dipoles on the neutron EDM: the neutron mixes into baryon octet states via weak interactions, and the following $c\to u \gamma$, $b \to d \gamma$ or $s \to d \gamma$ transition induce a contribution to the nEDM. These bounds are mostly limited by the poor knowledge of the relevant form factors, which dominate the uncertainties in the computation of the matrix elements.
Secondly, we reviewed the present measurements of CPV effects in $D$, $K$ and $B$ meson decays, and the corresponding effects of flavour violating dipoles. On one hand, QCD dipole transitions are  most strongly constrained by measurements of direct CP asymmetries ($A_{{\rm CP}}$). For $c\to u g$, this comes from the $A_{{\rm CP}}$ difference of $D^0 \to \pi^{+}\pi^{-}$ and $D^0 \to K^{+} K^{-}$ decays, while for $s\to d g$ and $b\to sg$ the best sensitivity is exhibited by $\epsilon^{\prime}/\epsilon $ and $A_{{\rm CP}}(B\to K (X_s)\g)$ respectively. On the other hand, EM dipoles are at present best bounded by their contributions to rare radiative (and semileptonic) meson decays: $D\to\rho\g$, $K_L\to\pi^0e^+e^-$ and $B^{}\to X\g$ ($X=K^*,\phi,X_s$), for $c\to u \gamma$, $s \to d \gamma$ and $b \to d \gamma$ respectively. 
In addition, we also considered bounds on CP and flavour violating couplings of the Higgs boson coming from neutral meson mixing. These  currently represent the most stringent probes of such NP effects, which in turn can be induced by high scale dipole operators through operator mixing. 

We performed our numerical analysis, both in the MFV and $U(2)^3$ schemes, by fixing the heavy NP scale at $\Lambda = 5$ TeV, computing the full SMEFT one-loop RG evolution and matching conditions to LEFT at the weak scale $\mu_W = 160$~GeV. The resulting bounds on single coefficients are summarized in Figs.~\ref{fig:MFV_dipoles_down}, \ref{fig:U2_dipoles_up} and \ref{fig:U2_dipoles_down} and in the tables in Appendix~\ref{app:AllBounds}. 
Our results demonstrate that flavour violating observables can serve as complementary probes to EDMs, giving in most cases similar or even stronger bounds. This is most evident in down-quark dipoles, where the most stringent limits are provided by precise measurements of CP asymmetries in $b\to s\g$ transitions. Similarly, the top-loop enhanced matching conditions of $\Q_{quqd}$ operators into dipoles lead to ${\cal O}(10)$ and ${\cal O}(10^2)$ limits on the Wilson coefficients from $b\to s\g$ in MFV and $U(2)^3$ schemes respectively.
In the up-quark sector, smaller quark masses and CKM factors entering $c\to u$ transitions generally lead to strongly suppressed contributions to low energy observables within MFV. Consequently,  the best $\sim{\cal O}(10^3)$ or larger bounds are obtained by two-loop induced EDMs. In the $U(2)^3$ scheme these suppressions are alleviated and can lead to stronger bounds, ${\cal O}(1-10)$, depending on the precise spurion structure considered.
{Finally we note that within the general structure of SMEFT, EDMs can receive  contributions from additional operators, which do not carry quark flavour indices and have thus been omitted in the present analysis. Possible cancellations among these various contributions can thus partially alleviate EDM constraints on dipole operators, making complementary flavour probes essential.}

The results presented in this paper can serve as a guideline for building models of heavy NP that violate CP at high scales, as well as a benchmark for the present status of CPV bounds in SMEFT. Future high precision experimental results will help improve the bounds considered in this analysis. Together with better limits on neutron and electron EDMs, precise measurements of CP asymmetries in $K$, $D$ and $B$ meson decays are thus highly anticipated. 

%
\section*{Acknowledgments}
%

We thank Luka Leskovec for discussions on the hadronic matrix elements and Ajdin Palavrić for useful discussions on the spurion expansions. The authors acknowledge the financial support from the Slovenian Research Agency (grant No. J1-3013 and research core funding No. P1-0035).

\begin{appendix}

\section{Numerical inputs}
\label{app:NumericalInputs}
Here we list the numerical inputs used in the analysis.
We employ the CKM matrix with the following values of rephasing invariants:
\begin{align}
&|V_{us}|=0.2243\,,\qquad |V_{cb}| = 0.04221\,,\qquad |V_{ub}| = 3.62\E{-3}\,,\\
&J = \Im \left[V_{ud} V_{us}^* V_{cd}^* V_{cs} \right] = 3.1872\E{-5}\,.\nonumber
\end{align}
These exactly reproduce the CKM matrix form in the \texttt{wilson} package, guaranteeing the consistency of flavour rotations that we use throughout the calculation. The above set corresponds to the standard Wolfenstein parameters
\beq 
\lambda = 0.2243\,, \qquad A = 0.8390\,, \qquad \bar{\rho} = 0.1105\,, \qquad \bar{\eta} = 0.3562\,.
\eeq
We assume that the running of CKM elements is negligible. For completeness, we give here the rotation matrices that transform the quarks from the ``primed'' basis \eqref{eq:U2scheme:Yukawa:spurions} to the quark mass basis in the standard CKM phase convention:
\begin{align}
W_u &= 
\begin{pmatrix}
 0.99634 & 0.085449 & -4.8847\E{-8} \\
 0.085372 & -0.99545 & 0.042365 \\
 -0.00362 & 0.04221 & 0.99910 
\end{pmatrix}\,\mathrm{diag}(e^{-1.872\, i},1,1)\,,\\
W_d &= 
\begin{pmatrix}
 -0.97681 & -0.21412 & 0 \\
 -0.012985-0.21373i & 0.059234+ 0.97501\,i & 0\\
 0 & 0 & 1 
 \end{pmatrix}\,\mathrm{diag}(e^{1.251\, i},e^{1.650\, i},1)
 \end{align}
for the rotations of the left-handed quarks, and 
\begin{align}
    V_u &= 
    \begin{pmatrix}        
     1 & 3.0642\E{-7} & 2.6680\E{-8} \\
3.0642\E{-7} & -(1-1.2076\E{-8}) & 0.00015541 \\
 -2.6727\E{-8} & 0.00015541 & 1-1.2076\E{-8}
 \end{pmatrix}\,\mathrm{diag}(e^{-1.872\, i},1,1)\,,\\
V_d &= \mathbb{1}\, \mathrm{diag}(e^{1.251\, i},e^{1.650\, i},1)\,,
\end{align}
for the right-handed quarks. Rephasings of the left- and right-handed quark components have to be the same to keep the quark mass terms real.

Next, we give the values of the three gauge couplings at the matching scale, $\mu_W = 160$ GeV. We have
\beq 
g_1(\mu_W) = 0.358\,, \qquad g_2(\mu_W) = 0.648\,, \qquad g_3(\mu_W) = 1.177\,.
\eeq
The running of the gauge couplings is decoupled from the other SM sectors~\cite{Mihaila:2012fm} and is not affected by new SMEFT operators in our framework. The evolution to the scale $\Lambda = 5$ TeV gives
\beq 
g_1(\Lambda) = 0.366\,, \qquad g_2(\Lambda) = 0.630\,, \qquad g_3(\Lambda) = 0.983\,.
\eeq
Turning our attention to the SM Yukawa couplings, at the matching scale we have
\beq
\begin{split}
    & y_u(\mu_W) = 6.967\times10^{-6}\,, \qquad y_c(\mu_W) = 3.477\times10^{-3}\,, \qquad y_t(\mu_W) = 9.413\times10^{-1}\,, \\
    & y_d(\mu_W) = 1.480\times10^{-5}\,, \qquad y_s(\mu_W) = 2.928\times10^{-4}\,, \qquad y_b(\mu_W) = 1.578\times10^{-2}\,.
\end{split}
\eeq
The running of Yukawa couplings to $\Lambda$ is with good approximation independent of SMEFT operators and driven by the QCD evolution as $y_i(\Lambda) = \left[\alpha_s(\Lambda)/\alpha_s(\mu_W)\right]^{14/21}y_i(\mu_W)\sim0.8~y_i(\mu_W)$, where $\alpha_s = g_3^2/(4\pi)$. We obtain the values
\beq
\begin{split}
    & y_u(\Lambda) = 5.612\times10^{-6}\,, \qquad y_c(\Lambda) = 2.801\times10^{-3}\,, \qquad y_t(\Lambda) = 7.601\times10^{-1}\,, \\
    & y_d(\Lambda) = 1.193\times10^{-5}\,, \qquad y_s(\Lambda) = 2.359\times10^{-4}\,, \qquad y_b(\Lambda) = 1.267\times10^{-2}\,.
\end{split}
\eeq
The parameters of the $U(2)^3$ spurions can be obtained with the procedure detailed in Section~\ref{subsec:U2scheme}. The parameter $\epsilon_q$, computed at the scale $\Lambda$, is
\beq 
\epsilon_q(\Lambda) = y_t(\Lambda) |V_{ts}| = 3.22\times10^{-2}\,.
\eeq
The phase and rotation angles in Eq.~\eqref{eq:CKMinvariants} are defined in terms of CKM element ratios and thus are scale invariant:
\begin{equation}
 \theta_d = 2.9258\,,\quad \theta_u = 0.085477\,, \quad \epsilon_q = 0.032230\,,\quad  \alpha_d = 1.63147\,.
\end{equation}

%
\section{Tables of bounds}
\label{app:AllBounds}
%
Here we list the three most stringent bounds for all the relevant coefficients in the operator flavour expansions. See Section~\ref{sec:flavourExpansion} for definitions of the terms in the MFV and $U(2)$ expansions. All bounds are referred to the respective Wilson coefficient taken at the New Physics scale $\Lambda = 5$ TeV, see Section~\ref{subsec:Numerics} for details on the numerical method.

%
\subsection{Tables of bounds: MFV scheme}
\label{app:AllBounds:MFV}
%

\begin{table}[!h]
\begingroup
\setlength{\tabcolsep}{16pt} 
\renewcommand{\arraystretch}{1.2} 
\centering
\begin{tabular}{c | c | c | c} 
 \hline\hline
 Coefficient & 1st &  2nd & 3rd  \\ 
   \hline
  \multirow{2}{2em}{ $F_{uH}^{(1,2)}$ } &  eEDM & nEDM & $B_d^0 - \bar B_d^0$
    \\ 
  & $4.6\times10^3$ & $1.1\times10^5$ & $1.4\times10^8$ 
  \\ 
    \hline
  \multirow{2}{2em}{ $F_{uG}^{(1,2)}$ } &  eEDM & nEDM & $\Delta A_{{\rm CP}}(D^0)$
    \\ 
  & $7.6\times10^3$ & $2.1\times10^5$ & $5.3\times10^5$ 
  \\ 
      \hline
  \multirow{2}{2em}{ $F_{uW}^{(1,2)}$ } &  eEDM & nEDM & $A_{{\rm CP}}(B\to X\g)$
    \\ 
  & $3.8\times10^5$ & $3.4\times10^6$ & $3.7\times10^6$ 
  \\ 
    \hline
  \multirow{2}{2em}{ $F_{uB}^{(1,2)}$ } &  eEDM & $A_{{\rm CP}}(B\to X\g)$ & nEDM
    \\ 
  & $1.0\times10^5$ & $2.0\times10^6$ & $2.3\times10^6$
  \\ 
  \hline \hline
    \multirow{2}{2em}{ $F_{dH}^{(2,1)}$ } & eEDM & $B_d^0 - \bar B_d^0$ & $B_s^0 - \bar B_s^0$ 
    \\ 
  & $3.6\times 10^2$& $1.2\times10^3$ & $2.4\times10^3$ 
  \\
     \hline
  \multirow{2}{2em}{ $F_{dG}^{(2,1)}$ } & $A_{{\rm CP}}(B\to X\g)$ & nEDM & ${\cal B}(B\to X\g)$
    \\ 
  & $2.0$ & $2.5\times10^1$ & $1.5\times10^2$
  \\
       \hline
  \multirow{2}{2em}{ $F_{dW}^{(2,1)}$ } & $A_{{\rm CP}}(B\to X\g)$ & ${\cal B}(B\to X\g)$ & nEDM
    \\ 
  & $8.2\times10^{-1}$ & $2.4\times10^1$ & $5.2\times10^1$
  \\
    \hline
  \multirow{2}{2em}{ $F_{dB}^{(2,1)}$ } & $A_{{\rm CP}}(B\to X\g)$  & ${\cal B}(B\to X\g)$ & nEDM
    \\ 
  & $6.7\times10^{-1}$ & $1.4\times10^1$ &  $3.4\times10^1$
  \\
  \hline \hline
    \multirow{2}{2em}{ $F_{quqd,1}^{(1,1)}$ } & $A_{{\rm CP}}(B\to X\g)$ & nEDM & eEDM 
    \\ 
  & $9.8\times10^{1}$ & $3.4\times10^2$ & $1.4\times10^3$
  \\
     \hline
  \multirow{2}{2em}{ $\tilde F_{quqd,1}^{(1,1)}$ }  & $A_{{\rm CP}}(B\to X\g)$ & nEDM & eEDM
    \\ 
  & $2.8\times10^2$ & $3.4\times10^2$ & $1.4\times 10^3$
  \\
       \hline
  \multirow{2}{2em}{ $F_{quqd,8}^{(1,1)}$ } & $A_{{\rm CP}}(B\to X\g)$ & nEDM & ${\cal B}(B\to X\g)$
    \\ 
  & $6.8\times10^{1}$ & $3.9\times10^2$ & $1.7 \times 10^3$
  \\
    \hline
  \multirow{2}{2em}{ $\tilde F_{quqd,8}^{(1,1)}$ } & nEDM & $A_{{\rm CP}}(B\to X\g)$ & eEDM
    \\ 
   & $3.9\times10^2$ & $1.3\times10^3$ & $8.3\times10^{3}$
  \\
 \hline\hline
\end{tabular}
\caption{Three strongest bounds on up-quark Yukawa, dipoles, and the considered four-fermion operators in the MFV scheme.}
\label{table:BestBounds:MFV}
\endgroup
\end{table}

\clearpage
%
\subsection{Tables of bounds: $U(2)^3$ scheme}
\label{app:AllBounds:U2}
%

%
%

\begin{table}[!h]
\begingroup
\setlength{\tabcolsep}{16pt} 
\renewcommand{\arraystretch}{1.1} 
\centering
\begin{tabular}{c | c | c | c} 
 \hline\hline
 Coefficient & 1st &  2nd & 3rd  \\ 
   \hline
  \multirow{2}{5em}{ $F_{uH}^{(1)}$ } &  eEDM & nEDM & $B_s^0 - \bar B_s^0$
    \\ 
  & $5.6\times10^{-1}$ & $1.3\times10^1$ & $1.6\times10^4$ 
  \\ 
    \hline
  \multirow{2}{5em}{ $F_{uG}^{(1)}$ } &  nEDM & eEDM & $\Delta A_{{\rm CP}}(D^0)$
    \\ 
  & $3.0\times10^{-2}$ & $9.3\times10^{-1}$ & $1.7\times 10^2$ 
  \\ 
      \hline
  \multirow{2}{5em}{ $F_{uW}^{(1)}$ } &  eEDM & $A_{{\rm CP}}(B\to X\g)$ & ${\cal B}(B\to X\g)$
    \\ 
  & $4.6\times10^1$ & $4.4\times10^2$ & $5.0\times10^2$ 
  \\ 
    \hline
  \multirow{2}{5em}{ $F_{uB}^{(1)}$ } &  eEDM & $A_{{\rm CP}}(B\to X\g)$ & nEDM
    \\ 
  & $1.2\times10^1$  & $2.4\times10^2$  & $2.9 \times 10^2$
  \\ 
     \hline\hline
  \multirow{2}{5em}{ $F_{uH}^{(V_q)}$ } &  eEDM & $B_d^0 - \bar B_d^0$ & $B_s^0 - \bar B_s^0$
    \\ 
  & $4.0\times10^2$ & $4.3\times10^4$  & $8.3\times10^4$ 
  \\ 
    \hline
  \multirow{2}{5em}{ $F_{uG}^{(V_q)}$ } &  $A_{{\rm CP}}(B\to X\g)$ & $\Delta A_{{\rm CP}}(D^0)$ & eEDM 
    \\ 
  & $4.4\times10^1$ & $2.2\times10^2$ & $6.6\times10^2$ 
  \\ 
      \hline
  \multirow{2}{5em}{ $F_{uW}^{(V_q)}$ } &  $A_{{\rm CP}}(B\to X\g)$ & $\Delta A_{{\rm CP}}(D^0)$ & $B_d^0 - \bar B_d^0$
    \\ 
  & $2.3\times10^1$ & $2.5\times10^3$ & $1.3\times10^4$ 
  \\ 
    \hline
  \multirow{2}{5em}{ $F_{uB}^{(V_q)}$ } &  $A_{{\rm CP}}(B\to X\g)$ & eEDM & $\Delta A_{{\rm CP}}(D^0)$
    \\ 
  & $2.0\times10^1$ & $8.7\times10^3$ & $9.5\times10^3$ 
  \\ 
     \hline\hline
  \multirow{2}{5em}{ $F_{uH}^{(\Delta_u)}$ } & nEDM & eEDM & $D^0 - \bar D^0$ 
    \\ 
  & $2.1\times10^1$ & $5.7\times10^2$   & $5.3\times10^3$ 
  \\ 
    \hline
  \multirow{2}{5em}{ $F_{uG}^{(\Delta_u)}$ } &  nEDM & $\Delta A_{{\rm CP}}(D^0)$ & $D^0\to\rho\g$
    \\ 
  & $5.8\times10^{-3}$ & $1.3\times10^1$ & $1.8\times10^4$
  \\ 
      \hline
  \multirow{2}{5em}{ $F_{uW}^{(\Delta_u)}$ } &  nEDM & $\Delta A_{{\rm CP}}(D^0)$ & $D^0\to\rho\g$
    \\ 
  & $1.7\times10^{-2}$ & $1.5\times10^2$ & $2.1\times10^3$
  \\ 
    \hline
  \multirow{2}{5em}{ $F_{uB}^{(\Delta_u)}$ } &  nEDM & $\Delta A_{{\rm CP}}(D^0)$ & $D^0\to\rho\g$
    \\ 
  & $1.8\times10^{-2}$ & $5.5\times10^2$ & $1.2\times10^3$
  \\ 
     \hline\hline
  \multirow{2}{5em}{ $F_{uH}^{(V_q,\Delta_u)}$ } &  $D^0 - \bar D^0$ & eEDM & nEDM
    \\ 
  & $1.2\times10^5$ & $4.1\times10^5$ & $9.2\times10^7$ 
  \\ 
    \hline
  \multirow{2}{5em}{ $F_{uG}^{(V_q,\Delta_u)}$ } &  $\Delta A_{{\rm CP}}(D^0)$ & $D^0\to\rho\g$ & nEDM
    \\ 
  & $2.9\times10^2$ & $3.6\times10^6$ & $8.1\times10^6$ 
  \\ 
      \hline
  \multirow{2}{5em}{ $F_{uW}^{(V_q,\Delta_u)}$ } &  $\Delta A_{{\rm CP}}(D^0)$ & $D^0\to\rho\g$ & nEDM
    \\ 
  & $3.6\times10^3$ & $1.6\times10^5$ & $3.6\times10^5$ 
  \\ 
    \hline
  \multirow{2}{5em}{ $F_{uB}^{(V_q,\Delta_u)}$ } &  $\Delta A_{{\rm CP}}(D^0)$ & $D^0\to\rho\g$ & $D^0 - \bar D^0$
    \\ 
  & $1.2\times10^4$ & $1.9\times10^7$ & $2.7\times10^7$ 
  \\ 
 \hline\hline
\end{tabular}
\caption{Three strongest bounds on up-quark Yukawa and dipoles coefficients in the $U(2)^3$ scheme. }
\label{table:BestBounds:U2:UpDipoles}
\endgroup
\end{table}

%
%

\begin{table}[!h]
\begingroup
\setlength{\tabcolsep}{16pt} 
\renewcommand{\arraystretch}{1.1} 
\centering
\begin{tabular}{c | c | c | c} 
 \hline\hline
 Coefficient & 1st &  2nd & 3rd  \\ 
   \hline
  \multirow{2}{5em}{ $F_{dH}^{(1)}$ } &  eEDM & $B_s^0 - \bar B_s^0$ & nEDM
    \\ 
  & $2.6$ & $1.6\times10^4$  & $2.5\times10^4$ 
  \\ 
    \hline
  \multirow{2}{5em}{ $F_{dG}^{(1)}$ } &  nEDM & $A_{{\rm CP}}(B\to X\g)$ & ${\cal B}(B\to X\g)$
    \\ 
  & $7.4\times10^{-2}$ & $1.5$ & $1.4\times10^1$ 
  \\ 
      \hline
  \multirow{2}{5em}{ $F_{dW}^{(1)}$ } & $A_{{\rm CP}}(B\to X\g)$ & ${\cal B}(B\to X\g)$ & eEDM
    \\ 
  & $1.1\times10^{-1}$ & $2.9$ & $1.7\times10^1$ 
  \\ 
    \hline
  \multirow{2}{5em}{ $F_{dB}^{(1)}$ } & ${\cal B}(B\to X\g)$ & $A_{{\rm CP}}(B\to X\g)$ & eEDM
    \\ 
  & $1.4$ & $2.1\times10^1$ & $3.2\times10^2$ 
  \\ 
     \hline\hline
  \multirow{2}{5em}{ $F_{dH}^{(V_q)}$ } &  $B_d^0 - \bar B_d^0$ & $B_s^0 - \bar B_s^0$ & nEDM
    \\ 
  & $1.2\times10^1$ & $2.4\times10^1$ & $1.2\times10^3$
  \\ 
    \hline
  \multirow{2}{5em}{ $F_{dG}^{(V_q)}$ } &  $A_{{\rm CP}}(B\to X\g)$ & nEDM & $\epsilon'/\epsilon$
    \\ 
  & $1.9\times10^{-2}$  & $4.2\times10^{-1}$ & $1.5\times10^2$ 
  \\ 
      \hline
  \multirow{2}{5em}{ $F_{dW}^{(V_q)}$ } &  $A_{{\rm CP}}(B\to X\g)$ & $B_d^0 - \bar B_d^0$ & $B_s^0 - \bar B_s^0$
    \\ 
  & $7.2\times10^{-3}$ & $2.5\times10^2$ & $4.7\times10^2$ 
  \\ 
    \hline
  \multirow{2}{5em}{ $F_{dB}^{(V_q)}$ } &  $A_{{\rm CP}}(B\to X\g)$ & $K_L\to\pi^0e^+e^-$ & $B_d^0 - \bar B_d^0$
    \\ 
  & $6.4\times10^{-3}$ & $2.9\times10^2$ & $1.4\times10^3$ 
  \\ 
     \hline\hline
  \multirow{2}{5em}{ $F_{dH}^{(\Delta_d)}$ }  & nEDM & $\epsilon_K$ &  eEDM
    \\ 
   & $1.0\times10^1$ & $3.4\times10^3$ & $6.5 \times 10^7$
  \\ 
    \hline
  \multirow{2}{5em}{ $F_{dG}^{(\Delta_d)}$ } &  nEDM & $\epsilon'/\epsilon$ & ${\cal B}(B\to X\g)$
    \\ 
  & $4.1\times10^{-3}$ & $7.5\times10^2$ & $1.8\times10^3$
  \\ 
      \hline
  \multirow{2}{5em}{ $F_{dW}^{(\Delta_d)}$ } &  nEDM  & $K_L\to\pi^0e^+e^-$ & ${\cal B}(B\to X\g)$
    \\ 
  & $5.2\times10^{-3}$  & $2.6\times10^2$ & $6.6\times10^2$
  \\ 
    \hline
  \multirow{2}{5em}{ $F_{dB}^{(\Delta_d)}$ } &  nEDM & ${\cal B}(B\to X\g)$ & $K_L\to\pi^0e^+e^-$
    \\ 
  & $7.8\times10^{-3}$ & $1.2\times10^2$ & $1.5\times 10^2$
  \\ 
     \hline\hline
  \multirow{2}{5em}{ $F_{dH}^{(V_q,\Delta_d)}$ } & nEDM & $B_s^0 - \bar B_s^0$ & $B_d^0 - \bar B_d^0$
    \\ 
 & $7.7\times10^4$ & $9.6\times10^4$   & $9.7\times10^5$
  \\ 
    \hline
  \multirow{2}{5em}{ $F_{dG}^{(V_q,\Delta_d)}$ } &  ${\cal B}(B\to X\g)$   & $K_L\to\pi^0e^+e^-$ & $\epsilon'/\epsilon$
    \\ 
  & $9.2\times10^1$  & $7.4\times10^5$ & $1.9\times10^6$
  \\ 
      \hline
  \multirow{2}{5em}{ $F_{dW}^{(V_q,\Delta_d)}$ } &  ${\cal B}(B\to X\g)$  & $K_L\to\pi^0e^+e^-$ & $\epsilon_{K}$ 
    \\ 
  & $1.9\times10^1$ & $3.3\times10^4$ & $1.4\times10^5$ 
  \\ 
    \hline
  \multirow{2}{5em}{ $F_{dB}^{(V_q,\Delta_d)}$ } &  ${\cal B}(B\to X\g)$  & $\epsilon'/\epsilon$ & $K_L\to\pi^0e^+e^-$
    \\ 
  & $8.9$ & $2.1\times10^6$ & $8.0\times10^6$
  \\ 
 \hline\hline
\end{tabular}
\caption{Three strongest bounds on down-quark Yukawa and dipoles coefficients in the $U(2)^3$ scheme. }
\label{table:BestBounds:U2:DownDipoles}
\endgroup
\end{table}

%
%

\begin{table}[!h]
\begingroup
\setlength{\tabcolsep}{16pt} 
\renewcommand{\arraystretch}{1.1} 
\centering
\begin{tabular}{c | c | c | c} 
 \hline\hline
 Coefficient & 1st &  2nd & 3rd  \\ 
   \hline
  \multirow{2}{5em}{ $F_{quqd,1}^{(1)}$ } &  $A_{{\rm CP}}(B\to X\g)$ & eEDM & ${\cal B}(B\to X\g)$
    \\ 
  & $7.9$ & $1.3\times10^1$ & $1.8\times10^2$ 
  \\ 
    \hline
  \multirow{2}{5em}{ $F_{quqd,1}^{(V_q)}$ } &  $A_{{\rm CP}}(B\to X\g)$ & $B_d^0 - \bar B_d^0$ & $B_s^0 - \bar B_s^0$
    \\ 
  & $1.1$ & $2.7\times10^2$ & $5.2\times10^2$ 
  \\ 
      \hline
  \multirow{2}{5em}{ $\bar{F}_{quqd,1}^{(V_q)}$ } & $A_{{\rm CP}}(B\to X\g)$ & $B_d^0 - \bar B_d^0$ & $B_s^0 - \bar B_s^0$
    \\ 
  & $3.5$ & $6.4\times10^1$ & $1.3\times10^2$  
  \\ 
    \hline
  \multirow{2}{5em}{ $F_{quqd,1}^{(V_q^2)}$ } & $A_{{\rm CP}}(B\to X\g)$ & $B_d^0 - \bar B_d^0$ & $B_s^0 - \bar B_s^0$
    \\ 
  & $7.3\times10^2$ & $4.0\times10^5$ & $7.8\times10^5$ 
  \\ 
     \hline
  \multirow{2}{5em}{ $F_{quqd,1}^{(\Delta_u)}$ } &  nEDM  & $B_d^0 - \bar B_d^0$ & $B_s^0 - \bar B_s^0$
    \\ 
  & $1.2\times10^1$ & $1.4\times10^5$ & $2.8\times10^5$ 
  \\ 
    \hline
  \multirow{2}{5em}{ $F_{quqd,1}^{(\Delta_d)}$ } & nEDM & ${\cal B}(B\to X\g)$ & $\epsilon_K$
    \\ 
  & $2.1\times10^{-1}$ & $8.6\times10^2$ & $1.9\times10^4$
  \\ 
      \hline
  \multirow{2}{5em}{ $\bar{F}_{quqd,1}^{(\Delta_u)}$ } &  $B_d^0 - \bar B_d^0$ & $B_s^0 - \bar B_s^0$ & $\Delta A_{{\rm CP}}(D^0)$
    \\ 
  & $1.3\times10^4$ & $2.7\times10^4$ & $3.4\times10^4$ 
  \\ 
    \hline
  \multirow{2}{5em}{ $\bar{F}_{quqd,1}^{(\Delta_d)}$ } &  nEDM & ${\cal B}(B\to X\g)$ & $K_L\to\pi^0 e^+e^-$
    \\ 
  & $2.6\times10^2$ & $1.6\times10^4$ & $2.3\times10^4$
  \\ 
    \hline
    \multirow{2}{5em}{ $F_{quqd,1}^{(V_q \Delta_u)}$ } &  $\Delta A_{{\rm CP}}(D^0)$ & eEDM & $D^0 - \bar D^0$
    \\ 
  & $6.8\times10^5$ & $3.1\times10^6$ & $1.5\times10^8$ 
  \\ 
    \hline
  \multirow{2}{5em}{ $\bar{F}_{quqd,1}^{(V_q \Delta_u)}$ } &  $A_{{\rm CP}}(B\to X\g)$ & $B_d^0 - \bar B_d^0$ & $B_s^0 - \bar B_s^0$
    \\ 
  & $1.3\times10^5$ & $2.2\times10^7$ & $4.3\times10^7$ 
  \\ 
      \hline
  \multirow{2}{5em}{ $F_{quqd,1}^{(V_q \Delta_d)}$ } & ${\cal B}(B\to X\g)$  & $\epsilon'/\epsilon$ &  nEDM
    \\ 
  & $1.2\times10^3$ & $2.3\times10^5$ & $3.4\times10^5$  
  \\ 
    \hline
  \multirow{2}{5em}{ $\bar{F}_{quqd,1}^{(V_q \Delta_d)}$ } & $\epsilon'/\epsilon$  & $K_L\to\pi^0e^+e^-$ &  nEDM
    \\ 
  & $2.2\times10^4$ & $2.2\times10^5$ & $2.6\times10^5$  
  \\ 
     \hline
  \multirow{2}{5em}{ $\hat{F}_{quqd,1}^{(V_q \Delta_u)}$ } &  $A_{{\rm CP}}(B\to X\g)$ & $B_d^0 - \bar B_d^0$ & $B_s^0 - \bar B_s^0$
    \\ 
  & $1.3\times10^5$ & $2.2\times10^7$ & $4.3\times10^7$ 
  \\ 
    \hline
  \multirow{2}{5em}{ $\hat{F}_{quqd,1}^{(V_q \Delta_d)}$ } & $\epsilon'/\epsilon$ & $\epsilon_K$ & nEDM
    \\ 
  & $1.9\times10^5$ & $3.6\times10^5$ & $2.6\times10^6$  
  \\ 
 \hline\hline
\end{tabular}
\caption{Three strongest bounds on four-fermion operators $\Q_{quqd}^{(1)}$ coefficients in the $U(2)^3$ scheme.}
\label{table:BestBounds:U2:QUQD1}
\endgroup
\end{table}

\begin{table}[!h]
\begingroup
\setlength{\tabcolsep}{16pt} 
\renewcommand{\arraystretch}{1.1} 
\centering
\begin{tabular}{c | c | c | c} 
 \hline\hline
 Coefficient & 1st &  2nd & 3rd  \\ 
   \hline
  \multirow{2}{5em}{ $F_{quqd,8}^{(1)}$ } &  $A_{{\rm CP}}(B\to X\g)$ & eEDM  & ${\cal B}(B\to X\g)$
    \\ 
  & $1.1\times10^1$ & $8.0\times10^1$ & $1.9\times10^2$ 
  \\ 
    \hline
  \multirow{2}{5em}{ $F_{quqd,8}^{(V_q)}$ } &  $A_{{\rm CP}}(B\to X\g)$ & $B_d^0 - \bar B_d^0$ & $B_s^0 - \bar B_s^0$
    \\ 
  & $8.3\times10^{-1}$ & $3.2\times10^2$ & $6.1\times10^2$ 
  \\ 
      \hline
  \multirow{2}{5em}{ $\bar{F}_{quqd,8}^{(V_q)}$ } & $A_{{\rm CP}}(B\to X\g)$ & $B_d^0 - \bar B_d^0$ & eEDM
    \\ 
  & $9.3$ & $4.5\times10^4$ & $7.5\times10^4$  
  \\ 
    \hline
  \multirow{2}{5em}{ $F_{quqd,8}^{(V_q^2)}$ } & $A_{{\rm CP}}(B\to X\g)$ & $B_d^0 - \bar B_d^0$ & $B_s^0 - \bar B_s^0$
    \\ 
  & $5.5\times10^2$ & $2.6\times10^5$ & $5.1\times10^5$ 
  \\ 
     \hline
  \multirow{2}{5em}{ $F_{quqd,8}^{(\Delta_u)}$ } &  nEDM & $B_d^0 - \bar B_d^0$ &  $\Delta A_{{\rm CP}}(D^0)$
    \\ 
  & $1.4\times10^2$ & $3.1\times10^5$ & $3.5\times10^5$ 
  \\ 
    \hline
  \multirow{2}{5em}{ $F_{quqd,8}^{(\Delta_d)}$ } & nEDM &  ${\cal B}(B\to X\g)$ & $K_L\to\pi^0 e^+e^-$
    \\ 
  & $4.5\times10^{-1}$ & $1.0\times10^3$ & $1.7\times10^5$
  \\ 
      \hline
  \multirow{2}{5em}{ $\bar{F}_{quqd,8}^{(\Delta_u)}$ } & nEDM & $\Delta A_{{\rm CP}}(D^0)$ & $B_d^0 - \bar B_d^0$
    \\ 
   & $1.9\times10^5$ & $2.6\times10^5$ & $7.9\times10^5$
  \\ 
    \hline
  \multirow{2}{5em}{ $\bar{F}_{quqd,8}^{(\Delta_d)}$ }  & nEDM & ${\cal B}(B\to X\g)$ & $K_L\to\pi^0 e^+e^-$
    \\ 
   & $2.9\times10^2$ & $7.0\times10^3$ & $2.2\times10^4$
  \\ 
    \hline
    \multirow{2}{5em}{ $F_{quqd,8}^{(V_q \Delta_u)}$ } &  $\Delta A_{{\rm CP}}(D^0)$ & eEDM & $D^0\to\rho\g$ 
    \\ 
  & $7.4\times10^6$ & $1.4\times10^8$ & $5.3\times10^8$ 
  \\ 
    \hline
  \multirow{2}{5em}{ $\bar{F}_{quqd,8}^{(V_q \Delta_u)}$ } &  $A_{{\rm CP}}(B\to X\g)$ & $B_d^0 - \bar B_d^0$ & $B_s^0 - \bar B_s^0$
    \\ 
  & $8.8\times10^4$ & $1.3\times10^8$ & $2.6\times10^8$ 
  \\ 
      \hline
  \multirow{2}{5em}{ $F_{quqd,8}^{(V_q \Delta_d)}$ } & ${\cal B}(B\to X\g)$ & nEDM & $\epsilon'/\epsilon$
    \\ 
  & $1.3\times10^3$ & $2.0\times10^6$ & $2.0\times10^6$
  \\ 
    \hline
  \multirow{2}{5em}{ $\bar{F}_{quqd,8}^{(V_q \Delta_d)}$ }  & $\epsilon'/\epsilon$ & $K_L\to\pi^0e^+e^-$ & nEDM
    \\ 
    & $1.7\times10^5$ & $2.1\times10^5$ & $4.4\times10^5$ 
  \\ 
     \hline
  \multirow{2}{5em}{ $\hat{F}_{quqd,8}^{(V_q \Delta_u)}$ } &  $A_{{\rm CP}}(B\to X\g)$ & $B_d^0 - \bar B_d^0$ & $B_s^0 - \bar B_s^0$
    \\ 
  & $8.8\times10^4$ & $1.3\times10^8$ & $2.6\times10^8$ 
  \\ 
    \hline
  \multirow{2}{5em}{ $\hat{F}_{quqd,8}^{(V_q \Delta_d)}$ } & nEDM & $\epsilon'/\epsilon$ & $K_L\to\pi^0e^+e^-$
    \\ 
   & $2.6\times10^5$ & $6.2\times10^5$ & $1.7\times10^6$ 
  \\ 
 \hline\hline
\end{tabular}
\caption{Three strongest bounds on four-fermion operators $\Q_{quqd}^{(8)}$ coefficients in the $U(2)^3$ scheme. }
\label{table:BestBounds:U2:QUQD8}
\endgroup
\end{table}

%
%

\begin{table}[!h]
\begingroup
\setlength{\tabcolsep}{16pt} 
\renewcommand{\arraystretch}{1.1} 
\centering
\begin{tabular}{c | c | c | c} 
 \hline\hline
 Coefficient & 1st &  2nd & 3rd  \\ 
   \hline
  \multirow{2}{5em}{ $F_{qu,1}^{(V_q)} $ } &  $\Delta A_{{\rm CP}}(D^0)$ &  $A_{{\rm CP}}(B\to X\g)$ & $D^0 - \bar D^0$
    \\ 
  & $3.9\times10^4$ & $3.7\times10^5$ & $8.0\times10^5$ 
  \\ 
    \hline
  \multirow{2}{5em}{ $\bar{F}_{qu,1}^{(V_q)} $ } & $A_{{\rm CP}}(B\to X\g)$ & $\Delta A_{{\rm CP}}(D^0)$ & nEDM
    \\ 
  & $1.0\times10^3$ & $3.9\times10^4$ & $1.1\times10^5$
  \\ 
      \hline
  \multirow{2}{5em}{ $F_{qu,1}^{(\Delta_u)} $ } & nEDM & $\Delta A_{{\rm CP}}(D^0)$ & eEDM
    \\ 
  & $3.8$ & $2.5\times10^3$  & $5.1\times10^3$
  \\ 
    \hline
  \multirow{2}{5em}{ $F_{qu,1}^{(V_q \Delta_u)} $ } & $\Delta A_{{\rm CP}}(D^0)$ & nEDM & $D^0 - \bar D^0$
    \\ 
   & $4.9\times10^4$ & $1.3\times10^5$ & $1.0\times10^6$
  \\ 
     \hline
  \multirow{2}{5em}{ $\bar{F}_{qu,1}^{(V_q \Delta_u)}$ }  & nEDM & $D^0 - \bar D^0$ & /
    \\ 
   & $5.0\times10^7$ & $2.7\times10^8$ & /
  \\ 
    \hline
  \multirow{2}{5em}{ $\hat{F}_{qu,1}^{(V_q \Delta_u)}$ } & nEDM & $D^0 - \bar D^0$ & /
    \\ 
   & $5.0\times10^7$ & $2.7\times10^8$ & /
  \\ 
      \hline
      \hline
   \multirow{2}{5em}{ $F_{qu,8}^{(V_q)} $ } &  $\Delta A_{{\rm CP}}(D^0)$ &  $A_{{\rm CP}}(B\to X\g)$ & $D^0 - \bar D^0$
    \\ 
  & $2.3\times10^5$ & $3.2\times10^5$ & $6.0\times10^5$ 
  \\ 
    \hline
  \multirow{2}{5em}{ $\bar{F}_{qu,8}^{(V_q)} $ } & $A_{{\rm CP}}(B\to X\g)$ & nEDM & $\Delta A_{{\rm CP}}(D^0)$
    \\ 
  & $2.2\times10^3$ & $7.6\times10^4$ & $2.4\times10^5$ 
  \\ 
      \hline
  \multirow{2}{5em}{ $F_{qu,8}^{(\Delta_u)} $ } & nEDM  & eEDM & $\Delta A_{{\rm CP}}(D^0)$
    \\ 
  & $4.1$ & $3.8\times10^3$ & $1.5\times10^4$
  \\ 
    \hline
  \multirow{2}{5em}{ $F_{qu,8}^{(V_q \Delta_u)} $ } & nEDM & $\Delta A_{{\rm CP}}(D^0)$ &  $D^0 - \bar D^0$
    \\ 
  & $1.0\times10^5$ & $2.9\times10^5$ & $7.5\times10^5$
  \\ 
     \hline
  \multirow{2}{5em}{ $\bar{F}_{qu,8}^{(V_q \Delta_u)}$ }  & nEDM &  $D^0 - \bar D^0$ & /
    \\ 
  & $3.7\times10^7$ & $2.0\times10^8$ & /
  \\ 
    \hline
  \multirow{2}{5em}{ $\hat{F}_{qu,8}^{(V_q \Delta_u)}$ } & nEDM &  $D^0 - \bar D^0$ & /
    \\ 
  & $3.7\times10^7$ & $2.0\times10^8$ & /
  \\ 
 \hline\hline
\end{tabular}
\caption{Three strongest bounds on four-fermion operators $\Q_{qu}^{(1,8)}$ in the $U(2)^3$ scheme.}
\label{table:BestBounds:U2:QU}
\endgroup
\end{table}

\clearpage

\end{appendix}

\bibliography{cedm}
\bibliographystyle{h-physrev}
\end{document}